\documentclass[12pt]{article}
\usepackage[psamsfonts]{amsfonts}
\usepackage{amsmath,amssymb,amsthm}
\usepackage[dvips]{graphicx}

\def\UseSection{
        \numberwithin{equation}{section}
	\theoremstyle{plain}
        \newtheorem{theorem}    {Theorem}[section]
        \DefineTheorems 
}

\def\DefineTheorems{
	
	\newtheorem{lemma}      [theorem] {Lemma}
	
	\newtheorem{prop}       [theorem] {Proposition}
	
	\newtheorem{cor}        [theorem] {Corollary}

	\theoremstyle{definition}
	\newtheorem{defn}       [theorem] {Definition}

	\newtheorem{rk} 	[theorem] {Remark}
	\theoremstyle{definition}

}

\newcommand{\bt}   {\begin{theorem}}
\newcommand{\et}   {\end  {theorem}}
\newcommand{\bl}   {\begin{lemma}}
\newcommand{\el}   {\end  {lemma}}
\newcommand{\bp}   {\begin{prop}}
\newcommand{\ep}   {\end  {prop}}
\newcommand{\bc}   {\begin{cor}}
\newcommand{\ec}   {\end  {cor}}
\newcommand{\bd}   {\begin{defn}}
\newcommand{\ed}   {\end  {defn}}

\newcommand{\ba}   {\begin{array}}
\newcommand{\ea}   {\end  {array}}
\newcommand{\be}   {\begin{enumerate}}
\newcommand{\ee}   {\end  {enumerate}}
\newcommand{\bi}   {\begin{itemize}}
\newcommand{\ei}   {\end  {itemize}}

\def\eq#1\en{\begin{equation}#1\end{equation}}  
\def\eqsplit#1\ensplit{
	\begin{equation}\begin{split}#1\end{split}\end{equation}
	}
\def\eqalign#1\enalign{
	\begin{align}#1\end{align}
	}
\def\eqmul#1\enmul{
	\begin{multline}#1\end{multline}
	}
\newcommand{\eqarrstar} {\begin{eqnarray*}} 
\newcommand{\enarrstar} {\end{eqnarray*}} 
\newcommand{\eqarray}   {\begin{eqnarray}} 
\newcommand{\enarray}   {\end{eqnarray}} 
\newcommand{\nnb}	{\nonumber \\} 

\newcommand{\lbeq}[1]  {\label{e:#1}}
\newcommand{\refeq}[1] {\eqref{e:#1}}    
\newcommand{\lbfg}[1]  {\label{fg: #1}}
\newcommand{\reffg}[1] {\ref{fg: #1}}

\makeatletter
\newcommand{\labelcounter}[2]{{%
	\stepcounter{#1}
	\protected@write\@auxout{}%
	{\string\newlabel{#2}{{\csname the#1\endcsname}{\thepage}}}%
	{\ref{#2}}
	}}
\makeatother
\newcommand{\Ebold} {{\mathbb E}}

\newcommand{\Pbold} {{\mathbb P}}

\newcommand{\Rbold} {{\mathbb R}}

\newcommand{\Zbold} {{\mathbb Z}}

\newcommand{\Acal}   {\mathcal{A}} 
 
\newcommand{\Ccal}   {\mathcal{C}} 
\newcommand{\Dcal}   {\mathcal{D}}

\newcommand{\Gcal}   {\mathcal{G}}

\newcommand{\Lcal}   {\mathcal{L}}

\newcommand{\Scal}   {\mathcal{S}} 
\newcommand{\Tcal}   {\mathcal{T}} 
\newcommand{\Ucal}   {\mathcal{U}} 
 
\newcommand{\Wcal}   {\mathcal{W}}

\newcommand{\Dhat} {{\hat{D} }}

\newcommand{\Rd}    {{ {\Rbold}^d}}
\newcommand{\Zd}    {{ {\Zbold}^d }}

\newcommand{\spose}[1] {{\hbox to 0pt{#1\hss}} }
\newcommand{\ltapprox} {\mathrel{\spose{\lower 3pt\hbox{$\mathchar"218$}}
 \raise 2.0pt\hbox{$\mathchar"13C$}}}
\newcommand{\gtapprox} {\mathrel{\spose{\lower 3pt\hbox{$\mathchar"218$}}
 \raise 2.0pt\hbox{$\mathchar"13E$}}}

\newcommand{\ddk}  {\frac{d^d k}{(2\pi)^d}}

\newcommand{\kx}   {{  k \cdot x }}

\newcommand{\nin}  {{ \not\in }}

\newcommand{\prodtwo}[2]{
	\prod_{ \mbox{ \scriptsize 
		$\begin{array}{c} 
		{#1} \\ 
		{#2}  
		\end{array} $ } 
		} 
	} 

\UseSection  
\renewcommand{\to}      {\rightarrow}
\newcommand{\Z}{{\mathbb Z}}

\renewcommand{\kx}  {k\hspace{-0.05ex}\cdot\hspace{-0.05ex}x}

\newcommand{\intddkpi}  {\int\limits_{[-\pi,\pi]^{d}}\!\ddk}

\newcommand{\Abegin}{A^{(0)}}
\newcommand{\oldB}{T}
\newcommand{\oldBfour}{H}
\newcommand{\Aend}{A^{(\text{end})}} 

\newcommand\kakenhi{Grant-in-Aid for Scientific Research (C)}
\newcommand\monbushou{the Ministry of Education, Science, Sports and Culture}

\title  {Critical two-point functions and the lace expansion
    for spread-out high-dimensional
    percolation and related models
        }

\author {Takashi Hara \\
        Graduate School of Mathematics \\
        Nagoya University \\
        Chikusa-ku \\
        Nagoya 464-8602, Japan \\
        {\small\tt hara@math.nagoya-u.ac.jp}\vspace*{-3pt}\\ 
 		{\footnotesize\tt http://www.math.nagoya-u.ac.jp/\~{}hara} \\
    \and
    Remco van der Hofstad\\
        Stieltjes Institute for Mathematics\\
        Delft University\\
        Mekelweg 4\\
        2628 CD Delft, The Netherlands\\
        {\small\tt R.W.vanderHofstad@its.tudelft.nl}\vspace*{-3pt}\\
 		{\footnotesize\tt http://ssor.twi.tudelft.nl/\~{}hofstad} \\
		\vspace{6pt}
        \and
        Gordon Slade \\
        Department of Mathematics \\
        University of British Columbia\\
        Vancouver, BC, Canada V6T 1Z2 \\
        {\small\tt slade@math.ubc.ca}\vspace*{-3pt}\\
 		{\footnotesize\tt http://www.math.ubc.ca/people/faculty/slade/index.html}
        }

\begin{document}

	\maketitle

\begin{abstract}
We consider spread-out models of self-avoiding walk, bond percolation,
lattice trees and bond lattice animals on $\Zd$,
having long finite-range connections, above their upper
critical dimensions $d=4$ (self-avoiding walk), $d=6$ (percolation)
and $d=8$ (trees and animals).  
The two-point functions for these
models are respectively the generating function for self-avoiding walks
from the origin to $x \in \Zd$, the probability of a connection from
$0$ to $x$, and the generating function for lattice trees or lattice
animals containing $0$ and $x$.  We use the lace expansion to
prove that for sufficiently spread-out models above the upper critical
dimension, the two-point function of each model decays, at the critical
point, as a multiple of $|x|^{2-d}$ as $x \to \infty$.
We use a new unified method to prove convergence of the lace expansion.
The method is based on $x$-space methods
rather than the Fourier transform.  Our results also yield unified and
simplified proofs of the bubble
condition for self-avoiding walk, the triangle condition for percolation,
and the square condition for lattice trees and lattice animals, for
sufficiently spread-out models above the upper critical dimension.
\end{abstract}

\tableofcontents

\section{Introduction}
\label{sec-intro}

\subsection{Critical two-point functions}
\label{sec-1.1}
In equilibrium statistical mechanical models at criticality,
correlations typically decay according to a power law, rather
than exponentially as is the case away from the critical point.
We consider models of self-avoiding walks, 
bond percolation, lattice trees and bond lattice
animals on the lattice $\Zd$.  Let $|x|$ denote
the Euclidean norm of $x \in \Zd$.
Assuming translation invariance, and denoting the critical
two-point function for any one of these models by
$U_{p_c}(x,y) = U_{p_c}(y-x)$, the power-law decay is traditionally written as
        \eq
        \lbeq{2ptdecay}
        U_{p_c}(x) \sim \mbox{const.}\frac{1}{|x|^{d-2+\eta}},
        \quad \mbox{as $|x| \to \infty$}.
        \en
The critical exponent $\eta$ is known as the {\em anomalous dimension},\/
and depends on the model under consideration.
Its value is believed to depend on $d$ but otherwise to be {\em universal}\/,
which means insensitive to
many details of the model's definition.

The above models have {\em upper critical dimensions}\/
        \eq
        \lbeq{dcdef}
        d_c = \left\{ \begin{array}{ll}
        4 & \mbox{for self-avoiding walk} \\
        6 & \mbox{for percolation} \\
        8 & \mbox{for lattice trees and lattice animals}
        \end{array} \right.
        \en
above which critical exponents cease to depend on the dimension.
Our purpose in this paper is to prove \refeq{2ptdecay}
for $d>d_c$, with $\eta = 0$, for certain
long-range models having a small parameter.  The small parameter is
used to ensure convergence of the {\em lace expansion}.  There is now
a large literature on the lace expansion, but proving
\refeq{2ptdecay} for $d>d_c$ remained an open question.
To make this paper more self-contained, a review of the basic steps
in the derivation of the lace expansion will be included.

All past approaches to the lace expansion have relied heavily on the Fourier
transform of the two-point function.  We present a new approach to
the lace expansion, based directly in $x$-space.  Our approach provides
a unified proof of convergence of the expansion, with most of the analysis
applying simultaneously to all the models under consideration.  
There is one model-dependent
step in the convergence proof, involving estimation of certain Feynman
diagrams.  The Feynman diagrams are model-specific, and converge
when $d>d_c$.  This is the key place where the assumption $d>d_c$ enters
the analysis.  We use a new method to estimate the relevant Feynman
diagrams, based in $x$-space rather than using the Fourier transform.

As we will explain in more detail below, weaker versions of \refeq{2ptdecay}
have been obtained previously, for the Fourier transform of the two-point
function.  These statements for the Fourier transform follow as corollaries
from our $x$-space
results.  In addition, our results immediately imply the bubble, square
and triangle conditions for sufficiently spread-out models of
self-avoiding walks, lattice trees and lattice animals, and percolation,
for $d>d_c$.
These diagrammatic conditions, which had been obtained previously
using Fourier methods, are known to imply existence (with mean-field
values) of various critical exponents.

For $d \leq d_c$, it remains an open question to prove the existence
of $\eta$.  In fact, it has not been proved for self-avoiding walk
nor for lattice trees or animals that $U_{p_c}(x)$ is even finite 
for $2 \leq d \leq d_c$.
For percolation, the two-point function is a probability, so it is
certainly finite.  However, it has not been proved for $2 \leq d \leq 6$
that it approaches zero
as $|x| \to \infty$, except for $d=2$
\cite{Kest82}.  Such a result is known to imply absence of percolation at
the critical point \cite{AKN87}, which, for general dimensions,
is an outstanding open problem in percolation theory. 

For self-avoiding walks, partial results suggesting that
$\eta = 0$ for $d=d_c=4$ have been obtained in \cite{BEI92} for
a hierarchical lattice and in \cite{IM94} for a variant of
the Edwards model.
Contrary to other critical exponents at the upper critical dimension,
no logarithmic factors appear to leading order.
It is believed that $\eta >0$ for self-avoiding walk for $2 \leq d <4$ 
\cite{MS93}.
Interestingly, there is numerical evidence that $\eta <0$ for
percolation when $3 \leq d <6$ \cite{AMAH90}, and it
has been conjectured that $\eta <0$ also
for lattice trees and lattice animals
when $2 \leq d < 8$ \cite{BFG86}  (see also \cite{PS81} for $d=3$
and \cite{LI79} for $d =8-\epsilon$).  
The exponent $\eta$ is believed to be related to the exponents $\gamma$
for the susceptibility and $\nu$ for the correlation length by
the scaling relation $\gamma =(2-\eta)\nu$.  Some exact but
nonrigorous values of 
$\gamma$ and $\nu$ have been predicted (see \cite{Grim99,Hugh96,MS93,PS81}), 
which lead to the exact predictions
$\eta = \frac{5}{24}$ for 2-dimensional self-avoiding walk and percolation,
and $\eta = -1$ for 3-dimensional lattice trees and animals.

\subsection{Main results}
\label{sub-1.2}
The spread-out models are
defined in terms of a function $D : \Zd \to [0,\infty)$,
which depends on a positive parameter $L$.  We will take $L$ to be large,
providing a small parameter $L^{-1}$.
We will consider only those $D$ which obey the conditions imposed
in the following definition.

\begin{defn}
\label{def-Dsp1}
Let $h$ be a non-negative bounded function on $\Rd$ which is
piecewise continuous,
symmetric under the lattice symmetries, supported in $[-1, 1]^{d}$, 
and normalised
so that $\int_{[-1, 1]^{d}}h(x) d^d x = 1$.  
Then for large $L$ we define
    \eq
    \lbeq{Dbasic2}
    D(x) = \frac{h(x/L)}{\sum_{x\in \Zd}h(x/L)}.
    \en
Since $\sum_{x\in \Zd}h(x/L) \sim L^{d}$ (using a Riemann sum
approximation to $\int_{[-1, 1]^{d}}h(x)d^dx$), 
the assumption that $L$ is large
ensures that the denominator of \refeq{Dbasic2}
is nonzero.
We also define $\sigma^2 = \sum_x |x|^2 D(x)$.
\end{defn}

The sum  $\sum_x |x|^p D(x)$ can be
regarded as a Riemann sum, and is asymptotic
to a multiple of $L^p$ for $p>0$.  In particular, $\sigma$ and $L$
are comparable.
A basic example obeying the conditions of Definition~\ref{def-Dsp1} is
given by the function $h(x)=2^{-d}$ for $x \in [-1,1]^d$, $h(x)=0$
otherwise, for which $D(x) = (2L+1)^{-d}$ for $x \in [-L,L]^d \cap \Zd$,
$D(x)=0$ otherwise.

\smallskip
Next, we define the models we consider.
Let $\Omega_D = \{ x \in \Zd : D(x) >0\}$.  By Definition~\ref{def-Dsp1},
$\Omega_D$ is finite and $\Zd$-symmetric.
A {\em bond}\/ is a pair of sites
$\{x,y\} \subset \Zd$ with $y-x \in \Omega_D$.  For $n \geq 0$,
an $n$-step {\em walk}\/ from $x$ to $y$ is a mapping
$\omega : \{0,1,\ldots, n\} \to \Zd$ such that $\omega(i+1)-\omega(i) \in
\Omega_D$ for $i=1,\ldots, n-1$.
We sometimes consider a walk to be a set of bonds,
rather than a set of sites.
Let $\Wcal(x,y)$ denote the set of walks
from $x$ to $y$, taking any number of steps.
An $n$-step {\em self-avoiding walk}
is an $n$-step walk $\omega$ such that
$\omega(i) \neq \omega(j)$ for each pair $i \neq j$.
Let $\Scal(x,y)$ denote the set of self-avoiding walks from $x$ to $y$,
taking any number of steps.
A {\em lattice tree}\/ is a finite connected set of bonds
which has no cycles.
A {\em lattice animal}\/
is a finite connected set of
bonds which may contain cycles.
Although a tree $T$ is defined as a set of bonds, we
write  $x \in T$ if $x$ is an endpoint of some bond of $T$, and similarly
for lattice animals.
Let $\Tcal(x,y)$ denote the set of lattice trees containing $x$ and $y$,
and let $\Acal(x,y)$ denote the set of lattice animals containing $x$ and $y$.

Given a finite set $B$
of bonds and a nonnegative parameter $p$, we define its {\em weight}\/ to be
\eq
    W_{p,D}(B) = \prod_{\{x,y\} \in B} pD(y-x).
\en
If $B$ is empty, we set $W_{p,D}(\varnothing)=1$.
The random walk and self-avoiding walk two-point functions are defined
respectively by
\eq
    S_p(x) = \sum_{\omega \in \Wcal(0,x)} W_{p,D}(\omega),
    \quad \quad
    \sigma_p(x) = \sum_{\omega \in \Scal(0,x)} W_{p,D}(\omega).
\en
For any $d>0$, $\sum_x S_p(x)$ converges for $p < 1$
and diverges for $p>1$, and $p=1$ plays the role of a critical point.
It is well-known \cite{Uchi98} that, for $d>2$,
\eq
\lbeq{Cxdecay}
    S_1(x) \sim \mbox{const.}\frac{1}{|x|^{d-2}},
    \quad \mbox{as $|x| \to \infty$},
\en
so that $\eta = 0$.  A standard subadditivity argument
\cite{HM54,Hugh95,MS93} implies that $\sum_x \sigma_p(x)$
converges for $p<p_c$ and diverges for $p>p_c$, for some finite positive
critical value $p_c$.

The lattice tree and lattice animal two-point functions are defined by
\eq
\lbeq{1.1}
    \rho_p (x) = \sum_{T \in \Tcal(0,x)} W_{p,D}(T) , \quad \quad
    \rho_p^a (x) = \sum_{A \in \Acal(0,x)} W_{p,D}(A) .
\en
A standard
subadditivity argument implies that there are positive finite
$p_c$ and $p_c^a$ such that $\sum_x \rho_p^{(a)}(x)$ converges
for $p<p_c^{(a)}$
and diverges for $p>p_c^{(a)}$ \cite{Klar67,Klei81}.

Turning now to bond percolation,
we associate independent Bernoulli random variables
$n_{\{x,y\}}$ to each  bond $\{x,y\}$, with
        \eq
        \Pbold(n_{\{x,y\}}=1)=p D(x-y)\qquad \Pbold(n_{\{x,y\}}=0)=1- p D(x-y),
        \en
where $p \in [0, ( \max_x D(x))^{-1}]$.  
(Note that $p$ is not a probability.)
A configuration is a
realisation of the bond
variables.  Given a configuration,
a bond $\{x,y\}$ is called {\em occupied}\/ if
$n_{\{x,y\}}=1$ and otherwise is called {\em vacant}.\/  Let $C(x)$ denote
the random set of sites $y$ such that there is a path from $x$ to $y$ 
consisting of occupied bonds.  The percolation {\em two-point function}\/
is defined by
\eq
    \tau_p(x) = \Pbold_p(x \in C(0)),
\en
where $\Pbold_p$ is the probability measure on configurations induced by
the bond variables.
There is a critical value $p_c \in (0,1)$ such that 
$\sum_{x} \tau_p(x) < \infty$
for $p \in [0,p_c)$
and $\sum_{x} \tau_p(x) = \infty$ for $p \geq p_c$.
This critical point can also be characterised by the fact that
the probability of existence of an infinite cluster of occupied bonds is
$1$ for $p>p_c$ and $0$ for $p<p_c$ \cite{AB87,Mens86}.

We use $U_p(x)$ to refer to the two-point function of all
models simultaneously.
We use $p_c$ to denote the critical points for the different models,
although they are, of course, model-dependent.  In what follows,
it will be clear from the context which model is intended.

Let
\eq
    a_d = \frac{d\Gamma(\frac{d}{2}-1)}{2\pi^{d/2}}.
\en
We write
$O(f(x,L))$ to denote a quantity bounded by $\mbox{const.} f(x,L)$,
with a constant
that is independent of $x$ and $L$ but may depend on $d$.
We define $\epsilon$ by
\eq
\lbeq{epsilon-def}
    \epsilon = \begin{cases}
    2(d-4) & (\text{self avoiding walk}) \\
    d-6     & (\text{percolation}) \\
    d-8     & (\text{lattice trees and animals})
    \end{cases}
\en
and write
\eq
	\epsilon_2 = \epsilon \wedge 2.
\en
Our main result is the following theorem.

\begin{theorem}
\label{thm-main}
Let $U_{p_c}(x)$ denote the
critical two-point function for self-avoiding walk, percolation,
lattice trees or lattice animals.
Let $d > d_c$, and fix any $\alpha >0$.
There is a constant $A= 1+O(L^{-2+\alpha})$ depending on $d$, $L$ and the model, 
and an
$L_0$ depending on $d$, $\alpha$ and the model,
such that for $L \geq L_0$ 
        \eq
        \lbeq{tauasy}
        U_{p_c}(x) = \frac{a_dA}{\sigma^2|x|^{d-2}} \left[
        1 +
        O\Big(\frac{L^{\epsilon_2 +\alpha}}
        {|x|^{\epsilon_2-\alpha} }\Big)    +
        O\Big(\frac{L^2}
        {|x|^{2-\alpha} }\Big)
        \right]
        \en
as $|x| \to \infty$.  Constants in the error terms for \refeq{tauasy} and
$A-1$ depend on $\alpha$.  
\end{theorem}

We expect that Threorem~\ref{thm-main} remains true with $\alpha = 0$,
but it is convenient in our analysis to allow a small power of $L$ or
$|x|$ to enter into error estimates.
Results closely related to Theorem~\ref{thm-main},
for nearest-neighbour models in very high dimensions,
are proved in \cite{Hara00} using a different method.

The leading asymptotics of the critical random walk two-point function
$S_1(x)$ are also given by \refeq{tauasy}, with $A=1$.  This will be
discussed in detail, in Proposition~\ref{prop-A} below.
The second error term in \refeq{tauasy} represents an error term in the 
expansion for random walk, while the first error term represents the difference
between random walk and the other models.
The fact that the power $|x|^{2-d}$ appears as the leading power
in \refeq{tauasy},
independent of the precise form of $D$ or
the value of large $L$, is an illustration of universality.

As was pointed out in Section~\ref{sec-1.1}, it is a consequence of
\refeq{tauasy} for percolation that there is no percolation at the critical
point.  In other words, for $d >6$ and for $L$ large, with probability $1$
there is no infinite cluster of occupied bonds when $p=p_c$.
There are, however, large emerging structures present at $p=p_c$ that are 
loosely referred to as the incipient infinite cluster.  
The result of Theorem~\ref{thm-main} for percolation provides a
necessary ingredient for work of Aizenman~\cite{Aize97} in this regard.
Roughly speaking, Aizenman showed that if a (then unproved)
weaker statement than
\refeq{tauasy} holds for $d>6$, then at $p_c$ the largest percolation clusters
present within a box of side length $M$ are of size approximately
$M^4$ and are approximately $M^{d-6}$ in number.  Details can be
found in \cite{Aize97}.  Equation~\refeq{tauasy} now implies that
Aizenman's conclusions do hold for sufficiently spread-out models with $d>6$.

The following corollary will follow immediately from Theorem~\ref{thm-main}.
The conclusion of the corollary was proved previously
in \cite{MS93}
for self-avoiding walk, in \cite{HS90a} for
percolation, and in \cite{HS90b} for lattice trees and lattice animals.
The corollary is known to imply existence (with mean-field values) 
of various critical exponents \cite{AN84,BA91,MS93,TH87}.

\begin{cor}
\label{cor-bub}
For $d > d_c$ and $L \geq L_0$, the self-avoiding walk bubble condition,
the percolation triangle condition, and the lattice tree and lattice animal
square conditions all hold.
These diagrammatic conditions are respectively the statements that
the following sums are finite:
\[
    \sum_{x \in \Zd} \sigma_{p_c}(x)^2, \quad
    \sum_{x,y \in \Zd} \!\!\tau_{p_c}(0,x)\tau_{p_c}(x,y)\tau_{p_c}(y,0),
    \quad
    \sum_{w,x,y \in \Zd}\!\!\!
    \rho^{(a)}_{p_c^{(a)}}(w)\rho^{(a)}_{p_c^{(a)}}(x-w)
    \rho^{(a)}_{p_c^{(a)}}(y-x)
    \rho^{(a)}_{p_c^{(a)}}(y).
\]
\end{cor}

Theorem~\ref{thm-main} implies a related result for the Fourier transform
of the critical two-point function.
Given an absolutely summable
function $f$ on $\Zd$, we denote its Fourier transform by
        \eq
        \lbeq{FTdef}
        \hat{f}(k) = \sum_{x \in \Zd} f(x) e^{ik\cdot x},
        \quad\quad k \in [-\pi,\pi]^d.
        \en
In general, \refeq{2ptdecay} can be expected to correspond to
\eq
\lbeq{etak}
    \hat{U}_{p_c}(k) \sim \mbox{const.} \frac{1}{|k|^{2-\eta}}, \quad
    \mbox{as $k \to 0$.}
\en
However, some care is required with this correspondence.
In particular, if $\eta=0$ then $U_{p_c}(x)$ is not summable,
and hence its Fourier transform is not
well-defined.  Moreover, the inverse Fourier transform of a function
asymptotic to a multiple of $|k|^{-2}$,
which does exist for $d>2$, is not necessarily asymptotic to a multiple of
$|x|^{2-d}$ without further assumptions.
A counterexample is given in \cite[page~32]{MS93}.

The situation is well-understood for random walk \cite{Spit76}.
For $d >2$, it is the
case that $S_1(x)$ is given by the inverse Fourier transform
\eq
\lbeq{FTinv}
    S_1(x) = \int_{[-\pi,\pi]^d} \frac{e^{-ik\cdot x}}{1-\hat{D}(k)}
     \frac{d^dk}{(2\pi)^d} .
\en
Therefore, it is reasonable to assert that
\eq
\lbeq{Chatdef}
    \hat{S}_1(k) = \frac{1}{1-\hat{D}(k)},
\en
even though $S_1(x)$ is not summable.
Our assumptions on $D$ imply that
$1-\hat{D}(k) \sim \sigma^2 |k|^2/(2d)$ as $k \to 0$.
Comparing with \refeq{etak} and \refeq{Chatdef},
this gives the $k$-space version of the statement that $\eta = 0$ for
random walk.

For the models of Theorem~\ref{thm-main}, we
have the following corollary.
A proof of the corollary will be given
in Section~\ref{sec-pfmain}.
The quantity $\hat{U}_{p_c}(k)$
appearing in the corollary represents
the Fourier transform of the corresponding
$x$-space two-point functions $U_{p_{c}}(x)$,
in the sense that the $x$-space two-point
functions are given by the inverse Fourier transform of the $k$-space
quantities.  It will be part of the proof to demonstrate this correspondence.
Recall that $\epsilon_2 = \epsilon \wedge 2$.

\begin{cor}
\label{cor-k}
For $d > d_c$  and $L \geq L_0$, the Fourier transforms of the
critical two-point functions of the models of Theorem~\ref{thm-main} obey
\eq
    \hat{U}_{p_c}(k)
    = \frac{2dA}{\sigma^2|k|^{2}} \left[ 1+\Delta_L(k) \right], \qquad 
	| \Delta_{L}(k) | \leq 
	\begin{cases}
	\mbox{const}. \,  |k|^{\epsilon_2} & (\epsilon \neq 2) \\
	\mbox{const}. \, |k|^{2} \,  \log |k|^{-1} & (\epsilon =
	2)
	\end{cases} 
\en
as $k \to 0$, with an $L$-dependent constant in the error term $\Delta_{L}$.
The constant $A$ is the same as the constant of Theorem~\ref{thm-main}.
\end{cor}

The conclusion of Corollary~\ref{cor-k} for self-avoiding walk
was established in \cite[Theorem~6.1.6]{MS93}, with $|\Delta_L(k)| 
\leq \mbox{const.}|k|^a$
for any
$a < \frac{d-4}{2}\wedge 1$.  For percolation, it was
proved in \cite[Theorem~1.1]{HS00b}
that, under the hypotheses of Corollary~\ref{cor-k},
$\lim_{k \to 0}|k|^2 \hat{\tau}_{p_c}(k) = A$, with no error
estimate but with joint control in the limit $(k,h) \to (0,0)$,
where $h$ is a magnetic field.
For lattice trees, the conclusion of Corollary~\ref{cor-k}
was implicitly proved in \cite{DS98}, with $|\Delta_L(k)| 
\leq \mbox{const.}|k|^a$
for some unspecified $a>0$.

The proof of Theorem~\ref{thm-main} also yields the following result
for the asymptotic behaviour of the critical points of self-avoiding
walk and percolation.  We do not obtain such a result for lattice trees
and lattice animals.
Much stronger results have
been obtained for nearest-neighbour self-avoiding walk and percolation
by pushing lace expansion methods further \cite{HS95}.
See \cite{Penr94} for related results obtained without using the
lace expansion, including for lattice trees.

\begin{cor}
\label{cor-pc}
Let $\alpha >0$.
For self-avoiding walk and percolation with $d > d_c$,
as $L \to \infty$
\eq
\lbeq{pc-saw}
    1 \leq p_c \leq 1 + O(L^{-2+\alpha}).
\en
\end{cor}

In \cite{HS01a,HS01b}, \refeq{pc-saw} 
is improved to $1 \leq p_c \leq 1 + O(L^{-d})$
for self-avoiding walk.

\subsection{Overview of the proof}
\label{sub-prfoverview}

In this section, we isolate four propositions which will be combined in
Section~\ref{sec-pfmain} to prove Theorem~\ref{thm-main}.

We define $I$ by $I(x) = \delta_{0,x}$, and denote the convolution of
two functions $f,g$ on $\Zd$ by
\eq
    (f*g)(x) = \sum_{y \in \Zd} f(x-y)g(y).
\en
Consider the random walk two-point function $S_z(x)$.
By separating out the contribution from the zero-step walk, and extracting
the contribution from the first step in the other walks, $S_z$ can be
seen to obey the convolution equation
\eq
\lbeq{sle}
    S_z = I + (z D*S_z).
\en
The lace expansion is a modification of this convolution equation,
for the models we are considering, that
takes interactions into account via a kind of inclusion-exclusion.

To state the lace expansion in a unified fashion, a change of variables
is required.  This change of variables is explained in Section~\ref{sec-le}.
To each $p \leq p_c$, we associate 
\eq\lbeq{zp}
    z = \begin{cases}
    p & (\text{self-avoiding walk and percolation})\\
    p \rho_{p}^{(a)}(0) & (\text{lattice trees and animals}).
    \end{cases}
\en
We denote by $z_{c}$ the value which corresponds to $p_{c}$ in the
above definition.  It is possible in principle that $z_c=\infty$ for
lattice trees and animals, but we will rule out this possibility in 
Section~\ref{sec-pfmain}, and we proceed in this section under the
assumption that $\rho_{p_c}^{(a)}(0) < \infty$.  
Since the right hand side of \refeq{zp} is increasing in $p$, it defines a
one-to-one mapping.  
For $p=p(z)$ given by \refeq{zp}, we also define
\eq
\lbeq{Gzx-def}
    G_z(x) = \begin{cases}
    \sigma_p(x) & (\text{self-avoiding walk})\\
    \rho_p^{(a)}(x)/\rho_p^{(a)}(0)
        & (\text{lattice trees and animals})\\
    \tau_p(x) & (\text{percolation}).
    \end{cases}
\en
We will explain in Section~\ref{sec-le} how the lace expansion gives rise to 
a function $\Pi_z$ on $\Zd$ and to the convolution equation
\eq
\lbeq{Gle}
    G_z = I + \Pi_z +  (z D*(I+\Pi_z) * G_z).
\en
The function $\Pi_z$ is symmetric under the symmetries of $\Zd$.
For self-avoiding walk, a small modification of the
usual analysis \cite{BS85,MS93} has been made to write the lace expansion 
in this form.  (A remainder term in the percolation expansion will be shown
to vanish, in Section~\ref{sec-pfmain}.)

The identity \refeq{Gle} reduces to \refeq{sle} when $\Pi_z \equiv 0$.
Our method involves treating each of the models as a small
perturbation of random walk, and the
function $\Pi_z(x)$ should be regarded as a small correction to
$\delta_{0,x}$.  As we will show in Section~\ref{sec-pfmain}, 
$\Pi_z(x)$ is small uniformly
in $x$ and $z \leq z_c$ for large $L$ and decays at least as fast as
$|x|^{-(d+2+\epsilon)}$, when $d > d_c$.
In particular, 
$\sum_x |x|^{2+s} |\Pi_z(x)|$ converges for $z \leq z_c$,
for any $s < \epsilon$, so $\Pi_z$ has a finite $(2+s)$ moment.
We assume the above bounds on $\Pi_z(x)$ in the remainder of this section.

Equations~\refeq{sle} and \refeq{Gle} can be rewritten as
\eq
\lbeq{Irep}
    I = (I - \mu D)*S_\mu = G_z - \Pi_z - (zD*(I+\Pi_z) * G_z) .
\en
Let $\lambda \in \Rbold$.
Writing
\eq
    G_z = \lambda S_\mu  +(I*G_z)  - \lambda (I*S_\mu)
\en
and using the first representation of \refeq{Irep} for $I$ in $I*G_z$ and the
second in $I*S_\mu$, we obtain
\eq
\lbeq{tr.man.6a}
    G_z =  \lambda ((I+ \Pi_z) * S_\mu)
    + (S_\mu * E_{z,\lambda, \mu} *G_z),
\en
with
\eq
\lbeq{Edef}
    E_{z,\lambda, \mu}
    = [I - \mu D] - \lambda [I -  z D*(I+ \Pi_{z})].
\en

By symmetry, odd moments of $E_{z,\mu,\lambda}(x)$
vanish.  We fix $\lambda$ and $\mu$ so that the zeroth and second
moments also vanish, i.e., 
\eq
\lbeq{zeroesE}
        \sum_{x \in \Zd}  E_{z,\lambda, \mu}(x) 
        = \sum_{x \in \Zd} |x|^2 E_{z,\lambda, \mu}(x)=0.
\en
Here we are assuming, as discussed above, that $\Pi_z$ has finite second
moment.  Thus we take
\eqarray
\lbeq{lampdef}
    \lambda & = & \lambda_{z}
    = \frac{1}
    {1 + z \sigma^{-2}\sum_x |x|^2 \Pi_{z}(x)}, \\
\lbeq{mupdef}
    \mu & = & \mu_z 
    = 1 - \lambda_z[1-z-z\sum_x \Pi_z(x)].
\enarray
For simplicity, we will write $E_{z}(x) = E_{z,\mu_{z},\lambda_{z}}(x)$.
Then \refeq{tr.man.6a} becomes
\eq
\lbeq{tr.man.6}
    G_{z}(x) = \lambda_{z} ((I+ \Pi_{z}) * S_{\mu_{z}})(x)
    + (S_{\mu_{z}} * E_{z} *G_{z})(x).
\en
The critical point obeys the identity
\eq
\lbeq{zcid}
	  1 - z_c - z_c \sum_x \Pi_{z_c}(x) = 0, 
\en
and hence $\mu_{z_c}=1$.
To see this, we sum \refeq{Gle} over $x$ to obtain
\eq
\lbeq{sus}
    \sum_{x} G_{z}(x) = \frac{1+ \sum_{x}\Pi_{z}(x)}
    {1 - z - z \sum_{x} \Pi_{z}(x)}.
\en
The left side is finite below the critical point, but diverges as
$z \uparrow z_c$ \cite{AN84,BFG86,MS93}.  
Under the assumption made above on $\Pi_z$, the critical point thus
corresponds to the vanishing of the denominator of \refeq{sus}.

Using the decay of $\Pi_z$ in $x$, we will argue that, at $z_c$,
the first term of \refeq{tr.man.6} gives
$\lambda_{z_c}[1+ \sum_y \Pi_{z_c}(y)] S_1(x)$ as the leading behaviour
of $G_{z_c}(x)$.  The second term will be shown to be an error term
which decays faster than $|x|^{-(d-2)}$.
In terms of the Fourier transform, we understand this
second term as follows.  By our choice of the
parameters $\lambda_z$ and $\mu_z$, $\hat{E}_{z_c}(k)$ should behave
to leading order as $k^{2+a}$ for some positive $a$.  Assuming
that $\hat{G}_{z_c}(k)$ behaves like $k^{-2}$, and since $\hat{S}_1(k)$
behaves like $k^{-2}$ by \refeq{Chatdef}, 
the second term of \refeq{tr.man.6}
would be of the form $k^{-2+a}$, which should correspond to
$x$-space decay of the form $|x|^{-(d-2+a)}$.
Our proof will be based on this insight.

The proof will require:
\renewcommand{\theenumi}{\roman{enumi}}
\begin{enumerate}
\item
information about the asymptotics of $S_\mu(x)$ (model-independent),
\item
an estimate providing bounds on the decay rate of
a convolution in terms of the decay of the functions being convolved
(model-independent),
\item
a mechanism for proving  that $\Pi_z(x)$ decays faster
than $|x|^{-(d+2)}$ (model-dependent), and
\item
given this decay of $\Pi_z(x)$,
an upper bound guaranteeing adequate decay of $(S_{\mu_z}*E_z)(x)$
(model-independent).
\end{enumerate}
The third item is the part of the argument that is model-dependent.
The restriction $d>d_c$ enters
here, in the bounding of certain Feynman diagrams that are specific to
the model under consideration.

The first ingredient in the above list, namely asymptotics for
the random walk generating
function, is provided by the following proposition.  The
proof of the proposition is deferred to Section~\ref{sec-G2ptfcn}.
More general results for the critical generating function $S_1(x)$
can be found in 
work of Uchiyama \cite{Uchi98}.  However, we are unable to apply
results of \cite{Uchi98} directly, since
we need control of the parameter $L$ in
our estimates that is not readily extractable from \cite{Uchi98}.
Our proof of Proposition~\ref{prop-A} will also make use of an analysis
of $S_\mu(x)$ for $\mu < 1$, which we will need in proving 
Proposition~\ref{lem-C} below.

\begin{prop}
\label{prop-A}
Let $d>2$, and suppose $D$ satisfies the conditions of
Definition~\ref{def-Dsp1}.
Then, for $L$ sufficiently large, $\alpha > 0$, $\mu \leq 1$ and $x \in \Zd$,
\eqarray
\lbeq{CL-uni}
    S_\mu(x)
    & \leq & \delta_{0,x}+ O\Big(\frac{1}{L^{2-\alpha}\, (|x|+1)^{d-2}}\Big),
\\
\lbeq{CL-uni2}
    S_1(x) & = & \frac{a_{d}}{\sigma^2} \frac{1}{(|x|+1)^{d-2}}
    +O\Big( \frac{1}{(|x|+1)^{d-\alpha}} \Big) .
\enarray
In \refeq{CL-uni}--\refeq{CL-uni2}, constants in 
error terms may depend on $\alpha$.
\end{prop}

For the second ingredient in the list above,
we will use the following proposition, whose
estimates show that the decay rate of functions implies a corresponding
decay for their convolution.
The elementary proof of the proposition will be given in Section~\ref{sec-conv}.

\begin{prop}
\label{lem-conv}
    {\hspace{10mm}}\\
(i) If functions $f, g$ on $\Zd$ satisfy
$| f(x) | \leq (|x|+1)^{-a}$ and
$| g(x) | \leq (|x|+1)^{-b}$
with $a \geq b >0$, then there exists a constant $C$ depending on $a, b, d$
such that
    \eq
    \lbeq{fg.2}
        \bigl | (f*g)(x) \bigr | \leq
   \begin{cases}
    C (|x|+1)^{-(a \land b)}
   & (a > d )
   \\
   C (|x|+1)^{d-(a + b)} &
       (a < d \mbox{ and } a+b>d) .
   \end{cases}
    \en
(ii)
Let $d>2$, and let
$f,g$ be functions on $\Zd$, where $g$ is $\Zd$-symmetric.
Suppose that there are $A,B,C >0$ and $s>0$ such that
    \eqarray
    \lbeq{GandH1}
        f(x) & = & \frac{A}{(|x|+1)^{d-2}}
    + O\Big(\frac{B}{(|x|+1)^{d-2+s}}\Big)
    \\
    \lbeq{GandH3}
    | g(x) | &\leq & \frac{C}{(|x|+1)^{d+s}}.
    \enarray
    Let $s_2 = s \wedge 2$.
    Then
    \eq
    \lbeq{GandH4}
        (f*g)(x) =
    \frac{A\sum_{y} g(y)}{(|x|+1)^{d-2}} + e(x)
    \en
    with
    \eq
    e(x) = \begin{cases}
    O(C(A+B)(|x|+1)^{-(d-2+s_2)}) & (s \neq 2) \\
    O(C(A+B)\log(|x|+2)(|x|+1)^{-d}) & (s = 2),
    \end{cases}
    \en
    where the constant in the error term depends on $d$ and $s$.
\end{prop}

For the third ingredient, we will use the following proposition.
The proof of the proposition involves model-dependent diagrammatic
estimates, and is given in Section~\ref{sec-Fd}.

\begin{prop}
\label{prop-diagbd}
Let $q<d$, and suppose that
    \eq
    \lbeq{taubd-ass1new}
    G_z(x)  \leq 
    \beta (|x|+1)^{-q} \quad (x \neq 0)
    \en
    with $\beta /L^{q-d}$ bounded away from zero.  
    Then for sufficiently small $\beta$ (which requires $L$ to be large)
    the following statements hold. 
    \\
(a) Let $z \leq 2$, and assume $\frac{1}{2}d<q <d$.  For self-avoiding walk,
there is a $c$ depending on $d$ and $q$ such that 
        \eq
        \lbeq{Pix-saw-bdnew}
        |\Pi_z(x)| \leq c \beta \, \delta_{0,x} +
        \frac{c \beta^{3}}{(|x|+1)^{3 q}} .
        \en
(b) Define $p=p(z)$ implicitly by \refeq{zp}, and
fix a positive constant $R$.
Let $z$ be such that $\rho_{p(z)}^{(a)}(0) \leq R$,
and assume $\frac{3}{4}d < q <  d$.  For lattice trees or lattice animals,
there is a $c$ depending on $d$, $q$ and $R$ such that
        \eq
        \lbeq{Pix-ltla-bd}
        |\Pi_z(x)| \leq c \beta \delta_{0,x} +
        \frac{c \beta^{2}}{(|x|+1)^{3 q -d}} .
        \en
(c) Let $z \leq 2$, and assume $\frac{2}{3}d < q <d$.  For percolation,
there is a $c$ depending on $d$ and $q$ such that
        \eq
        \lbeq{Pix-res-bd}
        |\Pi_z(x)|  \leq
        c \beta \delta_{0,x} +
        \frac{c \beta^{2}}{(|x|+1)^{2 q}} .
        \en
\end{prop}

The main hypothesis in Proposition~\ref{prop-diagbd} 
is an assumed bound on the
decay of the two-point function.  To motivate the form of the assumption,
we first note that $G_z(x)$ cannot be expected to decay faster than $D(x)$.
Let $\chi_L$ denote the indicator function of the cube $[-L,L]^d$.
By Definition~\ref{def-Dsp1}, 
\eq
\lbeq{Dchi}
	D(x) \leq O(L^{-d}) \chi_L(x) \leq O(L^{q-d}(|x|+1)^{-q}),
\en	 
and the upper bound is achieved when $|x|$ and $L$ are comparable.
This helps explain the assumption that $\beta /L^{q-d}$ is bounded away
from zero in the proposition.  Note that $G_z(0)=1$ for all $z \leq z_c$.

We will apply Proposition~\ref{prop-diagbd}
with $q = d-2$.  However, to do so,
we will have to deal with the fact
that \emph{a priori} we do not know that
\refeq{taubd-ass1new} holds for $z$ near $z_c$ with $q = d-2$.
Note that, for $q = d-2$, the conditions on $d$ in the above proposition
correspond to $d>d_c$, with $d_c$ given by \refeq{dcdef}.
Also, using $\epsilon$ defined in \refeq{epsilon-def},
all three bounds of the lemma can be unified (after weakening
the self-avoiding walk bound) in the form
\eq
\lbeq{Pix-unified}
    |\Pi_z(x)| \leq c \beta \, \delta_{0,x} +
        \frac{c \beta^2}{(|x|+1)^{d+2+\epsilon}}.
\en
Note that $\epsilon > 0$ if
and only if $d>d_c$.  It is at this stage of the analysis,
and \emph{only} here, that the upper critical dimension enters our
analysis.

Finally, the fourth ingredient is the following proposition.  Its proof
is model-independent and is given in Section~\ref{sec-CE}.

\begin{prop}
\label{lem-C}
Fix $z  \leq z_c$, $0<\gamma <1$, $\alpha >0$ and
$\kappa > 0$.  Let $\kappa_2 = \kappa \wedge 2$.
Assume that $z \leq C$
and that $|\Pi_z(x)| \leq \gamma(|x|+1)^{-(d+2+\kappa)}$.
Then there is a $c$ depending on $C, \kappa, \alpha$
but independent of $z,\gamma, L$ such that for $L$ sufficiently large
    \eq
    \lbeq{E2res.1}
         | (E_z * S_{\mu_z})(x)  |
        \leq
        \begin{cases}
        c\gamma L^{-d} & (x \neq 0) \\
        c \gamma L^{\kappa_2}
        (|x|+1)^{-(d+\kappa_2 -\alpha)} & (\mbox{\rm{all} $x$}).
        \end{cases}
    \en
\end{prop}

In \refeq{E2res.1}, we are interested in the case where $\alpha$ is close to
zero (and small compared to $\kappa_2$), so that the upper bound decays
faster in $|x|$ than $|x|^{-d}$.
It will be crucial in the proof of Proposition~\ref{lem-C} that $E_z$
is $\Zd$-symmetric,
and that we have chosen $\lambda_z$ and $\mu_z$ such that the zeroth and
second moments of $E_z$ vanish.
The coefficients of terms of order $|x|^{2-d}$ and $|x|^{-d}$, which
would typically be present in the convolution 
of a function decaying like $|x|^{-(d+2+\kappa)}$
with $S_{\mu_z}$, then vanish
and hence are absent in the upper bound of \refeq{E2res.1}.
This can partially be seen from the first term of \refeq{GandH4}, where the
leading term vanishes if and only if $\sum_y g(y)=0$.

\section{Proof of the main results}
\label{sec-pfmain}

In this section, we prove Theorem~\ref{thm-main} and
Corollaries~\ref{cor-bub}--\ref{cor-pc}, assuming
Propositions~\ref{prop-A}--\ref{lem-C}.
The proof will be based on the following elementary lemma.
The lemma states that under an appropriate continuity
assumption, if an inequality implies a stronger inequality, then in
fact the stronger inequality must hold.

\begin{lemma}
\label{lem-P4}
Let $f$ be a nonnegative
function defined on an interval $[z_1,z_c)$,
and let $a\in (0,1)$ be given.  Suppose that
\begin{enumerate}
    \item $f$ is continuous on the interval $[z_1,z_c)$.
    \item $f (z_1) \leq a$.
    \item for each $z \in (z_1,z_c)$, if $f(z) \leq 1$
    then in fact $f(z) \leq a$.
    (In other words, one inequality implies a stronger
    inequality.)
\end{enumerate}
Then $f(z) \leq a$ for all $z \in [z_1,z_c)$.
\end{lemma}

\begin{proof}
By the third
assumption, $f(z) \nin (a,1]$ for all $z \in (z_1,z_c)$.
By the first assumption,
$f(z)$ is continuous in $z \in [z_1,z_c)$.  Since
$f(z_1) \leq a$ by the second assumption, the above two facts
imply that $f(z)$ cannot enter the forbidden interval $(a,1]$ when
$z \in (z_1,z_c)$, and hence
$f(z) \leq a$ for all $z \in [z_1,z_c)$.
\end{proof}

\smallskip

We will employ Lemma~\ref{lem-P4} to prove the following proposition,
which lies at the heart of our method.
The proposition
provides a good upper bound on the critical two-point function
for nonzero $x$.
There is an additional detail required in the proof for lattice trees
and lattice animals, and we therefore treat these models separately
from self-avoiding walk and percolation.
The relevant difference between the models is connected with the fact
that $\sigma_z(0)=\tau_z(0)=1$, whereas $\rho_p^{(a)}(0) >1$ and we do not
know {\em a priori}\/ that $\rho_{p_c}^{(a)}(0) < \infty$.
In the proof, we establish the finiteness of
$\rho_{p_c}^{(a)}(0)$.
As usual, $\alpha$ should be regarded as almost zero in
Proposition~\ref{prop-P4P3}.

\begin{prop}
\label{prop-P4P3}
Fix $d> d_c$ and $\alpha > 0$. For $L$ sufficiently large depending on
$d$ and $\alpha$,
\eq
\lbeq{P4P3.1}
    G_{z_c}(x)
    \leq \frac{C}{L^{2-\alpha}(|x|+1)^{d-2}}
    \qquad (x \neq 0).
\en
In addition, $z_c \leq 1+O(L^{-2+\alpha})$, and, for lattice trees and lattice
animals, $\rho_{p_c}^{(a)}(0)<O(1)$.
The constants in all the above statements depend 
only on $d$ and $\alpha$, and not on $L$.

\end{prop}

\begin{proof}
We prove the desired bound for $\alpha <\frac{\epsilon \wedge 1}{2}$,
because the bound for
large $\alpha$ follows from that for small $\alpha$.
In the following, let $K$ denote the smallest constant that can be
used in the error bound of \refeq{CL-uni}, i.e.\ 
$K = \sup_{L \geq 1,x \neq 0} L^{2-\alpha} (|x|+1)^{d-2} S_{1}(x)
\in (0,\infty)$.

\smallskip \noindent {\em Self-avoiding walk and percolation.}
We will prove that $G_z(x)$ obeys the
upper bound of \refeq{P4P3.1} uniformly in $z < z_c$.  This is sufficient,
by the monotone convergence theorem.

Let
\eq
\lbeq{tr-P4def2}
    g_x(z) = (2K)^{-1}L^{2-\alpha} (|x|+1)^{d-2} G_z(x),
    \quad \quad
    g(z) = \sup_{x \neq 0} g_x(z).
\en
For self-avoiding walk and percolation, we will employ
Lemma~\ref{lem-P4} with $z_1=1$,
\eq
\lbeq{fzdef}
    f(z) = \max \{g(z) ,\textstyle{\frac{1}{2}}z\},
\en
and $a$ chosen arbitrarily in
$(\frac{1}{2},1)$.  We verify the assumptions of
Lemma~\ref{lem-P4} one by one, with the bound \refeq{P4P3.1}
then following immediately from Lemma~\ref{lem-P4}.  In the course of the
proof, the desired upper bound on $z_c$ will be shown to be a consequence
of a weaker bound than \refeq{P4P3.1}, in \refeq{pub}.  Since the proof
actually establishes \refeq{P4P3.1}, \refeq{pub} then follows.

\smallskip \noindent
{\it (i)}
Continuity of each $g_x$ on $[0,z_c)$ is immediate from the
fact that $\sigma_z(x)$ is a
power series with radius of convergence
$z_c$, and from the continuity in $z$ of $\tau_z(x)$ proved in
\cite{AKN87}.
We need to argue that the supremum of these continuous
functions is also continuous.  For this, it suffices to show that
the supremum is continuous on $[0,z_c-t)$ for every small $t >0$.
It is a standard result that $\sigma_z(x)$ and $\tau_z(x)$ decay
exponentially in $|x|$, with a decay rate that is uniform in
$z \in [0,z_c-t)$ (though not in $L$) \cite{Grim99,MS93}.
Thus $g_x(z)$ can be made less than any $\delta >0$,
uniformly in $z \in [0,z_c-t)$,
by taking $|x|$ larger than some $R=R(L,t,\delta)$.  However,
choosing $x_0$ such that $D(x_0) >0$, we see that
$g_{x_0}(z) \geq (2K)^{-1}L^{2-\alpha} (|x_0|+1)^{d-2} z D(x_0)
\geq (2K)^{-1}L^{2-\alpha} (|x_0|+1)^{d-2} D(x_0) \equiv \delta_0$.
Hence the supremum is attained for $|x| \leq R(L,t,\delta_0)$, which
is a finite set, and hence the supremum is continuous and the first
assumption of Lemma~\ref{lem-P4} has been established.

\smallskip \noindent {\it (ii)}
For the second assumption of the lemma, we note that
$\tau_1(x) \leq \sigma_1(x) \leq S_1(x)$
and apply the uniform bound of \refeq{CL-uni} to conclude that $g(1) \leq 1/2$. 
Since we have restricted $a$
to be larger than $\frac{1}{2}$, this implies $f(1) <a$.

\smallskip \noindent {\it (iii)}
Fix $z \in (1,z_c)$.
We assume that $f(z) \leq 1$, which implies
\eq
\lbeq{upass}
    G_z(x) \leq \frac{2 K L^{-2+\alpha}}{(|x|+1)^{d-2}}
    \qquad (x \neq 0).
\en
We will apply Proposition~\ref{prop-diagbd} with $q=d-2$
and $\beta = KL^{-2+\alpha}$.
Since we have taken $\alpha < \frac{1}{2}$, we have $\beta \ll 1$ 
and $\beta/L^{q-d} = KL^\alpha \gg 1$ for
sufficiently large $L$ depending on $\alpha$.
Proposition~\ref{prop-diagbd} then implies that
\eq
\lbeq{tr.st.Jbd1}
    | \Pi_z(x) | \leq c K  L^{-2+\alpha} \delta_{0,x}
    + \frac{c \, K^{2}\,  L^{-4+2\alpha}}{(|x|+1)^{d+2+\epsilon}}
    \leq \frac{c \, K  L^{-2+\alpha}}{(|x|+1)^{d+2+\epsilon}},
\en
where $\epsilon >0$ was defined in \refeq{epsilon-def}.
It addition, for percolation, as argued at the end of
Section~\ref{sec-percdiagrams}, the remainder term $R_z^{(N)}(x)$ vanishes
in the limit $N \to \infty$ under the assumption \refeq{upass}, yielding
the form \refeq{Gle} of the expansion.

Summing \refeq{Gle} over $x \in \Zd$ gives
\eq
\lbeq{sus.1}
    \sum_x G_z(x) = \frac{1+\sum_x \Pi_z(x)}{1-z - z\sum_x \Pi_z(x)} > 0,
\en
which is finite for $z<z_c$.  The numerator is positive by
\refeq{tr.st.Jbd1}, and hence the denominator is also positive.  Therefore,
since $z \leq 2$ by our assumption that $f(z) \leq 1$, \refeq{tr.st.Jbd1}
implies that
\eq
\lbeq{pub}
    z < 1- z \sum_x \Pi_z(x) \leq 1 +O(L^{-2+\alpha}).
\en
Since $a \in (\frac{1}{2},1)$, this implies that $z<2a$ for all $z<z_c$,
when $L$ is large.
Thus, to prove that $f(z) \leq a$, it suffices to show that $g(z) \leq a$.

The bound \refeq{tr.st.Jbd1} also implies that
$\lambda_z$ and $\mu_z$ are well-defined by \refeq{lampdef}--\refeq{mupdef},
and that $\lambda_z \to 1$ uniformly in $z \leq z_c$.
Using the convolution bound of
Proposition~\ref{lem-conv}(i),
\refeq{CL-uni}, and the first bound of
\refeq{tr.st.Jbd1}, it then follows that 
\eq
\lbeq{PsiC}
    |(\Pi_z * S_{\mu_z})(x)| 
    \leq \frac{O(L^{-4+2\alpha})}{(|x|+1)^{d-2}}
    = \frac{o(L^{-2 + \alpha})}{(|x|+1)^{d-2}} \quad\quad (x \neq 0).
\en
By Proposition~\ref{lem-C} with $\kappa = 2 \alpha < \epsilon$
and $\gamma = cKL^{-2+\alpha}$, for $L$ large we have
\eq
\lbeq{E2res.11}
    | (E_z * S_{\mu_z})(x) |
    \leq
    \begin{cases}
    O(L^{-2+\alpha -d}) & (x \neq 0) \\
    O( L^{-2+3\alpha})(|x|+1)^{-(d+\alpha)}
    & (\mbox{all $x$}).
    \end{cases}
\en
Using the first bound for $0<|x| \leq L$ and the second bound for 
$|x| \geq L$, we conclude from this that
\eq
\lbeq{ESnew}
	| (E_z * S_{\mu_z})(x) |
    \leq
    O(L^{-4+2\alpha})   (|x|+1)^{-(d-2)}
    \quad\quad (x \neq 0).
\en
By Proposition~\ref{lem-conv}(i), \refeq{upass}
and \refeq{E2res.11}, it then follows that
\eqarray
\lbeq{tr-convE2.1}
    | (E_z * S_{\mu_z}*G_z) (x) |
    & \leq &
    | (E_z * S_{\mu_z}) (x) | 
    + \sum_{y \neq 0}| (E_z * S_{\mu_z})(x-y)| \; |G_z(y) |
    \nnb
    & \leq & \frac{O(L^{-4+4\alpha})}{(|x|+1)^{d-2}}
    = \frac{o(L^{-2 + \alpha})}{(|x|+1)^{d-2}}
    \quad \quad (x \neq 0),
\enarray
where we have 
used \refeq{ESnew} to bound the first term in the first inequality.
Using the fact that $\lambda_z = 1+o(1)$ as $L \to \infty$,
and the definition of $K$, 
it then follows from the identity
\refeq{tr.man.6} that for $L$ sufficiently large we have
\eq
\lbeq{tr-convE2.2}
    G_z(x) \leq (1+o(1)) S_1(x) + \frac{o(L^{-2+\alpha})}{(|x|+1)^{d-2}}
    \leq \frac{2a \, K}{L^{2-\alpha}\, (|x|+1)^{d-2}}
    \quad \quad (x \neq 0).
\en
This yields $g(z) \leq a$, and completes the proof for self-avoiding walk
and percolation.

\smallskip \noindent {\em Lattice trees and lattice animals.}
We will first prove that $G_z(x)$ obeys the
upper bound of \refeq{P4P3.1} uniformly in $z < z_c$.

By \refeq{Dbasic2} and the fact that $h$ is bounded,
there is a $\delta_1 \geq 1$ such that
$D(x) \leq \delta_1 |\Omega_D|^{-1}$ for all $x$.
The number of $n$-bond
lattice trees or lattice animals containing the origin
is less than the number $b_n(L)$ of $n$-bond
lattice trees on the Bethe lattice of coordination
number $|\Omega_D|$ (the uniform tree of degree $|\Omega_D|$), which contain
the origin.  A standard subadditivity argument, together with the fact that,
as $L \to \infty$,
$\lim_{n \to \infty}b_n(L)^{1/n} \sim e|\Omega_D| \leq 3|\Omega_D|$ 
(see, e.g., \cite{Penr94}), implies that
$b_n(L) \leq (n+1)(3|\Omega_D|)^n$.  Therefore, for lattice trees or 
lattice animals,
\eq
\lbeq{ltlaBLbd}
    \rho_p^{(a)}(0) \leq \sum_{n=0}^\infty (n+1)(3\delta_1 p)^n
    = \frac{1}{(1-3\delta_1 p)^2}.
\en

Let $p_1 = \frac{1}{6\delta_1}$.
We use $z_1 = p_1 \rho_{p_1}^{(a)}(0)$ in Lemma~\ref{lem-P4}.
Note that $z_1$ is
well-defined, since \refeq{ltlaBLbd}
gives $\rho_p^{(a)}(0) \leq 4$ for $p \leq p_1$.  
In addition,
\refeq{ltlaBLbd} implies that $p_c \geq (3\delta_{1})^{-1} > p_1$,
so $z_c > z_1$.
We again fix $a \in (\frac{1}{2},1)$, and we
use the function $f(z)$ of \refeq{fzdef} in Lemma~\ref{lem-P4},
taking now
\eq
    g(z) = \sup_{x \neq 0} g_{x}(z) 
    \quad \mbox{with} \quad
    g_{x}(z) = \frac{1}{8K} L^{2-\alpha} (|x|+1)^{d-2} G_{z}(x) ,
\en
The desired bound on $G_z(x)$, for $z<z_c$, together with the 
desired bound on $z_c$, will follow once we verify the three conditions
of Lemma~\ref{lem-P4}.  We verify these conditions now.

\smallskip \noindent {\it (i)}  Continuity of $f(z)$ follows
from the exponential decay of $\rho_p^{(a)}(x)$ for $p<p_c$, 
as in the previous discussion, together
with the continuity of $\rho_p^{(a)}(0)$ for $p<p_c$.

\smallskip \noindent {\it (ii)}
By the remarks surrounding the definition of $z_1$, we have
$\frac{z_1}{2} \leq \frac{4}{2\cdot 6\delta_1} \leq \frac{1}{3} < a$.
Moreover, this implies $z_1 < \frac{2}{3} < 1$.
It remains to show that
\eq
\lbeq{Gz1bd}
	G_{z_1}(x) \leq \frac{8aK}{L^{2-\alpha}(|x|+1)^{d-2}} \quad \quad
	(x \neq 0).
\en
Since $\rho_{p_1}^{(a)} \geq 1$, we have $G_{z_1}(x) \leq \rho_{p_1}^{(a)}(x)$.
Each lattice tree or lattice animal containing $0$ and $x$ can be decomposed 
in a non-unique
way into a walk from $0$ to $x$ with a lattice tree or lattice
animal attached at each
site along the walk.  Therefore 
$\rho_{p_1}^{(a)}(x) \leq \rho_{p_1}^{(a)}(0) S_{z_1}(x)$.
Using $\rho_{p_1}^{(a)}(0) \leq 4$, it follows from Proposition~\ref{prop-A}
that $G_{z_1}(x) \leq 4KL^{-2+\alpha}(|x|+1)^{-(d-2)}$, which implies
\refeq{Gz1bd}.

\smallskip \noindent {\it (iii)}
Fix $z \in (z_1,z_c)$.
The assumption that $f(z) \leq 1$ implies the bound
$\rho_{p}^{(a)}(0) \leq z/p_1 \leq 12\delta_1$, and we take $R = 12\delta_1$
in Proposition~\ref{prop-diagbd}(b).  We then proceed as in the discussion
for self-avoiding walk and percolation.  We obtain \refeq{pub} as before,
so that $z < 2a$ as required.  The proof of \refeq{tr-convE2.2} also
proceeds as before.

\smallskip
The above discussion proves that $G_z(x)$ is bounded by the right side
of \refeq{P4P3.1} and that $\rho_p^{(a)}(0) \leq 4$, 
uniformly in $z<z_c$, and that $z_c \leq 1+O(L^{-2+\alpha})$.
The proof is then completed by observing that $\lim_{p\uparrow p_c}
\rho_p^{(a)}(x) = \rho_{p_c}^{(a)}(x)$, by monotone convergence.
\end{proof}

\smallskip
Proposition~\ref{prop-P4P3}
establishes the hypotheses of Proposition~\ref{prop-diagbd}, with 
$\beta$ proportional to $L^{-2+\alpha}$, $z=z_c$
and $q=d-2$.
Hence
the hypotheses of
Proposition~\ref{lem-C} are also now established, with $z=z_c$, 
$\kappa = \epsilon$,
and $\gamma =O(L^{-2+\alpha})$.  The conclusion
of Proposition~\ref{lem-C} has therefore also
been established.  Moreover, since Proposition~\ref{prop-P4P3} gives
a bound on $G_z(x)$ uniformly in $z \leq z_c$, the bounds of
Proposition~\ref{prop-P4P3} and Proposition~\ref{lem-C} hold uniformly
in $z \leq z_c$.  We will use this in the following.

\medskip \noindent {\em Proof of Theorem~\ref{thm-main}.}
Fix $z=z_c$, and recall the observation below \refeq{zcid} that $\mu_{z_c}=1$.
Define
\eq
\lbeq{Hdef}
    H(x) = \lambda_{z_c} \sum_{n=0}^\infty
    \left( (I+\Pi_{z_c})*(E_{z_c}*S_1)^{*n}\right) (x),
\en
where the superscript $*n$ denotes an $n$-fold convolution
and $(E_{z_c}*S_1)^{*0}=I$.  
Propositions~\ref{lem-conv}(i), \ref{prop-diagbd} and \ref{lem-C} 
guarantee that the series in \refeq{Hdef}
converges absolutely, and that
\eq
    H(x) = \lambda_{z_c} \delta_{0,x}
    + O\Big(
    \frac{L^{-2+\epsilon_2 +\alpha}}
    {(|x|+1)^{d+\epsilon_2 -\alpha}}
    \Big)
\en
with $\epsilon_2 = \epsilon \wedge 2$.
Iteration of \refeq{tr.man.6} then gives
\eq
    G_{z_c}(x) = (S_1*H)(x).
\en
By Proposition~\ref{lem-conv}(ii) and the asymptotic formula of \refeq{CL-uni2},
this yields
\eq
\lbeq{Gpcasy}
    G_{z_c}(x) = \frac{a_d A}{\sigma^2 (|x|+1)^{d-2}}
    + O\Big(
    \frac{L^{-2+\epsilon_2+\alpha}}
    {(|x|+1)^{d-2+\epsilon_2-\alpha}}\Big)
    + O\Big(
    \frac{1}
    {(|x|+1)^{d-\alpha}}
    \Big),
\en
with $A=\hat{H}(0)$.  
This proves Theorem~\ref{thm-main}, 
apart from the
assertion that $A=1+O(L^{-2+\alpha})$.

The constant $A=\hat{H}(0)$ can be evaluated as follows.  Since  \refeq{Hdef}
is absolutely summable over $x$, the Fourier transform
\eq
    \hat{H}(k) = \lambda_{z_c} (1+\hat{\Pi}_{z_c}(k))
    \sum_{n=0}^\infty
    [\hat{E}_{z_c}(k) \hat{S}_1(k)]^{n}
\en
is continuous in $k$.
Using \refeq{zeroesE}, the fact that $E_{z_c}(x)$ decays like
$|x|^{-(d+2+ \epsilon)}$, and dominated convergence, we have 
\eq
    \lim_{k \to 0} |k|^{-2} \hat{E}_{z_c}(k)
    = \lim_{k \to 0} 
    \sum_x E_{z_c}(x) |k|^{-2}
    \Big(\cos (k\cdot x) - 1 - \frac{|k|^2|x|^2}{2d}\Big)
    = 0.
\en
Since $\hat{S}_1(k)$ diverges like a multiple of $|k|^{-2}$ by \refeq{Chatdef},
we conclude from \refeq{lampdef} and the conclusion of
Proposition~\ref{prop-diagbd} that
\eq
\lbeq{Aform}
    A = \hat{H}(0) = \lim_{k \to 0}\hat{H}(k)
    = \lambda_{z_c} (1+\hat{\Pi}_{z_c}(0))
    =
    \frac{1+\hat{\Pi}_{z_c}(0)}
    {1 + z_c \, \sigma^{-2} \sum_x |x|^2 \Pi_{z_c}(x)} = 1 + O(L^{-2+\alpha}).
\en
\qed

\medskip \noindent
{\em Proof of Corollary~\ref{cor-bub}.}
The corollary follows immediately from Theorem~\ref{thm-main}
and the convolution bound of Proposition~\ref{lem-conv}(i).
\qed

\smallskip
To prove Corollary~\ref{cor-k}, the following lemma will be useful.

\begin{lemma}
\label{lem-Fourier1}
    Let $f(x)$ be a $\Zd$-symmetric function which obeys the bound 
	$| f(x) | \leq (|x|+1)^{-(d+2+\kappa)}$
	with $\kappa >0$.  Then 
    \eq
    \lbeq{fhatbd.1}
        \hat{f}(k)  = \hat{f}(0) + \frac{|k|^{2}}{2d} \nabla^2 \hat{f}(0) + e(k) 
	\en 
	with 
	\eq
	\lbeq{fhatbd.2}
        |e(k)| \leq  \begin{cases}
		const. \,  |k|^{2+(\kappa \wedge 2)} & (\kappa \neq 2) \\
		const. \, |k|^{4} \, \log |k|^{-1} 	&
		(\kappa = 2).
		\end{cases} 
    \en
\end{lemma}

\begin{proof}
    By the $\Zd$-symmetry of $f(x)$, 
    \eq
    \lbeq{fhatbd.11}
        \hat{f}(k) = \sum_{x}  f(x) \cos(k\cdot x) 
        = \hat{f}(0) + \frac{|k|^{2}}{2d} \nabla^2 \hat{f}(0)  
		+ \sum_{x} \Big( \cos (\kx) - 1  + \frac{(\kx)^{2}}{2}
        \Big) \, f(x) .
    \en
The expression in brackets of the third term 
	is bounded in absolute value
    both by $|k|^4|x|^4/24$ and $2+|k|^2 |x|^2/2$.  The third term of
\refeq{fhatbd.11} is therefore bounded by
\eqalign
\lbeq{fhatbd.13}
	\frac{|k|^{4}}{24}
        \sum_{x: |x| \leq |k|^{-1}}|x|^{4} \, | f(x)|
		+ \sum_{x: |x| > |k|^{-1}} 
		\Big( 2 +  \frac{|k|^{2}|x|^{2}}{2}  \Big) \, | f(x) |
		. 
\enalign
Using the assumed upper bound on $f(x)$ then   
gives \refeq{fhatbd.2} and completes the proof.
\end{proof}

\medskip \noindent
{\em Proof of Corollary~\ref{cor-k}.}
We first assume $\epsilon \neq 2$, and comment on the minor
modifications required for $\epsilon 
=2$ at the end of the proof. 

Let $\hat{F}_z(k) = 1 - z\hat{D}(k)(1+\hat{\Pi}_z(k))$. 
For $z<z_c$, as in \refeq{sus.1} 
we have
\eq
    \hat{G}_z(k)
    = \frac{1+ \hat{\Pi}_z(k)}{\hat{F}_z(k)}.
\en
As we have noted above, the bounds of Proposition~\ref{prop-diagbd} 
have been established with $q=d-2$, uniformly in $z \leq z_c$.
Therefore by Lemma~\ref{lem-Fourier1}, we have for $z \leq z_c$ 
\eqalign
\lbeq{cork1}
    1 + \hat{\Pi}_{z}(k) & =  1 + \hat{\Pi}_{z}(0)
    + O_L(|k|^{2}),
    \\
\lbeq{cork2}
    \hat{F}_{z}(k)
    & =    \hat{F}_{z}(0) +
    \frac{|k|^2}{2d} \nabla^2 \hat{F}_{z}(0)
    + O_L( |k|^{2 + (\epsilon\wedge 2)}),
\enalign 
with $L$-dependent error estimates.
Also, as observed in \refeq{pub}, 
$\hat{F}_z(0) > 0$ for $z<z_c$.
Thus we have the infra-red bound
\eq
\lbeq{irbd}
    0 < \hat{G}_z(k) \leq O_L(|k|^{-2})
\en
uniformly in $z < z_c$.

Since $G_{z_c}(x)$ behaves like $|x|^{-(d-2)}$, it is not summable over $x$
and hence the summation defining $\hat{G}_{z_c}(k)$ is not well-defined.
We define
\eq
\lbeq{cork3}
    \hat{G}_{z_c}(k) = \lim_{z \uparrow z_c}\hat{G}_z(k)
    = \frac{1+ \hat{\Pi}_{z_c}(k)}{ \hat{F}_{z_c}(k)}.
\en
This is a sensible definition, because $G_{z_c}(x)$ is then given by
the inverse Fourier transform of $\hat{G}_{z_c}(k)$.  In fact, using
monotone convergence in the first step, and
\refeq{irbd} and the dominated convergence theorem in the last step
(since $d\geq d_c +1 >2$), we have
\eq
    G_{z_c}(x) = \lim_{z \uparrow z_c} G_z(x)
    = \lim_{z \uparrow z_c}
    \int_{[-\pi,\pi]^d} \hat{G}_z(k) e^{-ik\cdot x} \frac{d^dk}{(2\pi)^d}
    =
    \int_{[-\pi,\pi]^d} \hat{G}_{z_c}(k)
    e^{-ik\cdot x} \frac{d^dk}{(2\pi)^d}.
\en

Since $\hat{F}_{z_c}(0)=0$ by \refeq{zcid}, \refeq{cork1}--\refeq{cork2}
then imply
\eq
\lbeq{cork12}
	\hat{G}_{z_c}(k) 
	= \frac{2d(1+\hat{\Pi}_{z_c}(0))}{\nabla^2 \hat{F}_{z_c}(0) |k|^2}
	\left[ 1 + O_L(|k|^{\epsilon\wedge 2}) \right]
	= \frac{2dA}{\sigma^2 |k|^2}
	\left[ 1 + O_L(|k|^{\epsilon\wedge 2}) \right].
\en
In the last equality, we used \refeq{zcid} and \refeq{Aform}.  

The case $\epsilon =2$ can be treated by adding an extra factor 
$\log|k|^{-1}$ to \refeq{cork2} and \refeq{cork12}. 
\qed

\medskip \noindent
{\em Proof of Corollary~\ref{cor-pc}.}
Recall the elementary fact that for self-avoiding walk and percolation,
$p_c=z_c \geq 1$.  The corollary then
follows immediately from Proposition~\ref{prop-P4P3}.
(The bound $z_c \leq 1+O(L^{-2+\alpha})$ is uninformative concerning $p_c$ 
for lattice trees and lattice animals, since we have proved only
that $\rho_{p_c}^{(a)}(0) \in [1,4]$.)
\qed

\medskip
It remains to prove Propositions~\ref{prop-A}--\ref{lem-C}.  After reviewing
the lace expansion in Section~\ref{sec-le}, these four propositions will
be proved in Sections~\ref{sec-G2ptfcn}, \ref{sec-conv}, \ref{sec-Fd}
and \ref{sec-CE} respectively.

\section{The lace expansion}
\setcounter{equation}{0}
\label{sec-le}

In this section, we review the key steps in the derivation of the lace
expansion.  In particular, we will describe 
how for each of our models the lace expansion
gives rise to the convolution equation \refeq{Gle}, which can be written as
\eq
\lbeq{Gle.2}
    G_{z}(x) = \delta_{0,x} + \Pi_{z}(x) + (zD * G_{z})(x)
    + (\Pi_{z}*zD*G_{z})(x).
\en
For self-avoiding walk, the lace expansion was introduced by
Brydges and Spencer in \cite{BS85}.  
Our treatment of the expansion for self-avoiding walk differs slightly
from the usual treatment, to allow for a simultaneous treatment of
lattice trees and animals.
For percolation, and for lattice trees
and lattice animals, the expansions
were introduced by Hara and Slade in \cite{HS90a,HS90b}.  For
overviews, see \cite{HS94,MS93}.
Proofs and further details can be found in the above references.

This section, together with Section~\ref{sec-Fd},
contains the model-dependent part of our analysis.

\subsection{Inclusion-exclusion}
\label{sec-le.i-e}

The expansion can be understood intuitively as arising from repeated use
of the inclusion-exclusion relation.  We describe this now in general terms,
postponing a more precise (but more technical) description to
Sections~\ref{sec-le.comb}--\ref{sec-le.perc}.

The two-point function for each of the
models under consideration is a sum, over geometrical objects, of weights
associated with these objects.  The geometrical objects are self-avoiding walks,
lattice trees, or lattice animals containing the two points $0$ and $x$.
This is the case also for percolation when $p<p_c$.  
For example, for the nearest-neighbour
model $\tau_p(x)=\sum_{A \in \Acal(0,x)}p^{|A|}(1-p)^{|\partial A|}$
for $p<p_c$,
where $\partial A$
represents the boundary bonds of $A$ and $|A|$ is the number of bonds in $A$.
We view these geometrical objects as a string of mutually-avoiding
beads, as depicted in Figure~\reffg{beads}.
For self-avoiding walk, the beads are simply lattice sites, whose mutual
avoidance keeps the walk self-avoiding.  For lattice trees,
the string represents the unique path, or {\em backbone}\/, in the tree
from $0$ to $x$,
and the beads represent lattice trees corresponding to branches along the
backbone.  These branches are mutually-avoiding, to preserve the overall
tree structure.

\begin{figure}
\begin{center}
\end{center}
\begin{center}
\includegraphics[scale = 0.4]{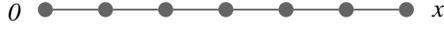}
\end{center}
\caption{\lbfg{beads}
A string of beads.}
\end{figure}

For lattice animals and percolation,
we need to introduce the notion of a pivotal bond.
A bond $\{a,b\}$ in $A \in \Acal(x,y)$ is called {\em pivotal}\/
for the connection
from $x$ to $y$ if its removal would disconnect the animal into two
connected
components, with $x$ in one component and $y$ in the other.
A lattice animal $A$ containing $x$ and $y$ is said to have a
\emph{double connection}
from $x$ to $y$ if there are two bond-disjoint
paths in $A$ between $x$ and $y$, or if $x=y$.
For lattice animals, the string in the string of beads represents the pivotal
bonds for the connection of $0$ and $x$.  The beads correspond to the
portions of the animal doubly-connected between pivotal bonds.  The mutual
avoidance of the beads
is required for consistency with the pivotal nature of the pivotal
bonds.  This picture is the same both for lattice animals and for percolation.

The basic idea of the lace expansion is the same in all four models.
It consists in approximating the two-point function by a sum of weights
of geometrical objects represented by a string of beads,
with the interaction between
the first bead and all subsequent beads neglected.  This treats the model
as if it were a Markov process.
The approximation causes configurations
which do not contribute to the two-point function to be included, and these
undesired contributions are then excluded
in a correction term.  The correction term is then subjected to repeated
and systematic further application of inclusion-exclusion.

Let $\Dcal(x,y)$ denote the
set of all animals having a double connection between $x$
and $y$, and, given a lattice animal,
let $\Delta(x)$ denote the set of sites that are doubly-connected
to $x$.  We define
\eq
    \psi_p^{(0)} (x) = \left\{
    \begin{array}{ll}
    0 & (\mbox{self-avoiding walk and lattice trees}) \\
    (1-\delta_{0,x})
    \sum_{A \in \Dcal(0,x)} W_{p,D}(A)  & (\mbox{lattice animals}) \\
    (1-\delta_{0,x})
    \Pbold_p( x \in \Delta(0) ) &
    (\mbox{percolation})
    \end{array}
    \right.
\en
and
\eq
    a_p = \left\{
    \begin{array}{ll}
    1 & (\mbox{self-avoiding walk and percolation}) \\
    \rho_p^{(a)}(0)  & (\mbox{lattice trees and lattice animals}).
    \end{array}
    \right.
\en
The procedure described in the preceding paragraph is implemented by
writing
\eq
\lbeq{up0}
    U_p(x) = a_p \delta_{0,x} + \psi_p^{(0)}(x) + a_p(pD*U_p)(x)
    + (\psi_p^{(0)}*pD*U_p)(x) + R_p^{(0)}(x).
\en
The terms on the right side can be understood as follows.
The first term is the contribution due to the case when the string
of beads consists of a single bead and $x=0$.  The term $\psi_p^{(0)}(x)$
is the contribution due to the case when the string of beads consists of
a single bead and $x \neq 0$.  The convolutions correspond
to the case where the string of beads consists of more than a single bead.
The factors $a_p$ and $\psi_p^{(0)}$ together give the contribution
from the first bead, the factor $pD$ is the contribution from the first
piece of string, and the factor $U_p$ is the contribution of the
remaining portion of the string
of beads.  These two terms neglect the interaction between the first bead
and the subsequent beads.  This is corrected by the correction term
$R_p^{(0)}(x)$, which is negative.

To understand the correction term, we first restrict attention to the
combinatorial models, which excludes percolation.  In this case,
the correction term simply involves the contributions from configurations
in which the first bead intersects some subsequent bead.  The contribution
due the case where the first such bead is actually the last bead is denoted
$-\psi_p^{(1)}(x)$.  If the first such bead is not the last bead, then
suppose it is the $j^{\rm th}$ bead.  The second through $j^{\rm th}$ beads
are mutually avoiding, and the $(j+1)^{\rm st}$ through last bead are
mutually avoiding, and these two sets of beads avoid each other.  We neglect
the mutual avoidance between these two sets of beads, making them independent
of each other, and add a correction term to exclude the undesired configurations
included through this neglect.  This leads to the identity
\eq
    R_p^{(0)}(x) = -\psi_p^{(1)}(x)
    - (\psi_p^{(1)}*pD*U_p)(x) + R_p^{(1)}(x).
\en
The inclusion-exclusion can then be applied to $R_p^{(1)}(x)$, and so on.
For percolation, the above procedure can also be applied, but more care
is needed in dealing with the probabilistic nature of the weights involved.
The form of the terms arising in the expansion for percolation is, however,
the same as the above.  When the process is continued indefinitely,
the result is
\eq
\lbeq{upsi}
    U_p(x) = a_p \delta_{0,x} + \psi_p(x) + a_p(pD*U_p)(x)
    + (\psi_p*pD*U_p)(x),
\en
with
\eq
\lbeq{psidef}
    \psi_p(x) = \sum_{N=0}^\infty (-1)^N \psi_p^{(N)}(x).
\en
The change of variables defined by \refeq{zp}--\refeq{Gzx-def} then gives
our basic identity \refeq{Gle.2}, once we define
\eq
\lbeq{Pipsi}
    \Pi_z(x) = a_p^{-1} \psi_p(x).
\en

Care is needed for convergence of \refeq{psidef}.  We require convergence
at $p=p_c$, which demands in particular that the individual terms in
the sum over $N$ are finite when $p=p_c$.  This will be achieved by taking
$d$ greater than the critical dimension $d_c$.  The role of large $L$ is
to ensure that the terms $\psi_{p_c}^{(N)}(x)$ are not only finite, but
grow small with $N$ sufficiently rapidly to be summable.
These issues are addressed in detail in Section~\ref{sec-Fd}.

The above discussion has been at an informal level, to establish
intuition for the expansion.  Our next goal is to make this more precise.

\subsection{Self-avoiding walk, lattice trees and lattice animals}
\label{sec-le.comb}

For the combinatorial models, an
elegant formalism introduced by Brydges and Spencer \cite{BS85}
can be used to make the discussion more precise, using the notion
of {\em lace}.\/  We discuss this now.

Let $R$ be an ordered set $R_0, R_1, \ldots, R_l$ of lattice animals, with 
$l$ arbitrary.  In particular, each $R_j$
may be simply a lattice
tree or a single site.  Given $R$, we define
\eq
\lbeq{Udef}
    \Ucal_{st}(R) \;\; = \left \{
    \begin{array}{rl}
    -1  &   \mbox{if $R_s \cap R_t \neq \varnothing$} \\
    0   &   \mbox{if $R_s \cap R_t = \varnothing$.}
    \end{array}
    \right .
\en
In \refeq{Udef}, the intersection is to be interpreted as the intersection
of sets of {\em sites}\/ rather than of bonds.
For $0 \leq a \leq b$, we also define
\eq
\lbeq{Kdef}
    K_R[a,b] = \prod_{a \leq s<t\leq b}(1+\Ucal_{st}(R)).
\en
The two-point functions for self-avoiding walk,
lattice trees, and lattice animals can be rewritten
in terms of $K_R$.

Given a finite set $B$ of bonds,
we let $|B|$ denote its cardinality.
For self-avoiding walk, we let $R$ consist of the sites along the walk
(the `beads'),
so that each $R_i$ is the single site $\omega(i)$.
Then the two-point function can be written
\eq
\lbeq{U2ptfcn}
    \sigma_p(x) = \sum_{\omega \in \Wcal(0,x)}
    W_{p,D}(\omega)
    K_{R}[0, |\omega|].
\en
The sum is over all walks, with or without self-intersections,
but $K_R$ is nonzero only for self-avoiding walks, for which $K_R=1$.
Thus $K_R$ provides the avoidance interaction.

For a lattice tree $T \ni 0,x$, we let the $R_i$ denote the branches
 (the `beads') along the backbone of $T$
joining $0$ to $x$.
The two-point function for lattice trees can be written
\eq
\lbeq{2.2}
    \rho_p(x) = \sum_{\omega \in \Wcal(0,x)}
    W_{p,D}(\omega)
    \left[ \prod_{i=0}^{|\omega|} 
    \sum_{R_i \in\Tcal(\omega(i),\omega(i))}W_{p,D}(R_i) \right]
    K_R[0,|\omega|]
    .
\en
The additional sums and product in \refeq{2.2}, compared with \refeq{U2ptfcn},
generate the branches attached along the backbone
$\omega$, and the factor $K_R$ ensures that the branches do not intersect.

For a lattice animal $A \in \Acal(x,y)$,
there is a natural order to the set of pivotal bonds for the connection from
$x$ to $y$, and each pivotal bond is directed in a natural way, as
in the left to right order in Figure~\reffg{beads}.
Given two sites $x,y$ and an animal $A$ containing $x$ and $y$, the
{\em backbone}\/
of $A$ is defined to be the ordered set of directed pivotal bonds
for the connection from $x$ to $y$.  In general this backbone
is not connected.  Let $R$  denote the set $R_0,R_1,\ldots$ of
connected components which remain after the removal of the backbone
from $A$ (the `beads').
Let $B= ( (u_1, v_1), ... , (u_{|B|}, v_{|B|}) )$
be an arbitrary finite ordered set of directed bonds.
Let $v_0 = 0$ and
$u_{|B|+1} = x$.  Then the two-point function for lattice animals can be
written as
\eq
\lbeq{2.2a}
    \rho_p^a (x) = \sum_{B: |B| \geq 0} W_{p,D}(B)
    \left[ \prod_{i=0}^{|\omega|} 
    \sum_{R_i \in\Dcal(u_i,v_i)}W_{p,D}(R_i) \right]
    K_R[0,|B|].
\en

The lace expansion proceeds by expanding out the product defining $K_R$,
in each of \refeq{U2ptfcn}--\refeq{2.2a}.  An
elementary but careful
partial resummation is then performed, which leads to a result equivalent
to that of the inclusion-exclusion procedure described in
Section~\ref{sec-le.i-e}.  We will review this procedure now, leading
to precise definitions for $\psi_p(x)$ and hence,
recalling \refeq{Pipsi}, also for $\Pi_z(x)$.

An essential ingredient is the following definition, in which the notion
of lace is defined.  It involves a definition of graph connectivity,
which for self-avoiding walk has been relaxed in the following
compared to the usual definition \cite{BS85,MS93},
to give a unified form of the expansion for
all the models.

\begin{defn}
\label{def-lace}
Given an interval $I = [a,b]$ of positive integers, we refer to a pair
$\{ s, t\}$ of elements of $I$ as an {\em edge}.\/  For $s<t$, we
write simply $st$ for $\{ s,t \}$.
A set of edges is called a {\em graph}.\/
The set of graphs on $[a,b]$ is denoted $\Gcal[a,b]$.  
A graph $\Gamma$ is said to be {\em connected}\/ if, as intervals,
$\cup_{st \in \Gamma}[s,t] = [a,b]$.
A {\em lace}\/ is a minimally connected graph, i.e., a connected graph for which
the removal of any edge would result in a disconnected graph.  The set of
laces on $[a,b]$  is denoted by $\Lcal [a,b]$.
Given a connected graph
$\Gamma$, the following prescription associates to $\Gamma$ a unique
lace ${\sf L}_\Gamma \subset \Gamma$:  The lace ${\sf L}_\Gamma$ consists of
edges $s_1 t_1, s_2 t_2, ...$ where, for $i \geq 2$,
\eq
    s_1 = a , \;\;\;\; t_1 = \max \{t : at \in \Gamma \}
    \quad
    t_{i} = \max \{ t: \exists st \in \Gamma, s \leq t_{i-1} \}
    \quad
    s_i = \min \{ s : st_i \in \Gamma \}  .
\en
The procedure terminates as soon as $t_N=b$.
Given a lace $L$, the set of all edges $st \nin L$
such that ${\sf L}_{L\cup \{st\} } = L $ is called the set of edges
{\em compatible}\/ with $L$ and is denoted  $\Ccal (L)$.
\end{defn}

For $0 \leq a<b$ we define
\eq
\lbeq{Jconngraph}
    J_R [a,b] =
    \sum_{L \in \Lcal [a,b] } \prod_{st \in L} \Ucal_{st}(R)
    \prod_{s't' \in \Ccal (L) } ( 1 + \Ucal_{s't'}(R) ) .
\en
This has a nice interpretation in terms of the beads
of Section~\ref{sec-le.i-e}.  In that language, the product over $\Ccal(L)$ in
\refeq{Jconngraph} is nonzero precisely
when pairs of beads compatible with the lace
$L$ avoid each other, as in the product defining $K_R$.
On the other hand, the product over $L$ is nonzero
precisely when the pairs of beads corresponding to lace edges do intersect
each other.  The number $N=N(L)$ of edges in $L$ corresponds to the
superscript in $\psi_p^{(N)}(x)$ in \refeq{psidef}.

The function $\psi_p(x)$ is defined, for the different models, by
\eqarray
\lbeq{Pidef}
    \psi^{{\rm saw}}_p (x)
    & = & \sum_{\substack{\omega : 0 \rightarrow x \\ |\omega| \geq 2}}
    W_{p,D}(\omega)
    J_\omega [0,|\omega|],
    \\
\lbeq{2.8}
    \psi^{{\rm lt}}_p (x) & = &
    \sum_{\substack{\omega : 0 \rightarrow x \\ |\omega| \geq 1}}
    W_{p,D}(\omega)
    \left[ \prod_{i=0}^{|\omega|} 
    \sum_{R_i \in\Tcal(\omega(i),\omega(i))}W_{p,D}(R_i) \right]
    J_R [ 0, |\omega| ]
    ,
    \\
\lbeq{2.11}
    \psi_p^{{\rm la}} (x) & = &
    (1-\delta_{0,x})\sum_{R \in \Dcal(0,x)} W_{p,D}(R)
    \nonumber \\
    && + \sum_{B : |B| \geq 1} W_{p,D}(B)
    \left[ \prod_{i=0}^{|\omega|} 
    \sum_{R_i \in\Dcal(u_i,v_i)}W_{p,D}(R_i) \right]
    J_R [0, |B| ]
    ,
\enarray
for any $p$ for which the right side converges.
We then define $z$ in terms of $p$ as in \refeq{zp}, and introduce $G_z(x)$
and $\Pi_z(x)$ as in \refeq{Gzx-def} and \refeq{Pipsi}.
The following theorem gives the basic convolution equation
\refeq{Gle.2} for the combinatorial models.  
For self-avoiding walk,
the proof involves a minor modification of the standard proof given in
\cite{BS85,MS93}, to account for the relaxed definition of connectivity.  
For lattice trees and lattice animals, the
proof is given in \cite{HS90b}.

\begin{theorem}
\label{thm-Gexp}
For any $p<p_c$
for which the series defining $\psi_{p}(x)$ is absolutely summable over $x$
(with absolute values taken inside the sums in \refeq{Pidef}--\refeq{2.11}),
the convolution equations \refeq{upsi} and \refeq{Gle.2} hold.
\end{theorem}

\noindent{\em Sketch of proof.}
The proof relies on the elementary identity
\eq
\lbeq{JK}
	K_R[0,b] = K_R[1,b] + J_R[0,b] + \sum_{a=1}^{b-1} J_R[0,a]K_R[a+1,b],
	\quad (b \geq 1).
\en
To prove \refeq{JK}, we first expand the product in \refeq{Kdef}
to obtain 
$K_R[0,b] = \sum_{\Gamma \in \Gcal[0,b]} \prod_{st \in \Gamma} U_{st}(R)$.
Graphs with no edge containing $0$ contribute $K_R[1,b]$.  Graphs with
an edge containing $0$ are then partitioned according to the interval
supporting the connected component containing $0$, and give rise to the
remaining two terms in the identity.

In the first term on the right side of \refeq{JK}, 
interactions between the first and 
subsequent beads do not occur, corresponding to the term
$a_p(pD*U_p)(x)$ of \refeq{upsi}.  The second term gives rise to the term
$\psi_p(x)$ of \refeq{upsi}.  
(For lattice animals, the first term of \refeq{2.11} arises from the case
of just one bead, which does not appear in \refeq{JK}.)
The last term represents an effective
decoupling of the interaction between beads $0$ to $a$ and
beads $a+1$ to $b$, and gives rise to the final
term of \refeq{upsi}.
\qed

For $N \geq 1$,
let $\Lcal^{(N)} [a,b]$ denote the set of laces in $\Lcal [a,b]$ consisting 
of exactly $N$ edges.  We define 
\eq 
\lbeq{2.12z} 
	J^{(N)}_R [a,b] = \sum_{L \in \Lcal^{(N)} [a,b]} \prod_{st \in L} 
	\Ucal_{st}(R) \prod_{s' t' \in \Ccal (L)} ( 1 + \Ucal_{s' t'}(R) ) .
\en 
For $N \geq 1$,
the quantity $\psi_p^{(N)}(x)$ discussed in Section~\ref{sec-le.i-e} then
corresponds to $(-1)^N$ 
times the contribution to \refeq{Pidef}--\refeq{2.11} arising
from the replacement of $J$ by $J^{(N)}$ in those formulas.  This
representation of $\psi_p^{(N)}(x)$ leads to the formula
$\psi_p(x) = \sum_{N=0}^\infty (-1)^N \psi_p^{(N)}(x)$ (with the $N=0$ term
arising only for lattice animals and given by the first term
of \refeq{2.11}).  This formula
was discussed via the inclusion-exclusion approach in the discussion
leading to \refeq{psidef}.

\subsection{Percolation}
\label{sec-le.perc}

The lace expansion discussed in Section~\ref{sec-le.comb}
is combinatorial in nature,
but the expansion for percolation is inherently probabilistic.
It relies entirely on inclusion-exclusion and does
not make use of an interaction
term $\Ucal_{st}$.  It is interesting that an expansion based on
such an interaction can be carried out for oriented percolation \cite{NY93},
which has an additional Markovian structure not present in ordinary
percolation.  However, this has not been done outside the oriented setting.
The expansion we present here, based on inclusion-exclusion,
applies to quite general
percolation models, including oriented percolation.
Before giving a precise statement of the expansion, we first
revisit the discussion of Section~\ref{sec-le.i-e}.

The discussion of the first application of inclusion-exclusion
can be recast as follows, in the
context of percolation.
Let $g_p^{(0)}(x) = \Pbold_p(x \in \Delta(0))$ denote the probability that
$0$ and $x$ are doubly connected.  If these two sites are not doubly connected,
then there is a first pivotal bond $(u,v)$ for the connection.
As in the discussion of lattice animals in Section~\ref{sec-le.comb},
we may regard this bond as being directed.  Let $F(0,u,v,x)$ denote the
event that $0$ and $x$ are connected, but not doubly connected, and that
$(u,v)$ is the first pivotal bond for the connection.  We would like
to approximate $\Pbold_p(F(0,u,v,x))$ by $(g_p^{(0)}*pD*\tau_p)(x)$,
which treats the first bead in the string of beads as independent of the
beads that follow.
To discuss the error in this approximation, we will use the 
following definitions.

\begin{defn}
\label{def-percterms1}
(a)
Given a set of sites $A \subset \Zd$ and a bond configuration, two sites
$x$ and $y$ are {\em connected in}\/ $A$ if there is an occupied path from
$x$ to $y$ having all of its sites in $A$, or if $x=y \in A$.
\newline
(b)
The {\em restricted two-point function}\/ is defined by
\[
    \tau_{p}^A(x,y) = \Pbold_{p} (x \; \mbox{and} \; y \; \mbox{are connected
    in} \; \Zd \backslash A ).
\]
(c)
Given a bond $\{u,v\}$ and a bond configuration, we define
$\tilde{C}^{\{u,v\}}(x)$
to be the set of sites which remain connected to $x$ in the new configuration
obtained by setting $\{u,v\}$ to be vacant.
\end{defn}

It can be shown \cite[Lemma~5.5.4]{MS93} that
\eq
    \Pbold_p(F(0,u,v,x)) = pD(v-u) \Ebold \big[ I[u \in \Delta(0)]
    \tau_p^{\tilde{C}^{\{u,v\}}(0)}(v,x) \big],
\en
where $\Ebold$ denotes expectation with respect to $\Pbold_p$.
The restricted two-point function in the above identity is a random variable,
since the set $\tilde{C}^{\{u,v\}}(0)$ is random.  The approximation discussed
above amounts to replacing the restricted two-point function
simply by $\tau_p(x-v)$, and gives
\eqarray
    \Pbold_p(F(0,u,v,x)) & = & g_p^{(0)}(u) p D(v-u) \tau_p(x-v)
    \nonumber \\ &&
\lbeq{perccort}
    - p D(v-u)\Ebold \big[ I[x \in \Delta(0)]
    \big( \tau_p(x-v) - \tau_p^{\tilde{C}^{\{u,v\}}(0)}(v,x) \big)\big].
\enarray

To understand the correction term in \refeq{perccort}, 
we introduce the following definition.
Two sites $x$ and $y$ are {\em connected
through}\/ $A$ if they are connected
in such a way that every occupied path from $x$ to $y$ has at least one bond
with an endpoint in $A$, or if $x=y \in A$.  Then, by definition,
\eq
    \tau_p(x-v) - \tau_p^{A}(v,x)
    =
    \Pbold_p ( \mbox{$v$ is connected to $x$ through $A$} ).
\en
Therefore
\eqarray
\lbeq{nest}
    \Pbold_p(F(0,u,v,x)) & = & g_p^{(0)}(u) p D(v-u) \tau_p(x-v)
    \\ \nonumber &&
    - p D(v-u)\Ebold \big[ I[x \in \Delta(0)]
    \big( \Ebold I[\mbox{$v$ is connected to $x$
    through $\tilde{C}^{\{u,v\}}(0)$}] \big) \big].
\enarray
In \refeq{nest},
we encounter the occurrence of a nested expectation, corresponding to
a pair of distinct percolation configurations.  This is the
analogue for percolation of the occurrence of independent strings of beads
in the combinatorial models.
The two percolation configurations interact with each other via the event
in the inner expectation, which requires a specific kind of intersection
between them.

An example of a pair of configurations contributing to this nested
expectation is depicted in
Figure~\reffg{Ctilde}.
In the figure, $(u',v')$ is the first pivotal bond for the connection
from $v$ to $x$ such that $v$ is connected to $u'$ through
$\tilde{C}^{\{u,v\}}(0)$.  It is possible that there is no such pivotal
bond, corresponding to a picture in which $u'=x$, and in that case no
further expansion is performed.  In the case where there is such a pivotal
bond, we perform the expansion again by treating the portion of
the cluster of $x$ following $u'$ as independent of the portion preceding
$u'$, in a manner similar to the first application of inclusion-exclusion
performed above.  This is discussed in detail in \cite{HS90a,MS93},
and we now just state the conclusion.

\begin{figure}
\begin{center}
\includegraphics[scale = 0.4]{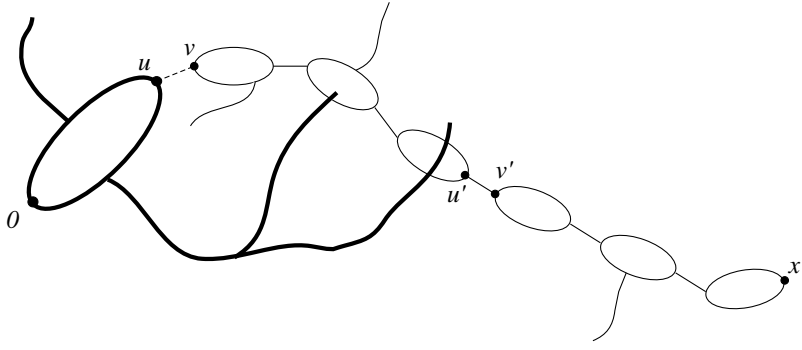}
\end{center}
\caption{\lbfg{Ctilde}
Schematic representation of the two configurations in a contribution to
the nested expectation of \refeq{nest}.  Bold lines represent occupied 
bonds in the outer expectation, while thin lines represent occupied 
bonds in the inner expectation.}
\end{figure}

In doing so, we will use subscripts to coordinate random
sets with the corresponding
expectations.  For example, we write
the subtracted term in \refeq{nest} as
\eq
\lbeq{nest1}
    p D(v-u) \Ebold_0 \big[ I[u \in \Delta(0)]
    \big( \Ebold_1 I[\mbox{$v$ is connected to $x$
    through $\tilde{C}^{\{u,v\}}_0(0)$}] \big) \big]
\en
to emphasise that the set occurring in the inner expectation is a random
set with respect to the outer expectation.
We will also make use of the following definition.
Given sites $x,y$ and a set of sites $A$, let
$E(x,y;A)$ be the event that $x$ is connected to $y$ through $A$ and there
is no directed pivotal bond for the connection from $x$ to $y$
whose first endpoint is connected to $x$ through $A$.
We make the abbreviation
$I_j = I[E(y_j',y_{j+1};\tilde{C}_{j-1})]$
with
$\tilde{C}_{j-1}  = \tilde{C}^{\{y_j,y_j'\}}_{j-1}(y_{j-1}')$
and $y_0'=0$,
and we write $p_{u,v} = pD(v-u)$.
In this notation, the situation with $u'=x$ discussed in the previous
paragraph makes a contribution to \refeq{nest1} equal to
\eq
    p_{u,v} \Ebold_0 \big[ I[u \in \Delta(0)]
    \big( \Ebold_1 I[E(v,x;\tilde{C}_{0})] \big) \big].
\en

Let
\eq
\lbeq{gpxdef}
    \psi_{p}^{(0)}(x) =  (1-\delta_{0,x}) \Pbold_{p} ( x \in \Delta(0) ).
\en
For $n \geq 1$, we define
\eqarray
    \psi_{p}^{(n)}(x) & =  &
    \sum_{(y_1,y_1')} p_{y_1,y_1'} \ldots \sum_{(y_n,y_n')}
    p_{y_n,y_n'}
    \Ebold_0 \big( I[y_1 \in \Delta(0)]
    \nonumber \\ && \quad \times
\lbeq{gpxdef.1}
    \Ebold_1 I_1 \Ebold_2 I_2 \Ebold_3 I_3 \ldots \Ebold_{n-1} I_{n-1}
    \Ebold_n I[E(y_n',x;\tilde{C}_{n-1})] \big) ,
\enarray
where the sums are over directed bonds and 
{\em all}\/ the expectations are nested.  Define
\eq
    \Psi_{p}^{(n)}(x) = \sum_{j=0}^n (-1)^j \psi_{p}^{(j)}(x)
\en
and
\eqarray
\lbeq{percrem}
    R_{p}^{(n)}(x) & = &  \sum_{(y_1,y_1')} p_{y_1,y_1'} \ldots
    \sum_{(y_{n+1},y_{n+1}')} p_{y_{n+1},y_{n+1}'}
    \Ebold_0 \Big( I[y_1 \in \Delta(0)]
    \nonumber \\ && \quad \times
    \Ebold_1 I_1 \Ebold_2 I_2 \ldots \Ebold_n \big(  I_{n}
    (\tau_{p}(x-y_{n+1}') - \tau_{p}^{\tilde{C}_n}(y_{n+1}',x) )\big)\Big).
\enarray
The following theorem is proved in \cite{HS90a}; see also \cite{HS94,MS93}.

\medskip
\begin{theorem}
\label{thm-percexp}
For $p<p_c$ and $N \geq 0$,
\eq
\lbeq{percexpx}
    \tau_{p}(x) = \delta_{0,x} + \Psi_{p}^{(N)}(x)
    + \left( p D * \tau_{p}\right) (x)
    + ( \Psi_{p}^{(N)} * p D * \tau_{p}) (x) +(-1)^{N+1}R_{p}^{(N)}(x).
\en
\end{theorem}

As we will show in Section~\ref{sec-percdiagrams},
the limit $N \to \infty$ can be taken
in \refeq{percexpx} under the hypotheses of Proposition~\ref{prop-diagbd}(c), 
with the remainder term vanishing in the limit.
Defining $\Pi_{p}(x) = \psi_p(x) = \sum_{j=0}^\infty (-1)^j \psi_{p}^{(j)}(x)$,
\refeq{percexpx} then becomes
\eq
\lbeq{percexpxinf}
    \tau_{p}(x) = \delta_{0,x} + \Pi_{p}(x) + \left( p D * \tau_{p}\right) (x)
    + \left( \Pi_{p} * p D * \tau_{p}\right) (x) .
\en
This is equivalent to \refeq{Gle.2}, with $z=p$ and $G_{z}(x) = \tau_{p}(x)$.

\section{Lace expansion diagrams}
\label{sec-Fd}

We begin in Section~\ref{sec-sawdiagrams} by recalling the well-established
procedure by which the lace expansion for self-avoiding walk gives rise
to diagrammatic upper bounds for $\psi_p^{(N)}(x)$ \cite{BS85,MS93}.  
We then bound these diagrams to prove
Proposition~\ref{prop-diagbd}(a).
For the other models, there are also diagrammatic upper bounds.
These bounds can be expressed in the form $\psi_{p}^{(N)}(x)
\leq M^{(N)}(x,x)$, where $M^{(N)}(x,y)$ is a recursively defined function
having a diagrammatic interpretation. 
In Section~\ref{subsub-general}, we prove Lemma~\ref{lem-Mbound},
a key lemma that will be used to
bound $M^{(N)}(x,y)$.
In Sections~\ref{sec-ltdiagrams}--\ref{sec-percdiagrams}, 
we recall the well-established
procedure by which the expansions of
Section~\ref{sec-le} give rise to diagrams \cite{HS90a,HS90b}
for lattice trees, lattice animals and percolation.
We will not provide complete
proofs here but attempt only to motivate the diagrams.
Once the diagrams have been identified, we estimate them 
using Lemma~\ref{lem-Mbound}.  This will provide a proof of
Proposition~\ref{prop-diagbd}(b-c).
In addition, in Section~\ref{sec-percdiagrams}, we will argue that,
for percolation, 
$\lim_{N\to \infty}R^{(N)}_p(x)=0$ under the hypotheses of 
Proposition~\ref{prop-diagbd}(c).

Our bounds here are 
in contrast to all previous diagrammatic estimates in lace expansion
analyses, which have been for $\sum_x \psi_p(x)$
rather than for fixed-$x$ quantities
\cite{BS85,HS90a,HS90b}.

\subsection{Self-avoiding walk diagrams}
\label{sec-sawdiagrams}

For self-avoiding walk, \refeq{2.13} 
simplifies to
\eq 
\lbeq{2.13a} 
	\psi_p^{(N)} (x) = (-1)^N
	\sum_{\substack{\omega : 0 \rightarrow x \\ |\omega| \geq 2}}
    W_{p,D}(\omega)
    J_R^{(N)} [ 0, |\omega| ] .
\en 
The diagrammatic representation of an expression of the form
\refeq{2.13a} has been discussed
many times in the literature, for example in \cite{BS85,MS93}.  
Here we focus on  the differences that arise because of the weakened
definition of connectivity used in Definition~\ref{def-lace}.

The factor $\prod_{st \in L}\Ucal_{st}$ in
$J^{(N)}$ imposes
$N$ bead intersections, which are self-intersections of the random walk.
These self-intersections divide the underlying time interval into 
subintervals, as illustrated in Figure~\reffg{sawlace}(a).  The factor
$\prod_{st \in \Ccal(L)}(1+\Ucal_{st})$ in $J^{(N)}$ is then bounded by
replacing each factor $1+\Ucal_{st}$ for which $s$ and $t$ lie in 
distinct subwalks by the factor $1$.  This produces a bound that can
be interpreted as involving a self-avoiding walk on each time subinterval,
with no interaction between the walks corresponding to distinct time intervals.
For example, the lace of Figure~\reffg{sawlace}(a) gives rise to the diagram of
Figure~\reffg{sawlace}(b).

\begin{figure}
\begin{center}
\includegraphics[scale = 0.4]{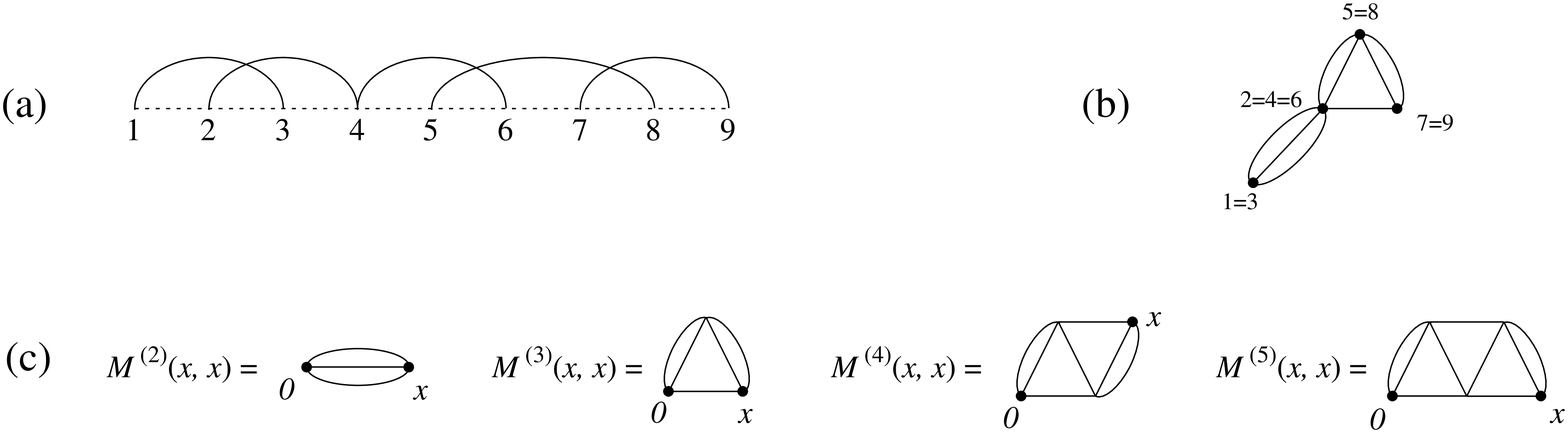}
\end{center}
\caption{\lbfg{sawlace}
(a) A lace with five edges, dividing the time interval into eight subintervals. 
(b) The self-avoiding walk diagram corresponding to this lace.
(c) Diagrams for $\pi_p^{(n)}(x)$.}
\end{figure}

A simplification for these diagrams occurs in the case
where two lace edges abut and do not overlap.  In this case, 
after discarding the interaction between distinct subwalks,
the interaction
decouples across the time coordinate where an abuttal occurs.
If we define $\pi_p^{(N)}(x)$ to be the contribution to the summation
in \refeq{2.13a} only 
from laces with no abuttal, then we are led to 
\eq
	0 \leq \psi_p^{(N)}(x)  
	\leq  
	\sum_{m=1}^N \sum_{\substack{(n_1,\ldots,n_m): \\ n_1+\cdots +n_m=N}}
	(\pi_p^{(n_1)} * \cdots * \pi_p^{(n_m)})(x).
\en
The quantity $\pi_p^{(n)}(x)$ is the quantity that has appeared in previous
lace expansion analyses \cite{BS85,MS93}.  We encounter $\psi_p$ instead,
because we have used a definition of graph connectivity in 
Definition~\ref{def-lace} that is relaxed compared to previous analyses,
to achieve a unified treatment with lattice trees and lattice animals.

We will bound $\psi_p^{(N)}(x)$ by combining a bound on $\pi_p^{(n)}(x)$
with Proposition~\ref{lem-conv}(i).  
To bound $\pi_p^{(n)}(x)$, we define
\eqalign
	\sigma'_{p}(x) & = \sigma_{p}(x) - \delta_{0,x}, \\
\lbeq{Asaw}
	A(u,v,x,y) & = \sigma_p'(v-u) \sigma_p(y-u) \delta_{v,x}, \\
\lbeq{M2sawdef}
	M^{(2)}(x, y) & = \sigma_{p}'(x)^2\sigma_p(y), \\
	M^{(n)}(x, y) & = \sum_{u,v\in \Zd} M^{(n-1)}(u,v)A(u,v,x,y)
	\qquad (n \geq 3).
	\enalign 
The standard bounds of \cite{BS85} can then be written as
\eqalign
\lbeq{pi1bd.0}
	0 & \leq \pi_p^{(1)}(x) \leq  
	\delta_{0,x}\sum_{v \in \Omega_D} pD(v) \sigma_p'(v) , \\
\lbeq{pinbd.0}
	0 & \leq \pi_p^{(n)}(x) \leq M^{(n)}(x,x) \qquad (n \geq 2).
\enalign
The power $3q$ in the desired decay $(|x|+1)^{-3q}$
can be understood
from the fact that there are three distinct routes from $0$ to $x$ in the
diagrams for $M^{(n)}(x,x)$; see Figure~\reffg{sawlace}(c). 

\smallskip
\noindent {\it Proof of Proposition~\ref{prop-diagbd}(a).} 
For self-avoiding walk, we have $z=p$, $G_z(x)=\sigma_p(x)$ and 
$\Pi_z(x)=\psi_p(x)$.  The hypotheses of the proposition are that
$\sigma_p'(x) \leq \beta(|x|+1)^{-q}$,
with $2q>d$, and that $p \leq 2$.  Since $\sigma_p(0)=1$, it follows
that $\sigma_p(x) \leq (|x|+1)^{-q}$.  (The lower bound on $\beta L^{q-d}$ 
assumed
in Proposition~\ref{prop-diagbd} is not
needed for self-avoiding walk.)
We must show that
\eq
\lbeq{psipisaw}
	\psi_p(x) \leq c\beta \delta_{0,x} + c\beta^3 (|x|+1)^{-3q}.
\en

By definition of $\sigma_p'$ and the hypotheses,
it follows from \refeq{pi1bd.0} that
\eq
\lbeq{pi1bd}
	0 \leq \pi_p^{(1)}(x) \leq 2^{1-q} \beta \delta_{0,x}.
\en
By \refeq{Asaw} and hypothesis, 
\eq
\lbeq{Abd.saw}
	A(u,v,x,y) \leq \frac{\beta}{(|v-u|+1)^q (|y-u|+1)^q} \delta_{v,x} . 
\en
Let 
\eq
	S(x) = \sum_{y \in \Zd} \frac{1}{(|y|+1)^q (|x-y|+1)^q},
	\quad \bar{S} = \sup_{x \in \Zd} S(x).
\en
Note that $\bar{S} < \infty$ if $2q>d$, by Proposition~\ref{lem-conv}(i).  
Diagrammatically, $S(x)$ corresponds
to an open bubble, and the condition 
$\bar{S}<\infty$ is closely related to the bubble 
condition \cite[Section~1.5]{MS93}.

We will show that \refeq{Abd.saw} implies there is a constant $C$ such that
\eq
\lbeq{Msawbd}
	M^{(n)}(x,y) \leq \beta ^n (C\bar{S})^{n-2} 
	\frac{1}{(|x|+1)^{2q}(|y|+1)^q}
	\qquad (n \geq 2).
\en 
This gives the conclusion of Proposition~\ref{prop-diagbd}(a), apart from
the fact that the $n=2$ term here has a factor $\beta^2$ rather than
the required $\beta^3$.  The missing factor of $\beta$ can be recovered
by noting that, by definition, $M^{(2)}(x,x)$ can be written as
$\sigma_p'(x)^3$, and we obtain one factor of $\beta$ for each factor
of $\sigma_p'$.  

To prove \refeq{Msawbd}, we use induction on $n$.  The case $n=2$ follows 
immediately from \refeq{M2sawdef} and the assumed bound on $\sigma_p'$.
To advance the induction, we assume \refeq{Msawbd} for $n-1$ and show
that it holds also for $n$.  The inductive hypothesis and 
\refeq{Abd.saw} then give
\eq
	M^{(n)}(x,y) \leq \sum_{u\in \Zd} 
	\frac{\beta^{n-1} (C\bar{S})^{n-3}}{(|u|+1)^{2q}(|x|+1)^q}
	\frac{\beta}{(|x-u|+1)^q (|y-u|+1)^q} 
	\quad (n \geq 3).
\en
It therefore suffices to show that there is a $C$ for which
\eq
\lbeq{sawlem}
	\sum_{u\in \Zd} 
	\frac{1}{(|u|+1)^{2q}}
	\frac{1}{(|x-u|+1)^q (|y-u|+1)^q} 
	\leq \frac{C\bar{S}}{(|x|+1)^{q}(|y|+1)^q}
	\quad (n \geq 3).
\en

To prove \refeq{sawlem}, we consider four cases.
\\
{\em Case 1}: $|u| \geq |x|/2$ and $|u| \geq |y|/2$.
In this contribution to \refeq{sawlem}, we may bound the factor
$(|u|+1)^{-2q}$ above by $2^{2q} (|x|+1)^{-q}(|y|+1)^{-q}$.  The remaining
summation over $u$ is then bounded above by $\bar{S}$, as required.  
\\
{\em Case 2}: $|u| \geq |x|/2$ and $|u| \leq |y|/2$.  The second inequality
implies that $|y-u| \geq |y|/2$.  We then argue as in Case 1.
\\
{\em Case 3}: $|u| \leq |x|/2$ and $|u| \geq |y|/2$. This is the same as 
Case 2, by symmetry.
\\
{\em Case 4}: $|u| \leq |x|/2$ and $|u| \leq |y|/2$.  This follows as
above, using $|y-u| \geq |y|/2$ and $|x-u| \geq |x|/2$.
\qed

\subsection{The diagram lemma}
\label{subsub-general}

In this section, we present a lemma that will be useful
for the diagrammatic estimates for lattice trees, lattice animals and
percolation. 
It involves a constant $\bar{S}$, which is defined for 
$q_1, q_2 >0$ by
\eq
\lbeq{S-def}
    S(x,y)= \sum_{u,v \in \Zd}
    \frac{1}{(|u-v|+1)^{q_1}(|y-u|+1)^{q_2}(|x-v|+1)^{q_1}}, 
	\qquad \bar{S} = \sup_{x,y \in \Zd} S(x,y).
\en
It is possible that $\bar{S}=\infty$, 
depending on the values of $d$ and the $q_i$.
However, if
\eq
\lbeq{q12}
        2q_1+d >  2q_1+q_2 > 2d  
\en
then $\bar{S}<\infty$.  In fact, given \refeq{q12}, it follows from 
Proposition~\ref{lem-conv}(i) that 
    \eq
    \lbeq{S1.7bd}
	S(x,y) 
    \leq \frac{C}{(|x-y|+1)^{2q_1+q_2-2d}}.
    \en
\begin{figure}
\begin{center}
\includegraphics[scale = 0.4]{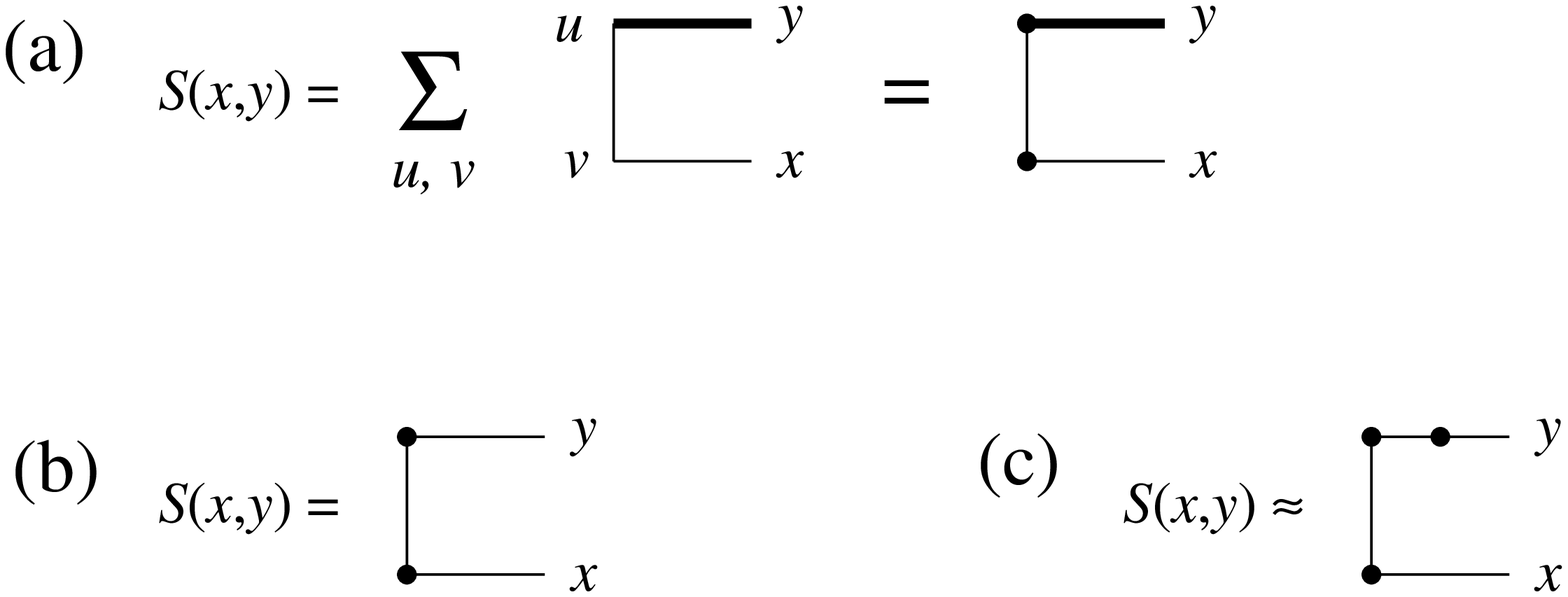}
\end{center}
\caption{\lbfg{S}
(a) The diagram for $S(x,y)$.  Thin lines decay with power $q_{1}$ and the
thick line decays with power $q_{2}$.  Unlabelled vertices are summed over. 
(b) The open triangle diagram for percolation 
($q_{1} = q_{2} = q$). 
(c) The open square diagram for lattice trees and lattice animals 
($q_{1} = q, q_{2} = 2q-d$). 
}
\end{figure}

Finiteness of $\bar{S}$ is related to the triangle condition for percolation
\cite{AN84} and to the square 
condition for lattice trees and lattice animals \cite{TH87}.  
To see this, we first
note the diagrammatic representation of $S(x,y)$ in figure~\reffg{S}(a).
When $q_1=q_2=q$, which is the relevant case for percolation,
this corresponds to the open 
triangle diagram depicted in Figure~\reffg{S}(b).  
When $q_1=q$ and $q_2=2q-d$, which is the relevant case for lattice
trees and lattice animals, $S$ corresponds 
to the open square diagram depicted in Figure~\reffg{S}(c).  To understand this
for the square diagram, we interpret the line decaying with power $q_2$
as arising from a convolution of two two-point functions decaying with
power $q$, in accordance with Proposition~\ref{lem-conv}(i).

The following lemma is the key lemma that will be used in bounding diagrams
for lattice trees, lattice animals and percolation.
Its statement involves functions $\Abegin: \Z^{2d}\to [0,\infty)$, 
$A^{(i)}:\Z^{4d}\to[0,\infty)$ for $i \geq 1$,  
$\Aend:\Z^{4d}\to[0,\infty)$,
and functions $M^{(N)}:\Z^{2d} \to [0,\infty)$
defined for $N \geq 1$ by 
    \eqalign\lbeq{ind-M}
    M^{(N)}(x,y)
	&=
	\sum_{u_1,v_1,\ldots,u_{N},v_{N} \in \Zd}
	\Abegin(u_{1},v_{1}) 
	\prod_{i=1}^{N-1} A^{(i)}(u_{i},v_{i},u_{i+1},v_{i+1})
	\Aend(u_{N},v_{N},x, y).
    \enalign
(For $N=1$, the empty product over $i$ is interpreted as $1$.)
The proof of Lemma~\ref{lem-Mbound} can be extended to $q_1<d$ and $q_2$
obeying \refeq{q12}, but since $q_2 \leq q_1$ in our applications, we 
add this assumption to simplify the proof.

\begin{lemma} 
\label{lem-Mbound} 
Fix $q_2\leq q_1 <d$ obeying \refeq{q12}, so that
$\bar{S} < \infty$.  
Let $K_0>0$.  Suppose that  
\eq
\lbeq{as-M1}
	\Abegin(x,y) \leq  K_0
    \left\{\frac{1}{(|x|+1)^{q_1}(|y|+1)^{q_2}}+
    \frac{1}{(|x|+1)^{q_2}(|y|+1)^{q_1}}\right\}, 
\en 
and suppose that $A^{(i)}$ for $i \geq 1$ and $\Aend$ satisfy 
    \eq\lbeq{as-A}
    A^{(*)}(u,v,x,y)
    \leq
    \frac{K_*}{(|u-v|+1)^{q_1}}
    \left\{\frac{1}{(|y-u|+1)^{q_2}(|x-v|+1)^{q_1}}+
    \frac{1}{(|y-u|+1)^{q_1}(|x-v|+1)^{q_2}}\right\}
    \en
with $K_*>0$.
Then there is a $C$ depending on $d,q_1,q_2$ such that for $N \geq 1$
    \eq
    \lbeq{MNbd}
    M^{(N)}(x,y) \leq ( C \bar{S})^{N-1} 
    \Big(\prod_{i=0}^{N-1}K_i \Big) K_{\rm end}
    \left\{\frac{1}{(|x|+1)^{q_1}(|y|+1)^{q_2}}+
    \frac{1}{(|x|+1)^{q_2}(|y|+1)^{q_1}}\right\}.
    \en
\end{lemma}

\begin{proof}
The proof is by induction on $N$.   
To deal with the fact that $M^{(N)}$ is not defined
literally by a convolution of $M^{(N-1)}$ with $\Aend$, we proceed as
follows.  Let $\tilde{M}^{(N)}$ be the 
quantity defined by replacing $\Aend$ by $A^{(N)}$ in the definition of 
$M^{(N)}$.  Because all the constituent factors in the definitions of 
$M^{(N)}$ and 
$\tilde{M}^{(N)}$ obey the same bounds, it suffices to prove that
$\tilde{M}^{(N)}$ obeys \refeq{MNbd} with $K_{\rm end}$
replaced by $K_N$.  We prove this by induction,
with the inductive hypothesis that $\tilde{M}^{(N-1)}$ obeys 
\refeq{MNbd} with $K_{\rm end}$ replaced by $K_{N-1}$ and
$N$ replaced by $N-1$ on the right side.

\begin{figure}
\begin{center}
\includegraphics[scale = 0.4]{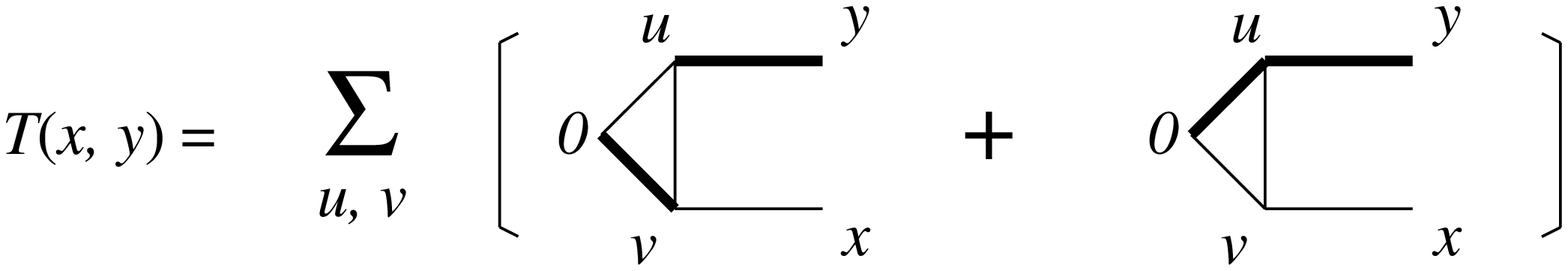}
\end{center}
\caption{\lbfg{prfLem41}
Diagram for $\oldB(x,y)$.  Thin lines decay 
with power $q_{1}$, while thick lines decay with power $q_{2}$. 
}
\end{figure}

For $x,y\in \Zd$, let
    \eqalign 
    \oldB(x,y) & =  \sum_{u,v\in \Zd} 
    \left\{\frac{1}{(|u|+1)^{q_1}(|v|+1)^{q_2}}+
    \frac{1}{(|u|+1)^{q_2}(|v|+1)^{q_1}}\right\}\lbeq{B-def}
    \nonumber \\
    &\quad \quad
    \times \frac{1}{(|u-v|+1)^{q_1}(|y-u|+1)^{q_2}(|x-v|+1)^{q_1}}.
    \enalign
This quantity is depicted in Figure~\reffg{prfLem41}. 
By definition, and using \refeq{as-M1}--\refeq{as-A},
    \eq
    \lbeq{Mind1}
    \tilde{M}^{(1)}(x,y) \leq K_0K_1  \big[\oldB(x,y)+\oldB(y,x)\big].
    \en
By the induction hypothesis, \refeq{ind-M} and \refeq{as-A},   
    \eq
    \tilde{M}^{(N)}(x,y) 
    \leq (C\bar{S})^{N-2} 
    \Big(\prod_{i=0}^{N}K_i \Big) \big[\oldB(x,y)+\oldB(y,x)\big].
    \en
It therefore suffices to show that
    \eq
    \lbeq{Bbound}
    \oldB(x,y) \leq 
    \frac{1}{2} C \bar{S} \left\{\frac{1}{(|x|+1)^{q_1}(|y|+1)^{q_2}}
    +\frac{1}{(|x|+1)^{q_2}(|y|+1)^{q_1}}\right\}.
    \en
    
To prove \refeq{Bbound}, we write $\oldB(x,y) \leq \sum_{i=1}^4 \oldB_i(x,y)$,
with $\oldB_i(x,y)$ defined to be the contribution to $T(x,y)$ arising
from each of the following four cases.  In the discussion of these four
cases, $C$ denotes a generic constant whose value may change from line to line.

\noindent {\it Case 1.} $|v| \geq |x-v|$ and $|u| \geq |u-y|$.
This implies $|v| \geq |x|/2$ and $|u| \geq |y|/2$, so that 
    \eq
	\lbeq{Bcase1}
    \oldB_1(x,y) \leq 
    C \bar{S}
    \left\{\frac{1}{(|x|+1)^{q_1}(|y|+1)^{q_2}}
	+\frac{1}{(|x|+1)^{q_2}(|y|+1)^{q_1}}\right\}.
    \en
    
\noindent {\it Case 2.} $|v| \geq |x-v|$ and $|u| \leq |u-y|$.
This implies $|v| \geq |x|/2$ and $|u-y| \geq |y|/2$.  Then  
    \eqalign
    \oldB_2(x,y) & \leq 
    \frac{C}{(|y|+1)^{q_2}}\sum_{u,v}
    \left\{
    \frac{1}{(|u|+1)^{q_1}(|x|+1)^{q_2}}
    + \frac{1}{(|u|+1)^{q_2}(|x|+1)^{q_1}}\right\}
    \nonumber \\
    \lbeq{Bcase2}
    &
    \qquad \times
    \frac{1}{(|u-v|+1)^{q_1}(|x-v|+1)^{q_1}}.
    \enalign
The second term of \refeq{Bcase2}
is bounded above by $C\bar{S}(|x|+1)^{-q_1}(|y|+1)^{-q_2}$, as required.
We bound the first term using Proposition~\ref{lem-conv}(i) to estimate
the two convolutions, obtaining a bound $C(|x|+1)^{-(3q_1+q_2-2d)}
(|y|+1)^{-q_2}$.  (Here we used the assumption $q_2 \leq q_1$ to ensure
that $3q_1-2d>0$, as required to apply Lemma~\ref{lem-conv}(i).)
It follows from \refeq{q12} that  
$3q_1+q_2-2d >q_1$,
which gives the desired result.

\noindent {\it Case 3.} $|v| \leq |x-v|$ and $|u| \geq |u-y|$.
This implies $|v-x| \geq |x|/2$ and $|u| \geq |y|/2$, and hence
    \eqalign
    \oldB_3(x,y) &\leq 
    \frac{C}{(|x|+1)^{q_1}(|y|+1)^{q_2}} \sum_{u,v}
    \left\{
    \frac{1}{(|y-u|+1)^{q_1-q_2}(|v|+1)^{q_2}}
    +\frac{1}{(|v|+1)^{q_1}}
    \right\}\\
    &\qquad\times 
    \frac{1}{(|u-v|+1)^{q_1}(|y-u|+1)^{q_2}}.
    \nonumber
    \enalign
Each term is bounded by
$C \bar{S}(|x|+1)^{-q_1}(|y|+1)^{-q_2}$, as required.

 \noindent {\it Case 4.} $|v| \leq
|x-v|$ and $|u| \leq |u-y|$. This implies $|v-x| \geq |x|/2$ and
$|u-y| \geq |y|/2$, and hence
    \eq
    \oldB_4(x,y) \leq 2C \bar{S} \frac{1}{(|x|+1)^{q_1}(|y|+1)^{q_2}}.
    \en
Adding the contributions in the four cases yields \refeq{Bbound}
and completes the proof.
\end{proof}

\begin{rk}
\label{rk-diag}
Let 
\eq
\lbeq{B4def}
	\oldBfour(z, w, x, y) = \sum_{u,v} \frac{1}{(|z-u|+1)^{q}} 
	\frac{1}{(|y-u|+1)^{q}} \frac{1}{(|w-v|+1)^{q}} \frac{1}{(|x-v|+1)^{q}} 
	\frac{1}{(|u-v|+1)^{q}} . 
\en 
By dividing into four cases according to whether  $|z-u|$ is greater than
or less than $|y-u|$ and
whether $|w-v|$ is greater than or less than $|x-v|$, the above proof can
be easily adapted to show that
\eq
\lbeq{Hbd.1}
	\oldBfour(w,z,x,y) \leq \frac{C\bar{S}}
	{(|y-z|+1)^{q} (|x-w|+1)^{q}} ,
\en 
where $\bar{S}$ is defined in 
\refeq{S-def} with now $q_{1} = q_{2} = q$.  
This will be used in Section~\ref{sec-percdiagrams} to analyse percolation.
\end{rk}

\subsection{Lattice tree diagrams}
\label{sec-ltdiagrams}
For lattice trees, the quantity 
$\psi_p^{(N)}(x)$ ($N \geq 1$) can be understood either as arising from
$N$ applications of inclusion-exclusion, along the lines discussed
in Section~\ref{sec-le.i-e}, or from the contribution
to \refeq{2.8} from laces having $N$ edges, as explained around
\refeq{2.12z}.
Explicitly,  
\eq 
\lbeq{2.13} 
	\psi_p^{(N)} (x) = (-1)^N
	\sum_{\substack{\omega : 0 \rightarrow x \\ |\omega| \geq 1}}
    W_{p,D}(\omega)
    \left[ \prod_{i=0}^{|\omega|}   
    \sum_{R_i \in \Tcal(\omega(i), \omega(i))} W_{p,D}(R_i) \right]
    J_R^{(N)} [ 0, |\omega| ]
    .
\en 
For a nonzero contribution
to $\psi_{p}^{(N)} (x)$, the factor $\prod_{st \in L} \Ucal_{st}$ in $J^{(N)}$
enforces intersections between the beads $R_s$ and $R_t$, for each $st \in L$.
This leads to bounds in which the contribution to $\psi_p^{(N)}(x)$ from
the $N$-edge laces can be bounded above by $N$-loop diagrams.  We 
illustrate this in detail only for the simplest case $N=1$.  

To bound $\psi_p^{(1)} (x)$, we proceed as follows.  There is a 
unique lace $0|\omega|$ consisting of a single edge, and all other edges
on $[0,|\omega|]$ are compatible with it.  Therefore
\eq
\lbeq{2.15}  
	\psi_p^{(1)} (x) = -
	\sum_{\substack{\omega : 0 \rightarrow x \\ |\omega| \geq 1}} 
	W_{p,D}(\omega)
	\left[ \prod_{i=0}^{|\omega|}   
    \sum_{R_i \in \Tcal(\omega(i),\omega(i))} W_{p,D}(R_i) \right] 
	\Ucal_{0 |\omega|}
	\prodtwo{0 \leq s < t \leq |\omega|}{(s,t) \neq (0, |\omega|)} 
	(1 + \Ucal_{st} ) . 
\en
After relaxing the last product to 
$\prod_{1 \leq s < t \leq |\omega|} (1 + \Ucal_{st})$,
the trees $R_{1}, \ldots, R_{l}$, together with the 
bonds of $\omega$ connecting them, can be considered as a single lattice  
tree connecting $\omega(1)$ and $x$.  
Writing this tree as $T_{1}$, writing $v = \omega(1)$, 
and stating the constraint 
imposed by $\Ucal_{0|\omega|}$ in words, we obtain  
\eqalign 
	0 \leq \psi_{p}^{(1)}(x) 
	& \leq  \sum_{v \in \Omega_D} pD(v)
	\sum_{R_{0} \in \Tcal(0,0)} W_{p,D}(R_0) 
	\sum_{T_{1} \in \Tcal( v, x)} W_{p,D}(T_1) \nnb 
	& \hspace{10mm} \times
	I [ \text{$R_{0}$ and the bead at $x$ of $T_{1}$ share a common site}] 
	\nnb 
\lbeq{R0T1}
	& \leq 
	\sum_{y\in \Zd} \sum_{v \in \Omega_D} pD(v)
	\sum_{R_{0} \in \Tcal(0,y)} W_{p,D}(R_0) 
	\sum_{T_{1} \in \Tcal( v, x)} W_{p,D}(T_1) \, 
	I [ (\text{bead at $x$ of $T_{1}$}) \ni y] .  
\enalign 
In \refeq{R0T1}, the summations over $R_{0}$ and $T_{1}$ can be performed 
independently.  The summation over $R_0$ simply gives $\rho_p(y)$.
For the summation over $T_{1}$, we  
note that there must be disjoint connections from $v$ to $x$ and
from $x$ to $y$, because $y$ is in the last bead of $T_{1}$.  Therefore the
sum over $T_1$ is bounded above by $\rho_p(x-v)\rho_p(y-x)$.  Define
\eq
\lbeq{rhotildef}
	\tilde{\rho}_p(x) 
	= (pD*\rho_p)(x) 
	= \sum_{v \in \Omega_D} pD(v)\rho_p(x-v).
\en
Then the above bound gives
\eq
\lbeq{lttri}
	0 \leq \psi_p^{(1)}(x)  
	\leq \sum_{y \in \Zd} \tilde{\rho}_p(x) \rho_p(y-x) \rho_p(y).
\en

\begin{figure}[h]
\begin{center}
\includegraphics[scale = 0.4]{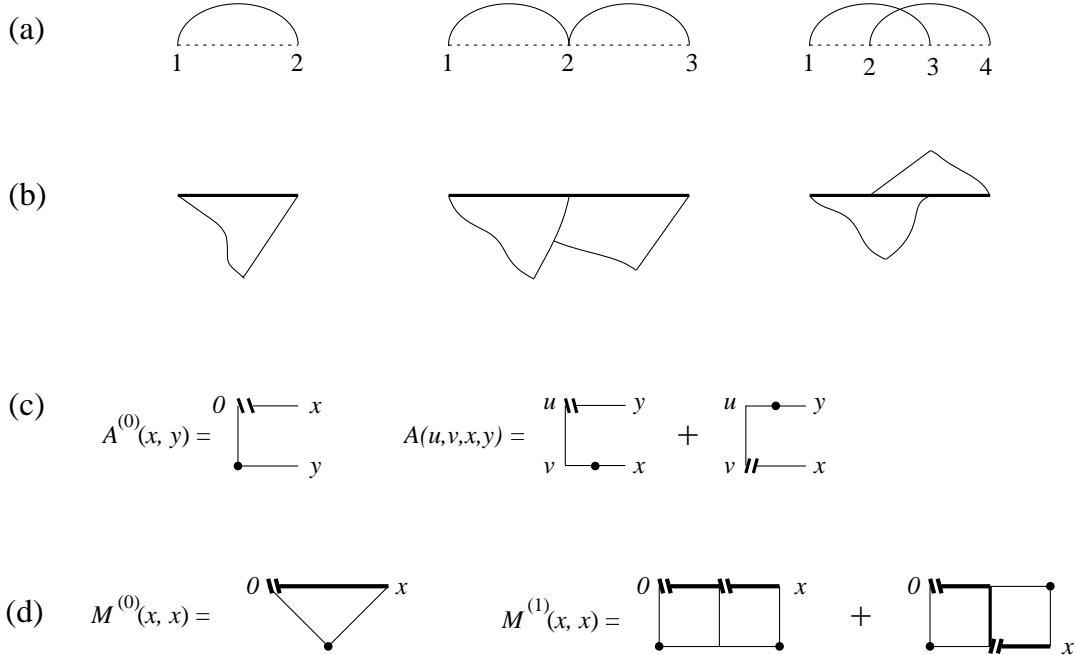}
\end{center}
\caption{\lbfg{treelace}
(a) The laces with one and two edges.  
(b) Bead intersections imposed by the laces.  
(c) Constituents for constructing $M^{(N)}$, where $A$ stands for 
both $A^{(i)}$ and $\Aend$.    Lines ending with double bars represent
$\tilde{\rho}$-lines.
(d) The one-loop and two-loop lattice tree diagrams, with 
lines corresponding to the backbone drawn in bold.
}
\end{figure}

For $N \geq 2$, a similar analysis can be performed, along the lines
discussed in \cite{HS90b}.
To state the resulting bound, we define  
\eq
\lbeq{M0ltdef}
	M^{(0)}(x,y) = \Abegin(x, y) 
	= \tilde{\rho}_p(x) \sum_{v}\rho_{p}(y-v) \rho_{p}(v)
\en 
and  
\eqalign
\lbeq{Adeflt}
        A^{(i)}(u,v,x,y) & 
		= \rho_p(v-u) \biggl [ \tilde{\rho}_p(y-u) \sum_{a \in \Zd}
    \rho_p(a-v) \rho_p(x-a) 
    + \tilde{\rho}_p(x-v) \sum_{a \in \Zd}
    \rho_p(a-u)\rho_p(y-a) \biggr ],
\enalign 
with $\Aend = A^{(i)}$.
We define
$M^{(N)}(x,y)$ ($N \geq 1$) recursively by \refeq{ind-M}. 
Then, for $N \geq 1$, the resulting bound is  
\eq
\lbeq{ltpsibd}
	0 \leq \psi_p^{(N)}(x) \leq M^{(N-1)}(x,x).
\en   
The first few diagrams are depicted in Figure~\reffg{treelace}.
The upper bound \refeq{ltpsibd} differs from the bound of \cite{HS90b},
which uses $\rho_p$ in place of $\tilde{\rho}_p$ in \refeq{Adeflt}.  We could
also use the bounds of \cite{HS90b} here, but the bounds with $\tilde{\rho}_p$
are easier to derive and lead ultimately to the same conclusion.

\smallskip
\noindent {\it  
Proof of Proposition~\ref{prop-diagbd}(b) for lattice trees.}
For lattice trees, we have $z=p\rho_p(0)$, 
$G_z(x)=\rho_p(x)/\rho_p(0)$ and 
$\Pi_z(x)=\psi_p(x)/\rho_p(0)$.  
The hypotheses of the proposition are that
$G_z(x) \leq \beta(|x|+1)^{-q}$ for $x \neq 0$,
with $\frac{3}{4}d < q < d$, that there is a constant $R$
such that $\rho_p^{(a)}(0) \leq R$, and that $\beta L^{q-d}$ is bounded
away from zero.
It follows that $\rho_p(x) \leq R\beta (|x|+1)^{-q}$ for $x \neq 0$. 
Since $\rho_p(0) \geq 1$, it is sufficient to conclude that
\eq
\lbeq{psibdlt}
	\psi_p(x) \leq c\beta \delta_{0,x} + c\beta^2 (|x|+1)^{d-3q},
\en
where $c$ may depend on $R$.  

By definition 
\eq
\lbeq{rhotilbd}
	\tilde{\rho}_p^{(a)}(x) = 
	pD(x) \rho_p^{(a)}(0)  
	+ \sum_{v \in \Omega_D : v \neq x} pD(v) \rho_p^{(a)}(x-v).
\en
Note that $p = \sum_{v \in \Omega_D}pD(v) < \rho_p^{(a)}(0)
\leq R$.  The first term on the right side can be bounded
as in \refeq{Dchi}, 
while the second term can be estimated by considering separately the
contributions due to $|x| \geq 2L$ and $|x|<2L$.  The result is
\eq
\lbeq{rhotilbd.1}
	\tilde{\rho}_p(x) \leq 
	\frac{C}{L^{d-q}(|x|+1)^q} + \frac{C\beta}{(|x|+1)^q} 
	\leq \frac{C\beta}{(|x|+1)^q},
\en
where we have invoked the hypothesis that $\beta L^{q-d}$ is bounded away
from zero.   

Therefore, by definition and by 
Proposition~\ref{lem-conv}(i),
\eq
\lbeq{M0lt}
	M^{(0)}(x,x) \leq c\beta (|x|+1)^{d-3q}.
\en
Similarly, $A^{(0)}(x,y)$ of \refeq{M0ltdef} obeys the bound of
\refeq{as-M1} with $q_1=q$ and $q_2=2q-d$.
Moreover, the factor $\beta$ on the right side 
of \refeq{M0lt} can be replaced by $\beta^2$
when $x \neq 0$,
since at least one of the two lower lines in the first diagram of
Figure~\reffg{treelace}(b) must make a nonzero displacement when $x \neq 0$.

For $N \geq 1$, we will show that the hypotheses imply 
    \eq
    \lbeq{MNltla}
    M^{(N)}(x,x) \leq  \frac{\beta^{N+1} C_1^N}{(|x|+1)^{3q-d}},
    \en
where $C_1$ is a constant.
By \refeq{ltpsibd} and \refeq{M0lt}, this will complete the proof.
The remainder of the proof is devoted to proving \refeq{MNltla}.

By Proposition~\ref{lem-conv}(i) and the above remarks, the function 
$A^{(i)}$ defined in \refeq{Adeflt} obeys 
    \eq
    A^{(i)}(u,v,x,y) \leq \frac{C\beta}{(|u-v|+1)^q}
    \left[
    \frac{1}{(|y-u|+1)^q (|x-v|+1)^{2q-d}} +
    \frac{1}{(|y-u|+1)^{2q-d}(|x-v|+1)^{q}}
    \right].
    \en
Hence, \refeq{as-A} applies with $q_1=q$, $q_2=2q-d$. By our assumption
on $q$, it follows that $q_2 \leq q_1 <d $ and \refeq{q12} is satisfied.  
Therefore, by Lemma~\ref{lem-Mbound}, there is a constant
$C_1$ such that
    \eq
    M^{(N)}(x,y) \leq  \beta^{N+1} C_1^N\left\{
    \frac{1}{(|x|+1)^q (|y|+1)^{2q-d}}
    +
    \frac{1}{(|x|+1)^{2q-d} (|y|+1)^q}
    \right\}
    .
    \en
This implies \refeq{MNltla} and completes the proof for lattice trees.
\qed

\subsection{Lattice animal diagrams}
\label{sec-ladiagrams}

The determination of the lattice animal diagrams is similar to that
for lattice trees.  It makes use of \cite[Lemma~2.1]{HS90b}, which can
be rephrased in our present context as follows.

\begin{lemma}
\label{lem-ladisj}
Given sets of lattice paths $E_i$ ($i = 1,\ldots , n$), let $\Acal_i$
denote the set of lattice animals which contain a path in $E_i$,
and let $\Acal$ denote the set of lattice animals which contain disjoint
paths in each of $E_1,\ldots, E_n$.  Then
\eq
	\sum_{A \in \Acal} W_{p,D}(A) \leq \prod_{i=1}^n \left[
	\sum_{A_i \in \Acal_i} W_{p,D}(A_i).
	\right]
\en
\end{lemma}

We denote the first term on the right side of \refeq{2.11} by 
$\psi_p^{(0)}(x)$ and denote the contribution to the second term due to
$J_R^{(N)}[0,|B|]$ by $\psi_p^{(N)}(x)$.
By Lemma~\ref{lem-ladisj}, 
\eq
\lbeq{psi0bd}
	\psi_p^{(0)}(x) 
	= (1-\delta_{0,x}) \sum_{A \in \Dcal(0,x)} W_{p,D}(A)
	\leq (1- \delta_{0,x})\rho_p^a(x)^2 .
\en  
By definition, 
\eq
\lbeq{psi1ltdef}
	\psi_{p}^{(1)}(x) = - \sum_{|B|:|B|\geq 1} 
	W_{p,D}(B) \, \biggl [
	\prod_{i=0}^{|B|} \sum_{R_{i} \in \Dcal(v_{i}, u_{i+1})} W_{p,D}(R_i) 
	\biggr ] \Ucal_{0 |B|} 
	\prod_{\substack{0 \leq s < t \leq |B|\\ st \neq 0|B|}} 
	(1 + \Ucal_{st}) ,
\en 
where the sum over $B$ is a sum over $|B|$ bonds
$(u_i,v_i)$ with $v_i-u_i \in \Omega_D$, where $v_0=0$ and $u_{|B|+1}=x$.

After relaxing the avoidance constraint in \refeq{psi1ltdef} to 
$\prod_{1 \leq s < t \leq |B|} (1 + \Ucal_{st})$, the
beads $R_{1}, \ldots, R_{|B|}$, together with the
pivotal bonds connecting them, can be considered as a single lattice  
animal connecting $v_{1}$ and $x$.  
Writing this animal as $A_{1}$, and stating the constraint 
imposed by $\Ucal_{0|B|}$ in words, we obtain  
\eqalign 
	0 \leq \psi_{p}^{(1)}(x) 
	& \leq  \sum_{(u, v)} pD(v-u)
	\sum_{R_{0} \in \Dcal(0, u)} W_{p,D}(R_{0}) 
	\sum_{A_{1} \in \Acal( v, x)} W_{p,D}(A_{1}) \nnb
	& \hspace{10mm}\times 
	I [ \text{$R_{0}$ and the last bead of $A_{1}$ share a common site}] 
	\nnb 
\lbeq{D0A1}
	& \leq 
	\sum_{y} \sum_{(u, v)} pD(v-u)
	\sum_{R_{0} \in \Dcal(0, u) : R_0 \ni y}  W_{p,D}(R_{0})  
	\sum_{A_{1} \in \Acal( v, x)} W_{p,D}(A_{1})  
	I [(\text{last bead of $A_{1}$}) \ni y] .  
\enalign 
In \refeq{D0A1}, the summations over $R_{0}$ and $A_{1}$ can be performed 
independently.  For the summation over $R_{0}$, we note that there must 
be a site $w$, and four disjoint connections joining $0$ to $w$, $w$ to $u$, 
$u$ to $0$, and $w$ to $y$.  For the summation over $A_{1}$, 
there must be disjoint connections joining $x$ to $v$ and $x$ 
to $y$, because $y$ is in the last bead of $A_{1}$.  
This is illustrated in Figure~\reffg{ltla1}, where on the left 
we show a typical contribution to the one-loop diagram, and on the 
right we show the connections used to bound it.  
Therefore, using Lemma~\ref{lem-ladisj} we obtain
\eq
\lbeq{la1}
	0 \leq \psi_p^{(1)}(x)
	\leq
	\sum_{u,w,y \in \Zd}
	\rho^a_p(u)\rho^a_p(w)\rho^a_p(u-w)\rho^a_p(y-w)\rho^a_p(x-y)
	\tilde{\rho}^a_p(x-u),
\en
where $\tilde{\rho}^a_p(x) = (pD*\rho^a_p)(x)$ as in \refeq{rhotildef}.
This diagram is depicted in Figure~\reffg{ltla2}.  The contribution
arising from the term with $u=w=0$ equals $\rho_p^a(0)^3$ times
the triangle diagram of \refeq{lttri}.  Taking the full
sum into account, 
the right side of \refeq{la1} corresponds diagrammatically to
the triangle diagram \refeq{lttri} with its vertex at the origin replaced
by a triangle.   

\begin{figure}[h]
\begin{center}
\includegraphics[scale = 0.5]{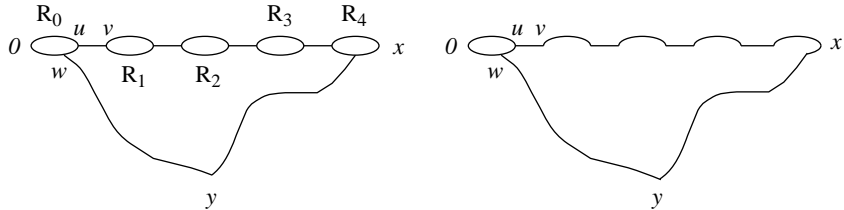}
\end{center}
\caption{
\lbfg{ltla1}
Configuration for the lattice animal one-loop diagram.}
\end{figure}

\begin{figure}[h]
\begin{center}
\includegraphics[scale = 0.4]{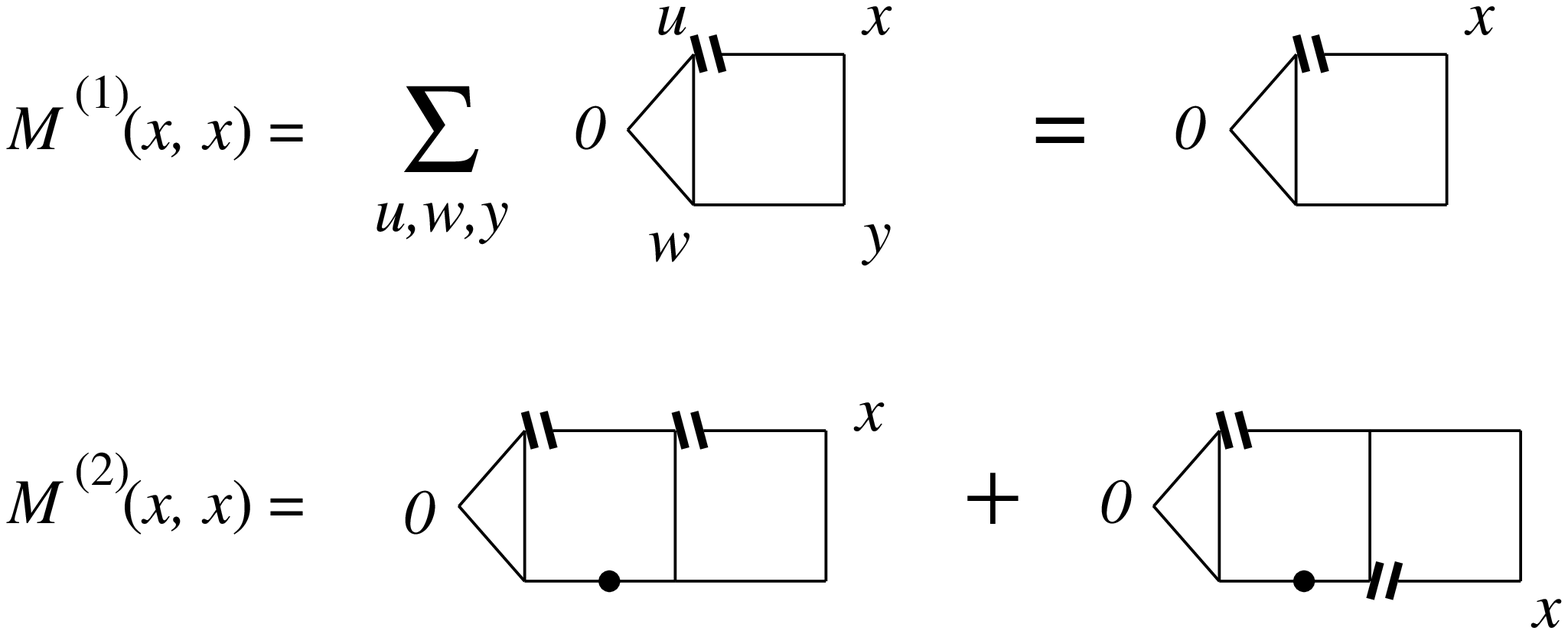}
\end{center}
\caption{
\lbfg{ltla2}
The one-loop and two-loop diagrams for lattice animals.  Lines 
ending with double bars represent $\tilde{\rho}^{a}$-lines.
}
\end{figure}

The above procedure can be extended to bound the higher-order terms.
The resulting diagrams are the lattice tree diagrams, with an extra initial
triangle as observed for $\psi_p^{(1)}(x)$.  
Now we define 
\eq
	\Abegin(x,y) 
	= \rho_{p}^{a}(x) \rho_{p}^{a}(y)
\en 
and use the $A^{(i)}=A^{\rm end}$ of \refeq{Adeflt} (with 
$\rho$ replaced by  $\rho^{a}$) to define $M^{(N)}$ 
recursively by \refeq{ind-M} for $N \geq 1$. 
Then for $N \geq 1$ we have 
\eq
\lbeq{laN}
	0 \leq \psi_p^{(N)}(x) \leq M^{(N)}(x,x) . 
\en
The cases $N=1,2$ are depicted in Figure~\reffg{ltla2}.

The bounds described above for lattice 
animals differ from those of \cite{HS90b}
in two respects.  One difference is that the diagrams of \cite{HS90b} involve
additional small triangles that make no significant difference and need not be
included.  A second difference is that here we are using $\tilde{\rho}^a$
whereas only $\rho^a$ was used in \cite{HS90b}.  It is in fact possible
to avoid the use of $\tilde{\rho}^a$, but a more involved argument than the
one provided in \cite{HS90b} is necessary for this.  However, the use
of $\tilde{\rho}^a$ poses no difficulties and is simpler, 
so we will use it here.

\smallskip
\noindent {\it  
Proof of Proposition~\ref{prop-diagbd}(b) for lattice animals.}
The proof proceeds in the same way as for lattice trees.
One minor difference for lattice animals is the presence of the term
$\psi_p^{(0)}$, for which \refeq{psi0bd} implies
\eq
	0 \leq \psi^{(0)}_p(x) \leq
	\frac{R^2\beta^2}{(|x|+1)^{2q}}(1-\delta_{0,x}).
\en
Since $2q > 3q -d$ by assumption, this is smaller than what is
required (second term of
\refeq{psibdlt}).
A second minor change is that the extraction of the extra factor
$\beta$ from the bound on $\psi_p^{(1)}$ is slightly different.
\qed

\subsection{Percolation diagrams}
\label{sec-percdiagrams}

For percolation, the BK inequality \cite{Grim99} plays the role that
Lemma~\ref{lem-ladisj} played for lattice animals.  In particular,
application of the BK inequality to \refeq{gpxdef} immediately gives
\eq
\lbeq{pd.1}
	\psi_p^{(0)}(x) \leq (1-\delta_{0,x}) \tau_p(x)^2.
\en
Higher order contributions can also be bounded using the BK inequality.
For example, application of BK to the contribution to Figure~\reffg{Ctilde}
when $u'=x$ leads to the bound
\eq
\lbeq{pd.2}
	\psi_p^{(1)}(x) \leq 
	\sum_{u,w,y,z \in \Zd}
	\tau_p(u)\tau_p(w)\tau_p(u-w) \tilde{\tau}_p(y-u) \tau_p(y-z)
	\tau_p(z-w) \tau_p(x-y) \tau_p(x-z),
\en
where
\eq
        \tilde{\tau}_p(x) = \sum_{v \in \Omega_D} pD(v) \tau_p(x-v).
\en
The right side of \refeq{pd.2} is depicted in Figure~\reffg{perc}.  It
involves the two distinct routes $0 \to u \to y \to x$
and $0 \to w \to z \to x$ from $0$ to $x$, 
which is suggestive of the fact that 
$\psi_p^{(1)}(x)$ could decay, like \refeq{pd.1}, twice as rapidly as the
two-point function.

To state bounds on $\psi_p^{(N)}(x)$ for general $N$, we define
\eqalign
	\Abegin(x, y) 
	& = \sum_{a,b \in \Zd}
    \tau_{p}(a)\tau_{p}(b)\tau_{p}(a-b)
	\tilde{\tau}_p(x-a)
    \tau_{p}(y-b), \\
\lbeq{A1def}
    A_1(u,v,x,y) & =  \tau_{p}(u-v) \sum_{a,b \in \Zd}
    \tau_{p}(u-a)\tau_{p}(v-b)\tau_{p}(a-b)
	\tau_{p}(y-a) \tilde{\tau}_p(x-b),
    \\
    A_2(u,v,x,y) & =  \tau_{p}(y-u) \sum_{a,b \in \Zd}
    \tau_{p}(u-a)\tau_{p}(v-a)\tau_{p}(a-b)\tau_{p}(v-b) \tilde{\tau}_p(x-b),
    \\
    A^{(i)}(u,v,x,y) & =  A_1(u,v,x,y) + A_2(u,v,x,y) \quad (i \geq 1),
    \\
	\Aend(u, v, x, y) & = \tau_{p}(u-v) 
    \tau_{p}(u-x)\tau_{p}(v-y). 
\enalign
The above quantities are depicted in Figure~\reffg{perc}.

We define $M^{(N)}$ for $N \geq 1$ by \refeq{ind-M}. 
It then follows from \cite[Proposition~2.4]{HS90a} that, for $N \geq 1$, 
\eq
    0 \leq \psi_p^{(N)}(x)  \leq M^{(N)}(x,x).
    \lbeq{hbd}
\en
Consistent with \refeq{pd.1}, we define $M^{(0)}(x,x) =
(1-\delta_{0,x}) \tau_p(x)^2$.
We also recall from \cite[Proposition~2.4]{HS90a} that for $N \geq 1$ 
the expansion
remainder term $R_p^{(N)}(x)$ of \refeq{percrem} obeys
\eq
    0 \leq 
    R_p^{(N)}(x) \leq \sum_{u \in \Zd} M^{(N)}(u,u) \tilde{\tau}_p(x-u).
    \lbeq{Rbd}
\en
We will use this below to conclude that
$\lim_{N \to \infty} R_p^{(N)}(x) =0$, assuming the hypotheses
of Proposition~\ref{prop-diagbd}(c).  The vanishing of this limit
was claimed below Theorem~\ref{thm-percexp} and used under \refeq{tr.st.Jbd1}.

\begin{figure}
\begin{center}
\includegraphics[scale = 0.4]{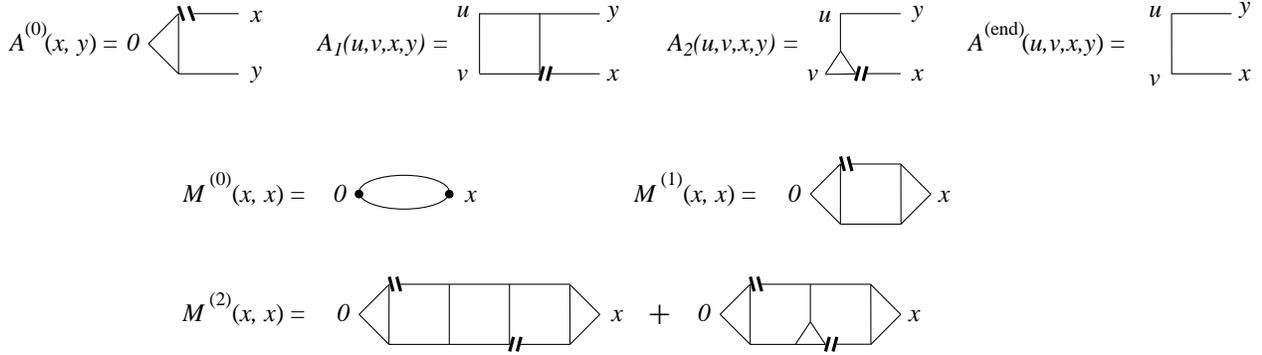}
\end{center}
\caption{\lbfg{perc}
The diagrams for percolation.  Lines 
ending with double bars represent $\tilde{\tau}^{a}$-lines.
}
\end{figure}

\smallskip
\noindent {\it Proof of Proposition~\ref{prop-diagbd}(c).}
For percolation, we have $z=p$, 
$G_z(x)=\tau_p(x)$ and 
$\Pi_z(x)=\psi_p(x)$.  
The hypotheses of the proposition are that
$\tau_p(x) \leq \beta(|x|+1)^{-q}$ for $x \neq 0$
with $\frac{2}{3}d < q < d$, that $\beta L^{q-d}$ is bounded away
from zero, and that $p \leq 2$.
It suffices to show that 
\eq
\lbeq{psibdperc}
	\psi_p(x) \leq c\beta \delta_{0,x} + c\beta^2 (|x|+1)^{-2q}.
\en
It follows immediately from \refeq{pd.1} that the contribution
to $\psi_p$ from $\psi_p^{(0)}$ does obey \refeq{psibdperc}, and we 
concentrate now on $N \geq 1$.

By the assumed bound on $\tau$, we
conclude as in \refeq{rhotilbd.1} that 
        \eq
        \lbeq{tautilbd}
        \tilde{\tau}_p(x) \leq \frac{C\beta}{(|x|+1)^{q}}.
        \en
We will apply Lemma~\ref{lem-Mbound} with $q_1=q_2=q$.  Our assumption
on $q$ implies that \refeq{q12} is satisfied.
We also need to verify that $\Abegin, A^{(i)}, \Aend$ obey the 
assumptions of Lemma~\ref{lem-Mbound}.  

It is clear that $\Aend$ obeys \refeq{as-A} with $q_{1} = q_{2} = q$
and $K_{\rm end} = O(1)$.  
For $\Abegin$, we note the decomposition  
\eq
\lbeq{Abegin-pec-dec}
	\Abegin(x,y) = \sum_{u,v} \biggl [ \tau_{p}(u) \tau_{p}(v) \biggr ] 
	\, \biggl [ \tau_{p}(u-v) \tau_{p}(y-v) \tilde{\tau}_{p}(x-u) \biggr ] 
	. 
\en
We can then apply Lemma~\ref{lem-Mbound}, considering the first factor 
as $\Abegin$ and the second factor as $A^{({\rm end})}$, to conclude
that $\Abegin$ obeys \refeq{as-M1} with $K_0=C\beta$.  

To check that 
$A^{(i)} = A_{1} + A_{2}$ obeys \refeq{as-A} with $K=C\beta$, we begin with 
$A_{2}$.  Define 
$a(u, v, x) = A_{2}(u,v,x,y)/\tau_{p}(y-u)$.  This 
quantity is nothing but $\Abegin(x-v,u-v)$ of \refeq{Abegin-pec-dec}.
Therefore, 
\eq
	a(u,v, x) \leq \frac{C\beta}{(|v-u|+1)^{q} (|x-v|+1)^{q}} , 
\en 
and  
$A_{2}$ obeys \refeq{as-A} 
with $q_{1} = q_{2} = q$ and $K=C\beta$. 
For $A_{1}$, recalling the definition of $\oldBfour$ in
\refeq{B4def}, we see that
$A_{1}(u,v,x,y)$ obeys the same bound as
$C\beta \tau_{p}(u-v) \oldBfour(u, v, x, y)$.
By \refeq{Hbd.1}, $A_1$ obeys \refeq{as-A} with $K=C\beta$.

It then follows from Lemma~\ref{lem-Mbound} that for $N \geq 1$ 
    \eq\lbeq{finalbdM}
    0 \leq \psi_p^{(N)}(x) \leq 
    M^{(N)}(x,x) \leq \frac{(\beta C)^{N}}{(|x|+1)^{2q}}.
    \en
The factor $\beta^N$ here arises from the $\beta$'s present in $A^{(0)}$ and
in each of the $N-1$ factors of $A^{(i)}$.
This gives an adequate bound for $N \geq 2$.
To complete the proof, it suffices to argue that for $N=1$ the power
of $\beta$ in \refeq{finalbdM} can be replaced by $\beta^2$ when $x \neq 0$.
This follows from the observation that for $N=1$ and $x\neq 0$, at least
two diagram lines in $M^{(1)}(x,x)$ 
must undergo a nontrivial displacement, and each of
these lines contributes a factor $\beta$.
\qed

\medskip \noindent {\em Proof that\/ $\lim_{N\to \infty} R_p^{(N)}(x)=0$
under hypotheses of Proposition~\ref{prop-diagbd}(c)}.
This is an immediate consequence of \refeq{Rbd}, \refeq{tautilbd}, 
\refeq{finalbdM}, 
and Proposition~\ref{lem-conv}(i).
\qed

\section{Convolution bounds}
\label{sec-conv}
\setcounter{equation}{0}

In this section, we prove Proposition~\ref{lem-conv}.

\smallskip \noindent{\em Proof of Proposition~\protect\ref{lem-conv}.}
(i) By definition,
\eq
\lbeq{fg.7}
    | (f*g)(x) | \leq
    \sum_{y: |x-y| \leq |y|}
    \frac{1}{(|x-y|+1)^{a}} \frac{1}{(|y|+1)^{b}}
    +
    \sum_{y: |x-y| > |y|}
    \frac{1}{(|x-y|+1)^{a}} \frac{1}{(|y|+1)^{b}}.
\en
Using $a \geq b$ and the change of variables $z=x-y$ in the second
term, we see that
\eq
\lbeq{fg.8}
    | (f*g)(x) |  \leq
    2\sum_{y: |x-y| \leq |y|}
    \frac{1}{(|x-y|+1)^{a}} \frac{1}{(|y|+1)^{b}}
    .
\en
In the above summation, $|y| \geq \frac{1}{2}|x|$.
Therefore, for $a>d$ we have 
\eq
\lbeq{fg.9}
    |(f*g)(x)| \leq
    \frac{2^{b+1}}{(|x|+1)^b}\sum_{y: |x-y| \leq |y|}
    \frac{1}{(|x-y|+1)^{a}}
    \leq 
    C (|x|+1)^{-b} . 
\en

Suppose now that $a<d$ and $a+b>d$.  In this case, we divide the sum
in \refeq{fg.8} according to whether 
$\frac{1}{2}|x| \leq |y| \leq \frac{3}{2}|x|$ or
$|y| \geq \frac{3}{2}|x|$.  The contribution to \refeq{fg.8} due to the
first range of $y$ is bounded above, as in \refeq{fg.9}, by
\eq
	\frac{2^{b+1}}{(|x|+1)^b}\sum_{y: |x-y| \leq 3|x|/2}
    	\frac{1}{(|x-y|+1)^{a}}
    	\leq 
    	\frac{C}{(|x|+1)^b} (|x|+1)^{d-a},
\en
as required.  
When $|y| \geq \frac{3}{2}|x|$, we have $|y-x| \geq |y|-|x| \geq |y|/3$.
Therefore, the contribution to \refeq{fg.8} due to the second range
of $y$ is bounded above by
\eq
	3^b \cdot 2 \sum_{y: |y| \geq 3|x|/2}
    	 \frac{1}{(|y|+1)^{a+b}}
    	\leq \frac{C}{(|x|+1)^{a+b-d}}.
\en
This completes the proof.

\smallskip \noindent (ii)
By (i), the convolution of $g$ with the error
term of $f$ gives a result that is $O(BC (|x|+1)^{-(d-2+s)})$.
This leaves us with the convolution of the main term with $g$, which is
given by
\eq
\lbeq{cor25.1}
    \sum_{y\in \Zd} g(y) \frac{A}{(|x-y|+1)^{d-2}}
    = \frac{A\sum_{y} g(y) }{(|x|+1)^{d-2}}
    + \sum_{y\in \Zd} g(y)
    \biggl [
    \frac{A}{(|x-y|+1)^{d-2}} - \frac{A}{(|x|+1)^{d-2}}
    \biggr ] .
\en
The first term is the desired main term, so it remains to prove that the
second term is an error term.  We denote the second term by $X$.

We consider separately the contributions to $X$
due to $|y|>\frac{1}{2}|x|$ and $|y| \leq \frac{1}{2}|x|$,
beginning with the former.  This contribution
to $X$, which we call $X_1$, is bounded above by
\eq
\lbeq{cor25.4}
    |X_1| \leq AC\sum_{y: |y| > |x|/2}
    \frac{1}{(|y|+1)^{d+s}}
    \biggl [
    \frac{1}{(|x-y|+1)^{d-2}} + \frac{1}{(|x|+1)^{d-2}}
    \biggr ] .
\en
Using part (i) of the proposition, the first term is bounded above by
\eq
	\frac{2^{2+s}AC}{(|x|+1)^{2+s}} 
	\sum_{y \in \Zd}
    	\frac{1}{(|y|+1)^{d-2}}
    	\frac{1}{(|x-y|+1)^{d-2}}
    	\leq
    	O \Bigl (\frac{AC}{(|x|+1)^{d-2+s}} \Bigr ).
\en
The second term of \refeq{cor25.4} obeys
\eq
    \frac{AC}{(|x|+1)^{d-2}} \sum_{y: |y| > |x|/2}
        \frac{1}{(|y|+1)^{d+s}}
    = O \Bigl (
    \frac{AC}{(|x|+1)^{d-2+s}}
    \Bigr ) .
\en
Combining these gives $X_1 =
O (AC(|x|+1)^{-(d-2+s)} )$, so $X_1$ is an error term.

Next, we consider the contribution to $X$ due to $|y| \leq \frac{1}{2}|x|$,
which we denote by $X_2$.  We estimate this term by expanding the
difference $\frac{1}{(|x-y|+1)^{d-2}} - \frac{1}{(|x|+1)^{d-2}}$ into
powers of $y$.  
Because of the $\Zd$-symmetry of $g(y)$, odd powers of $y$ in
the expansion 
give no contribution.
Let $h(t) = [|x|(1+t)+1]^{2-d}$, for $|t|<1$.
Using the Fundamental Theorem of
Calculus, it can be seen that $|h(t)-h(0)-h'(0)t| \leq c|t|^2 |x|^2
(|x|+1)^{-d}$ when $|t| \leq \frac 12$.  Applying this with
$t=|x|^{-1}|x-y|-1$ (the case $x=0$ does not contribute), 
we conclude that
\eq
    | X_{2} | \leq \sum_{y: |y| \leq |x|/2} | g(y) |
     \frac{cA|y|^{2}}{(|x|+1)^{d}} .
\en
Therefore, recalling that $s_2=s\wedge 2$, we have 
\eq
\lbeq{cor25.10}
    |X_2|  \leq
    \frac{cAC}{(|x|+1)^{d}}
    \sum_{y : |y| \leq |x|/2}
    \frac{|y|^2}{(|y|+1)^{d+s}}
    \leq
    \begin{cases}
    cAC(|x|+1)^{-d-2+s_2} & (s \neq 2) \\
    cAC(|x|+1)^{-d}\log(|x|+2) & (s=2).
    \end{cases}
\en
This completes the proof.
\qed

\section{The random walk two-point function}
\label{sec-G2ptfcn}
\setcounter{equation}{0}

In this section, we prove Proposition~\ref{prop-A}. 
We begin in Section~\ref{sec-Cub} with an elementary proof of the 
bound
\eq
\lbeq{CL-uniA1}
         \delta_{0,x} \leq S_\mu(x) \leq  \delta_{0,x} + O(L^{-d} ) ,
\en
which is uniform in $\mu \leq 1$ and $x \in \Zd$.
In Section~\ref{sec-intrep}, an integral
representation for $S_\mu(x)$ is introduced, which is analysed
in Sections~\ref{sb-ab-t>T}--\ref{sb-ab-t<T} for $|x|$ large compared with
$L$. The proof of the asymptotic formula
\refeq{CL-uni2} is then given in Section~\ref{sec-Casy}.

Once \refeq{CL-uni2} and \refeq{CL-uniA1} 
have been proved, the bound \refeq{CL-uni} then follows
easily.   In fact, it suffices to prove \refeq{CL-uni} for $\mu=1$, since
$S_{\mu}(x)$ is increasing in $\mu$.
However, \refeq{CL-uni} for $\mu=1$ follows immediately, by
using \refeq{CL-uniA1} for $|x| \leq L$ and \refeq{CL-uni2} for
$|x| >L$.

\subsection{Proof of the uniform bound}
\label{sec-Cub}
In this section, we prove the uniform bound
\refeq{CL-uniA1}.
The lower bound of \refeq{CL-uniA1}
follows immediately from the facts that $S_\mu(x)$
is positive for all $x$ by definition and that $S_\mu(0)$ receives
a contribution $1$ from the zero-step walk.  So it remains to
prove the upper bound.  For this, it suffices to consider $\mu=1$,
because $S_{\mu}(x)$ is increasing in $\mu$.

In preparation, and for later use, we first note some
properties of $D$.   By Definition~\ref{def-Dsp1}, $D(x) \leq O(L^{-d})$
and $\sigma \sim \mbox{const.}L$.  In addition, it is proved in 
\cite[Appendix~A]{HS01a} that there are constants $\delta_2$ and
$\delta_3$, such that for $L$ sufficiently large,
\eqalign
    \lbeq{Dsp-prop2}
    1 - \Dhat(k)
    & \geq \delta_{2}\, L^{2} \, |k|^{2}
    \qquad \mbox{for  }  |k| \leq L^{-1} ,
\\
    \lbeq{Dsp-def-cnd3}
    1 - \Dhat(k)
    & \geq \delta_{3} \qquad \mbox{for $k \in [-\pi,\pi]^d$ with
    $|k| \geq  L^{-1}$} .
\enalign

To prove the upper bound of \refeq{CL-uniA1}, we rewrite $[1-\Dhat(k)]^{-1}$ as
$1 + \Dhat(k) + \Dhat(k)^{2}[1 - \Dhat(k)]^{-1}$,
to obtain
\eq
    S_1 (x)
    = \intddkpi   \frac{e^{-i \kx}}{1 - \Dhat(k)}
    = \delta_{0,x} + D(x) +
    \intddkpi \frac{\Dhat(k)^{2} e^{-i \kx}}{1 - \Dhat(k)} .
\en
The second term is $O(L^{-d})$, so it remains to
prove that the last term is also $O(L^{-d})$.  We first estimate the absolute
value of the last term by taking absolute values inside the integral.
We then divide the integral
over $k$ into two parts, according to whether $|k|$ is greater than or
less that $L^{-1}$.

For the integral over small $k$, we note that
in general $|\Dhat(k)|\leq 1$.  Using \refeq{Dsp-prop2} yields
\eq
    \int_{|k|< L^{-1}} \ddk \frac{\Dhat(k)^{2}}{1 - \Dhat(k)}
    \leq
    \int_{|k|< L^{-1}} \ddk \frac{1}{\delta_2 L^{2}|k|^{2}}
    = O( L^{-d}) .
\en
Also, using \refeq{Dsp-def-cnd3},
the integral over large $k$ is bounded by
\eq
    \int_{\{k\in [-\pi,\pi]^d : |k| \geq L^{-1}\}}
    \ddk \frac{\Dhat(k)^{2}}{1 - \Dhat(k)}
    \leq
    \frac{1}{\delta_{3}}
    \intddkpi \Dhat(k)^{2}
    = \frac{1}{\delta_{3}} \sum_{y} D(y)^{2}
    =
    O( L^{-d}).
\en
This proves \refeq{CL-uniA1}.

\subsection{The integral representation}
\label{sec-intrep}

To prove the asymptotic
formula \refeq{CL-uni2}, which states that
        \eq
        \lbeq{CL-decay}
        S_1(x) = \frac{a_{d}}{\sigma^2} \frac{1}{(|x|+1)^{d-2}}
        +O\Big( \frac{1}{(|x|+1)^{d-\alpha}} \Big),
        \en
we will use an integral representation for $S_\mu(x)$.
By \refeq{CL-uniA1}, $S_1(x) \leq O(L^{-d})$ for $x \neq 0$.
This immediately implies \refeq{CL-decay} for $|x| \leq L^{1+\alpha/d}$.
It therefore suffices, in what follows, to restrict attention
to $|x| \geq L^{1+\alpha/d}$.

Although it is sufficient to consider only $\mu =1$ to prove
\refeq{CL-decay}, we consider also $0 \leq \mu <1$, as this will be used in
the proof of the main error estimate in Section~\ref{sec-CE}.
Let
\eq
\lbeq{Idef}
    I_{t, \mu} (x) =
    \int\limits_{[-\pi, \pi]^{d}}
    \ddk e^{-i \kx} \, e^{-t[1 - \mu\Dhat(k)]}.
\en
It follows from \refeq{sle} that $\hat{S}_\mu(k) = [1-\mu\hat{D}(k)]^{-1}$.
Thus, for $0 \leq \mu \leq 1$ we have the integral representation
\eqalign
\lbeq{ab-int-rep.1}
    S_\mu(x)
    & = \int\limits_{[-\pi, \pi]^{d}}
    \ddk \frac{e^{-i \kx} }{1 - \mu \Dhat(k)}
    = \int\limits_{[-\pi, \pi]^{d}}
    \ddk e^{-i \kx}  \int_{0}^{\infty} dt \, e^{-t[1 - \mu \Dhat(k)]}
    = \int_{0}^{\infty} dt \, I_{t,\mu}(x) .
\enalign

The integration variable $t$ plays the role of a time variable,
with the dominant contribution to $S_1(x)$ due to $t \approx
|x|^2/\sigma^{2}$.  With this in mind, we write
$S_\mu(x) = S_\mu^<(x;T) +S_\mu^>(x;T)$ with
\eq
\lbeq{G><-def}
    S_\mu^{<}(x;T)  =  \int_{0}^{T} dt \, I_{t, \mu}(x), \qquad
    S_\mu^{>}(x;T)  =  \int_{T}^{\infty} dt \, I_{t, \mu}(x) ,
\en
and choose $T$ to be equal to
\eq
\lbeq{ab-T-choice}
    T_{x} = \Bigl ( \frac{|x|}{\sigma} \Bigr )^{2 - 2\alpha/d},
\en
where $\alpha$ is the small parameter of Proposition~\ref{prop-A}.
With this choice,
it will turn out that $S_1^>(x)$ contains the leading
term of \refeq{CL-decay}, whereas $S_1^<(x)$ is an error term.
Our analysis will make use of different estimates on $I_{t,1}(x)$ for each
of these two terms.

\subsection{Integration over $[T,\infty]$}
\label{sb-ab-t>T}

In this section, we prove that for $|x| \geq L^{1+\alpha/d}$
and $L$ sufficiently large depending on $\alpha$, we have
\eq
\lbeq{Cmusmallt}
    S_1^>(x; T_{x})
    = \int_{T_{x}}^\infty I_{t,1}(x) \, dt
    = \int_{T_{x}}^\infty p_t(x) \, dt
        +O\Big( \frac{L^{-\alpha}}{(|x|+1)^{d-\alpha}} \Big),
\en
where 
\eq
\lbeq{ftdef}
        p_t(x) = \left ( \frac{d}{2 \pi \sigma^{2} t} \right )^{{d/2}}
        \exp \biggl ( - \frac{d \, |x|^{2}}{2 t \sigma^{2}}
        \biggr).
\en
The proof will make use of the following lemma, which extracts the
leading term from $I_{t,1}(x)$.

\begin{lemma}
    \label{lem-ab-large-t}
Let $d>2$, and
suppose $D$ obeys Definition~\ref{def-Dsp1}.  Then there are finite
$L$-independent constants $\tau$ and $c_1$ such that
for $t \geq \tau$
\eq
\lbeq{ab-Ibd.1}
        I_{t,1}(x) = p_t(x)  +r_t(x)
\en
with
\eq
\lbeq{Etbd}
        |r_t(x)| \leq c_1 \,   L^{-d} t^{-d/2-1} + e^{-t \delta_{3} } .
\en
\end{lemma}

Before proving Lemma~\ref{lem-ab-large-t}, we show how integration of
its bound leads to a proof of \refeq{Cmusmallt}.
It suffices to show that
the integral of the error term $r_t(x)$ in \refeq{ab-Ibd.1}
is bounded by the error term of \refeq{Cmusmallt}.
By \refeq{Etbd} we have
\eq
\lbeq{ab-RB1bd-1}
    \biggl |
    \int_{T_{x}}^{\infty} dt \, r_t(x) \biggr |
    \leq
    c L^{-d} \,  T_{x}^{-d/2}
    +
    \delta_{3}^{-1} \, e^{-\delta_{3} T_{x}} .
\en
The second term can be absorbed into the first term.
In fact, since $T_{x} \geq cL^{2(\alpha/d)(1-\alpha/d)}$, for any
positive $N$ we have
\eq
    e^{-\delta_3 T_{x}} \leq \frac{c_N}{T_{x}^N}
    = c_N \frac{L^{-d}}{T_{x}^{(d+2)/2}} \frac{L^d}{T_{x}^{N-(d+2)/2}},
\en
with the last factor less than $1$ for $L$ and $N$
sufficiently large depending on $\alpha$.  In addition, 
since $|x| > L$ and $d>\alpha$, the first term of \refeq{ab-RB1bd-1}
is equal to
\eq
\lbeq{ab-RB1bd-2}
        c L^{-d} \, T_{x}^{-d/2} =
    c \frac{L^{-\alpha}}{(|x|+1)^{d-\alpha}}.
\en
This proves \refeq{Cmusmallt}.

\bigskip
\noindent {\em Proof of Lemma~\ref{lem-ab-large-t}.}
By Taylor's theorem and symmetry,
for $k \in [-\pi, \pi]^{d}$ we have
\eq
     \lbeq{Dsp-prop3}
        1 - \Dhat(k) = \frac{\sigma^{2}|k|^{2}}{2d}  + R (k)
\en
with
\eqalign
      \lbeq{Dsp-prop7}
    |R(k)|
    & \leq  \frac{1}{4!} \sum_{x} D(x) (x \cdot k)^{4}
    \leq \mbox{const.} L^{4} |k|^{4} .
\enalign
Let
$k_t^{2} =4d \sigma^{-2}t^{-1}\log{t}$.
We write
$I_{t,1}(x)  = \sum_{j=1}^{4} I_{t}^{(j)}(x)$
with
\eqalign
\lbeq{ab-Itdiv.1}
    I_{t}^{(1)}(x) & =
    \int_{\Rd} \ddk e^{-i \kx - t \sigma^{2} |k|^{2}/(2d)}, \qquad
    \\
\lbeq{ab-Itdiv.2}
    I_{t}^{(2)}(x) & =
    - \int_{k_t <|k| <\infty} \ddk e^{-i \kx - t \sigma^{2} |k|^{2}/(2d)},
    \\
\lbeq{ab-Itdiv.5}
    I_{t}^{(3)}(x) & =
    \int\limits_{|k| < k_{t}} \ddk e^{-i \kx - t \sigma^{2} |k|^{2}/(2d)} \,
        \Bigl( e^{-t R(k)} - 1 \Bigr), \qquad
    \\
\lbeq{ab-Itdiv.7}
    I_{t}^{(4)}(x) & =
    \int\limits_{k \in [-\pi,\pi]^d : \, |k| > k_{t} } \ddk e^{-i \kx} \,
        e^{-t [1 - \Dhat(k)]} .
\enalign
The integrals $I_{t}^{(1)}(x)$ through $I_{t}^{(3)}(x)$ combine to give
the contribution to $I_{t,1}(x)$ due to $|k| \leq k_{t}$,
while $I_{t}^{(4)}(x)$ represents the contribution
from $|k| > k_{t}$.  

The first integral can be evaluated exactly to give
\eq
\lbeq{ab-I1r}
    I_{t}^{(1)}(x) = p_t(x)  .
\en
We therefore set $r_t(x) = \sum_{j=2}^4 I_{t}^{(j)}(x)$ and show that
$r_t(x)$ obeys \refeq{Etbd}.
By definition, 
\eqalign
\lbeq{ab-I2r}
    | I_{t}^{(2)}(x) |
    & \leq
    \int\limits_{k_t < |k| <\infty} \ddk
    e^{- t \sigma^{2} |k|^{2}/(2d)} 
    \leq
    c(t\sigma^{2})^{-d/2}
    e^{- t \sigma^{2} \, k_{t}^{2} /(4d)}
    \leq
    c L^{-d} t^{-d/2 -1} .
\enalign

The integral $I_{t}^{(3)}(x)$ is bounded as follows.
First, we note that for $|k|< k_{t}$ it follows from
\refeq{Dsp-prop7} and the definition of $k_t$ that 
$| t R(k) |  \leq  c (\log t)^{2}/t$, which 
is less than $1$ for sufficiently large $t$.
Using the bound $|e^x - 1 | \leq |x|$
for $|x| \leq 1$, 
and increasing the integration domain to $\Rd$ in
the last step, we have
\eqalign
\lbeq{ab-I5r}
    | I_{t}^{(3)}(x) |
    & \leq
    c \int\limits_{|k|< k_{t}} \ddk 
    \, e^{-t \sigma^{2} |k|^{2}/(2d)} \, |t R(k)|
    \leq
    c tL^4 \int\limits_{|k|< k_{t}} \ddk \, e^{-t \sigma^{2} |k|^{2}/(2d) } \,
    |k|^{4}
    \leq
   c L^{-d} t^{-d/2-1} .
\enalign

Finally we estimate $I_{t}^{(4)}(x)$.  We divide the integration domain
according to whether $|k|$ is greater than or less than $L^{-1}$.
By \refeq{Dsp-prop2}, as in \refeq{ab-I2r} the contribution due to
$|k|\leq L^{-1}$ is at most
\eq
\lbeq{ab-I7.1}
    \int\limits_{k_{t} < |k| < \infty } \ddk \,
    e^{-t  \delta_2 L^{2} |k|^{2}}
    \leq
    c L^{-d} t^{-d/2-1} .
\en
By \refeq{Dsp-def-cnd3},
the contribution due to $|k| > L^{-1}$ is at most
\eq
\lbeq{ab-I7.2}
    \int\limits_{k \in [-\pi,\pi]^d : \, |k| > L^{-1} } \ddk \,
    e^{-t \delta_{3} }
    \leq  e^{-t \delta_{3} } .
\en

Combining \refeq{ab-I2r}--\refeq{ab-I7.2} then gives the desired bound
\eq
\lbeq{ab-I234r1}
    |r_t(x) |
    \leq \sum_{j=2}^4 |I_{t}^{(j)}(x)|
    \leq
    c\,  L^{-d} t^{-d/2-1} + e^{-t \delta_{3} } .
\en
\qed

\subsection{Integration over $[0,T]$}
\label{sb-ab-t<T}

In this section, we prove the following lemma, which will also be used
in the main error estimate of Section~\ref{sec-CE}.
\begin{lemma}
    \label{lem-Smu<bd}
Let $|x|\geq L^{1+\alpha/d}$ and $T\leq T_{x}$.  Then for $0 \leq \mu \leq 1$
and sufficiently large $L$ depending on $\alpha$
\eq
\lbeq{Cmusmalltbd}
    S_\mu^<(x;T) = \int_{0}^{T}
    I_{t, \mu} (x) \, dt
        \leq \frac{1}{(|x|+1)^{d+2}} .
\en
\end{lemma}

We will prove this using the following lemma, whose proof involves a
standard large deviations argument.

\begin{lemma}
\label{lem-StIt}
For $x \in \Zd$, $t\geq 0$ and $t_0 = dL\|x\|_\infty /(2\sigma^{2})$,
\eq
\lbeq{Itbd9}
    0 \leq
    I_{t,1}(x) \leq \left\{
    \begin{array}{ll}
    \exp [ - \|x\|_\infty/L + \sigma^{2} t/(dL^2) ] \;
    & (0 \leq t < \infty)
    \\
    \exp [ - d\|x\|_\infty^2/(4\sigma^{2} t)] & (t \geq t_0)
    .
    \end{array}
    \right.
\en
\end{lemma}

\noindent {\em Proof of Lemma~\ref{lem-Smu<bd} assuming
Lemma~\ref{lem-StIt}.}
We first note that for fixed $x \in \Zd$,
$I_{t, \mu}(x)$ is nonnegative and is monotone increasing in $\mu$.
In fact,
expanding the exponential $e^{t\mu \hat{D}(k)}$ in \refeq{Idef} and
interchanging the integral and the sum 
(justified by absolute  convergence), gives
\eq
\lbeq{St-rep11}
    I_{t, \mu}(x) 
    = e^{-t} \, \sum_{n=0}^{\infty} \frac{(t\mu)^{n}}{n!}
    \intddkpi \, e^{-i \kx} \Dhat(k)^{n}
    = e^{-t} \, \sum_{n=0}^{\infty} \frac{(t\mu)^{n}}{n!}
    D^{*n}(x),
\en
where $D^{*n}$ denotes the $n$-fold $x$-space convolution. Because
$D^{*n}(x)$ is nonnegative, this
representation immediately implies the non-negativity of $I_{t,\mu}(x)$,
together with its monotonicity in $\mu$.
Therefore $S_\mu^{<}(x;T)$ is increasing in $\mu$ and in $T$,
and it suffices to prove \refeq{Cmusmalltbd} for
$\mu=1$ and $T = T_{x}$.  

In this case,
\refeq{Itbd9} gives
\eq
    \int_{0}^{T_{x}} dt \, I_{t,1}(x)
    \leq
    \int_{0}^{t_{0}} dt \,  \exp \Bigl [ -  \frac{\|x \|_{\infty}}{L}
    + \frac{\sigma^{2} t}{dL^2}
    \Bigr ]
    +
    \int_{t_{0}}^{T_{x}} dt \,
    \exp \Bigl [ - \frac{d \|x \|_{\infty}^{2} }{4 \sigma^{2} t }
    \Bigr ].
\en
The first integral can be performed exactly.  For the second, we
use the fact that for $a\geq T$,
\eq
    \int_0^{T} dt \, e^{-a/t} = a \int_{a/T}^\infty
    du \, u^{-2} e^{-u}\leq \frac{T^2}{a} e^{-a/T}.
\en
Now choose $T= T_{x}$ and $a=d\|x\|_\infty^{2} /(4\sigma^{2}) \geq T_{x}$,
for $|x| \geq L^{1+\alpha/d}$.  This gives
\eqarray
    \int_{0}^{T_{x}} dt \, I_{t,1}(x)
    & \leq &
    c
    \exp \Bigl [ -  \frac{\|x \|_{\infty}}{2L}
    \Bigr ]
    +
    \frac{4 \sigma^{2}}{d \|x\|_{\infty}^{2}}
    T_{x}^{2}
    \exp \Bigl [ -  \frac{d \|x \|_{\infty}^{2}}{4 \sigma^{2}T_{x}}
    \Bigr ]
    \nonumber \\
    & \leq &
    c \, \exp \Bigl ( - c \frac{|x|}{L}  \Bigr )
    + \, c \Bigl( \frac{|x|}{L} \Bigr)^{2-4\alpha/d}
    \exp \Bigl ( - c \bigl (\frac{|x|}{L} \bigr )^{2\alpha/d} \Bigr )
    .
\enarray
For $|x| \geq L^{1+\alpha/d}$, we have
$|x|/L \geq |x|^{(\alpha/d)/(1+\alpha/d)}$.  The integral
$\int_{0}^{T_{x}} dt \, I_{t,1}(x)$
therefore decays at least as fast as a constant
multiple of an exponential of a power of $|x|$, and hence 
eventually decays faster
than $|x|^{-(d+2)}$.
This completes the proof of \refeq{Cmusmalltbd}.
\qed

\bigskip
\noindent{\em Proof of Lemma~\ref{lem-StIt}.}
Since $D$ is supported only on $\|x\|_\infty \leq L$, we have
$D^{*n}(x) = 0$ for $\|x\|_\infty > n L$.  Since
$0 \leq D^{*n}(x) \leq 1$ for all $n$,
we can therefore bound \refeq{St-rep11}
using the inequality
\eq
    \sum_{n=N}^{\infty} \frac{t^{n}}{n!}
    \leq e^{t} \frac{t^{N}}{N!}
    \leq e^{t} \Bigl ( \frac{et}{N} \Bigr )^{N}  
\en
as
\eq
\lbeq{St-rep15}
    I_{t,1}(x)
    = e^{-t} \sum_{n\geq \|x\|_\infty/L} \frac{t^{n}}{n!}
    D^{*n} (x)
    \leq
    e^{-t} \sum_{n\geq \|x\|_\infty/L} \frac{t^{n}}{n!}
    \leq
    \Bigl ( \frac{et}{\|x\|_\infty/L} \Bigr )^{\|x\|_\infty/L} .
\en
Thus, $I_{t,1}(x)$ decays in $|x|$ more rapidly than any exponential,
and we may define the quantity
    \eq
    \lbeq{sb-Stdef}
        \phi_{t}(s) = \sum_{x\in \Zd}
        e^{s x_{1}} \, I_{t,1}(x)  \qquad (s \in \Rbold).
    \en

We claim that
\eq
\lbeq{St-rep2}
        \phi_{t}(s) = \exp \Bigl [
    t \, \sum_{x} D(x) \bigl [ \cosh (s x_{1}) -1 \bigr ]
    \Bigr ] .
\en
This follows by interchanging sums
in \refeq{St-rep11} to obtain
\eq
    \phi_{t}(s)
    = e^{-t}
    \sum_{n=0}^{\infty} \frac{t^{n}}{n!}
    \sum_{x} e^{s x_{1}} D^{*n} (x)
    =
    e^{-t}
    \sum_{n=0}^{\infty} \frac{t^{n}}{n!}
    \Bigl [ \sum_{x} e^{s x_{1}} D(x) \Bigr ]^{n}
    = \exp \Bigl [
    - t  + t \sum_{x} e^{s x_{1}} D(x) \Bigr ] ,
\en
and then using $D(x)=D(-x)$ and
$\sum_{x} D(x) = 1$.

By definition,
$\phi_{t}(s)= \sum_{y} e^{s |y_{1}|} I_{t,1}(y) \geq e^{s |x_{1}|}
I_{t,1}(x)$ for any $x \in \Zd$,
and therefore $I_{t,1}(x) \leq e^{- s |x_{1}|} \phi_{t}(s)$.
The $\Zd$-symmetry and the formula \refeq{St-rep2} for $\phi_{t}(s)$ then give
    \eq
    \lbeq{StIt-1}
        I_{t,1}(x) \leq \exp \Bigl [
    - s \|x \|_{\infty}
    + t  \sum_{y} D(y) \bigl [ \cosh (s y_{1}) -1 \bigr ]
    \Bigr ] .
    \en
When $s \leq L^{-1}$, we have $s|y_1| \leq 1$ for any $y$ contributing
to $\sum_{y} D(y) [ \cosh (s y_{1}) -1 ]$.
Since $\cosh x \leq 1 + x^{2}$  for $|x| \leq 1$, we obtain
\eq
\lbeq{StIt-6}
    0 \leq \sum_{y} D(y) \bigl [ \cosh (s y_{1}) -1 \bigr ]
    \leq s^{2} \sum_{y} D(y) y_{1}^{2}
    = s^{2} \frac{\sigma^{2}}{d} .
\en
Thus, for $s \leq L^{-1}$ we have
    \eq
    \lbeq{StIt-2}
        I_{t,1}(x) \leq \exp \Bigl [
    - s \|x \|_{\infty} + \sigma^{2} t d^{-1} s^{2}
    \Bigr ] .
    \en
Putting $s=L^{-1}$ in \refeq{StIt-2} gives the first bound of \refeq{Itbd9}.

The minimum of
the right side of \refeq{StIt-2} is attained at
$s = d \| x \|_{\infty}/(2 \sigma^{2} t)$,
but we may use \refeq{StIt-2} only for $s \leq
1/L$.  This condition will be valid provided
$t \geq t_0$.  Using the minimal value of $s$ in
\refeq{StIt-2} gives the second bound of \refeq{Itbd9}.
\qed

\subsection{Proof of the asymptotics}
\label{sec-Casy}

We now prove \refeq{CL-decay}.  As discussed below \refeq{CL-decay},
it suffices to consider $|x| \geq L^{1+\alpha/d}$.
For $|x| \geq L^{1+\alpha/d}$,
we have already proved a version of \refeq{CL-decay} having lower limit
of integration $T_x$ rather than $0$, since the combination of \refeq{Cmusmallt}
and \refeq{Cmusmalltbd} gives
\eq
    S_1(x) = S_1^<(x;T_{x}) + S_1^>(x;T_{x})
    = \int_{T_x}^\infty p_t(x) \, dt +
    O\Big(\frac{L^{-\alpha}}{(|x|+1)^{d-\alpha}} \Big)
    \quad\quad (|x| \geq L^{1+\alpha/d}).
\en

We will show that \refeq{CL-decay} holds as stated
with lower limit of integration $0$,
if $|x| \geq L^{1+\alpha/d}$.  For this, it suffices to show that the
quantity $R(x) = \int_0^{T_{x}} p_t(x)dt $ obeys the bound
\eq
\lbeq{smt}
    R(x) \leq O\Big(\frac{1}{(|x|+1)^{d}} \Big) 
    \quad\quad (|x| \geq L^{1+\alpha/d}).
\en
By definition, 
\eq
    | R(x) | \leq
    c L^{-d} \int_{0}^{T_x} dt \,  t^{-d/2} e^{-c'|x|^{2}/(tL^{2})} .
\en
To estimate the right side, we use the fact that
\eq
\lbeq{aa-int.1.2}
        \int_{0}^{T} dt \, t^{-b} \, e^{-a/t}
        =
        a^{1-b} \int_{a/T}^\infty u^{b-2} e^{-u} du
    \leq  
    \begin{cases}
    a^{-1} T^{2-b} e^{-a/T} & (1<b\leq 2) \\
    C(b)a^{1-b}e^{-a/2T} & (b >2)
    \end{cases}
\en
if $a \geq T >0$ and $b>1$, where $C(b)$ is a $b$-dependent constant.  
This inequality can be proved for $1<b\leq 2$ using $u^{b-2} \leq (a/T)^{b-2}$.
For $b >2$, it can be proved using $e^{-u} \leq e^{-u/2} e^{-a/2T}$.

We apply \refeq{aa-int.1.2} with $a = c|x|^{2}/L^{2}$, which is
greater than $T_{x}$ when $|x|\geq L^{1+\alpha/d}$ and $L$ is large.  The
exponent $b$ equals $d/2$, which is greater
than $1$ for $d>2$.  The result is that
\eq
    | R(x) |
    \leq c L^{-d} \, e^{-c''(|x|/L)^{2\alpha/d}} \,
    \left(\frac{|x|}{L}\right)^{q(d)},
\en
for some power $q(d)$.
For $|x| \geq L^{1+\alpha/d}$ sufficiently large, we therefore
have
\eq
\lbeq{ab-lt-cor10}
    | R(x) |
    \leq cL^{-d} \left(\frac{L}{|x|}\right)^{d} \leq
    \frac{c }{(|x|+1)^{d}} .
\en
This completes the proof of \refeq{CL-decay} if $|x| \geq L^{1+\alpha/d}$,
and hence for all $x$.
\qed

\section{The main error estimate}
\label{sec-CE}
\setcounter{equation}{0}

In this section, we prove Proposition~\ref{lem-C}.  We first obtain bounds
on $E_z(x)$ and $\hat{E}_z(k)$ in Section~\ref{sec-Ex}, 
and then 
complete the proof of Proposition~\ref{lem-C} in Section \ref{sub-errormain}.

\subsection{Bounds on $E_z$}
\label{sec-Ex}

First we derive the following  bound on $E_{z}(x)$ from the assumed bound on
$\Pi_{z}(x)$.

\begin{lemma}
    \label{lem-Epbd}
    Under the assumptions of Proposition~\ref{lem-C},
    \eq
    \lbeq{Epbd.1}
        | E_{z}(x)| \leq
        \begin{cases}
            c \gamma            & (x = 0) \\
            c \gamma L^{-d}     & (0 < |x| < 2L) \\
            c \gamma |x|^{-(d+2+\kappa)}  & (|x| \geq 2L).
        \end{cases}
    \en
\end{lemma}

\begin{proof}
By virtue of its definition in \refeq{Edef}, we can write
\eq
\lbeq{5-Edef}
    E_{z}(x) = (1-\lambda_z)\delta_{0,x}
    -(D*N_z)(x).
\en
with
\eq
\lbeq{5-Ndef}
    N_{z}(x) = [(1-\lambda_z)+\lambda_zz\hat{\Pi}_z(0)]\delta_{0,x}
    -\lambda_zz\Pi_z(x).
\en
To derive bounds on $N_{z}$ and thus on $E_{z}$, we first derive bounds
on $\Pi_{z}$ and $\lambda_{z}$.
Assuming $| \Pi_{z}(x) | \leq \gamma (|x|+1)^{-(d+2+\kappa)}$, we have
\eq
\lbeq{5-Pibd.1}
    \sum_y | \Pi_{z}(y) | \leq c \gamma,
    \qquad
    \sum_y |y|^{2} \, | \Pi_{z}(y) | \leq c \gamma .
\en
Also, by the formula for $\lambda_z$ of \refeq{lampdef}
and our assumption that $z \leq C$, 
it follows that
\eq
\lbeq{5-lambdabd.1}
    \lambda_{z} =O(1), \qquad \lambda_{z} -1 = O(\gamma ).
\en

The bounds \refeq{5-Pibd.1} and \refeq{5-lambdabd.1} imply
\eq
\lbeq{5-Npbd}
    N_{z}(x) = O(\gamma) \delta_{0,x}
    + O\Big( \frac{\gamma}{(|x|+1)^{d+2+\kappa}} \Big) ,
    \qquad
    \sum_{x} N_{z}(x) = O(\gamma),
\en
and hence
\eq
    (D*N_{z})(x) = \sum_{|y|\leq L} N_{z}(x-y) D(y)
    = \sum_{|y|\leq L} | N_{z}(x-y) | \, O(L^{-d})
    = O(\gamma L^{-d}).
\en
By \refeq{5-Edef}, this proves \refeq{Epbd.1} for
$0 \leq |x| < 2 L$.  For $|x|\geq 2L$, we note that $|x-y| \geq |x|/2$ when
$|y| \leq L$.  For such $y$, \refeq{5-Npbd} implies
$| N_{z}(x-y) | = O( \gamma |x|^{-(d+2+\kappa)})$, and
therefore
\eq
    (D*N_{z})(x) = \sum_{|y|\leq L} N_{z}(x-y) D(y)
    = O \bigl ( \frac{\gamma}{|x|^{d+2+\kappa}} \bigr )
    \sum_{|y|\leq L}  D(y)
    = O \bigl ( \frac{\gamma}{|x|^{d+2+\kappa}} \bigr ) .
\en
\end{proof}

Next, we use the above bound on $E_z(x)$ to derive a bound on 
$\hat{E}_z(k)$.

\begin{lemma}
    \label{lem-Ezhatbd}
    Let $\kappa_2 = \kappa \wedge 2$.  
    As $k \to 0$, under the assumptions of Lemma~\ref{lem-C},
    \eq
    \lbeq{Ezhatbd.99}
        | \hat{E}_{z}(k) | \leq 
		\begin{cases}
        c \gamma L^{2+\kappa_2} |k|^{2+\kappa_2} 
		& (\kappa \neq 2) \\
		c \gamma |k|^{4} \,  (L^{4} + \log |k|^{-1})
		& ( \kappa = 2).
		\end{cases} 
    \en
\end{lemma}

\begin{proof}
The proof proceeds as in the proof of Lemma~\ref{lem-Fourier1}.  
Since $\hat{E}_z(0)=\nabla^2\hat{E}_z(0)=0$, as in 
\refeq{fhatbd.11}--\refeq{fhatbd.13} we have
\eq
	|\hat{E}_{z}(k)| \leq 
	c|k|^{4}
        \sum_{x: |x| \leq |k|^{-1}}|x|^{4} \, | E_{z}(x)|
        +
        c  \sum_{x: |x| > |k|^{-1}} (1 + |k|^{2}|x|^{2}) \, | E_{z}(x) |.
\en
A calculation using Lemma~\ref{lem-Epbd} then implies that for 
$|k| \leq (2L)^{-1}$, 
\eq
        | \hat{E}_{z}(k) | \leq \begin{cases}
        c \gamma [ L^{4} |k|^{4} + |k|^{2+\kappa_2} ]
        & (\kappa \neq 2 )\\
        c \gamma [ L^{4} |k|^{4} + |k|^{4} \log |k|^{-1} ]
        & (\kappa = 2 ).
        \end{cases}
\en
The above bounds imply \refeq{Ezhatbd.99} for $|k| \leq (2L)^{-1}$.  
The case $|k| > (2L)^{-1}$ is bounded simply as
\eq
		|\hat{E}_{z}(k)| \leq \sum_{x} | E_{z}(x) | = O(\gamma), 
\en 
which satisfies \refeq{Ezhatbd.99} for $|k| > (2L)^{-1}$. 
\end{proof}

\subsection{Proof of Proposition~\ref{lem-C}}
\label{sub-errormain}

In this section, we prove Proposition~\ref{lem-C}.
It suffices to consider the case of small $\alpha$.  
We begin by proving the $x$-dependent bound, which is valid for all $x$.
The uniform bound, valid for $x \neq 0$, will follow immediately from this
proof.
The proof is divided into three cases,
according to the value of $x$.  We prove the $x$-dependent bound first assuming  
$\kappa \neq 2$, and comment on the minor modifications for $\kappa =2$ 
at the end.  

\bigskip\noindent \emph{Case~1:  $x = 0$.}
The uniform bound \refeq{CL-uniA1} on $S_{\mu_{z}}(x)$ implies that
\eq
\lbeq{5-Epbd.0}
    |(E_{z}*S_{\mu_{z}})(0)| \leq |E_{z}(0)| + O(L^{-d}) \sum_y |E_{z}(y)| .
\en
Lemma~\ref{lem-Epbd} implies $\sum_{y} |E_{z}(y)| = O(\gamma)$, and
hence \refeq{5-Epbd.0} is $O(\gamma)$ and satisfies
\refeq{E2res.1}.

\bigskip\noindent \emph{Case~2:  $0 < |x| \leq L^{1+\alpha/(d+\kappa_2)}$.}
For arbitrary $x \neq 0$, it follows from 
Lemma~\ref{lem-Epbd} and \refeq{CL-uniA1}
that
\eqalign
\lbeq{5-Epbd.2}
    (E_{z}*S_{\mu_{z}})(x)
    & = E_{z}(x) S_{\mu_{z}}(0)
    + \sum_{y: y \neq x} E_{z}(y) S_{\mu_{z}}(x-y)
    \nnb
    &
    = O(\gamma L^{-d}) +
    \sum_{y} |E_{z}(y)| \, O(L^{-d})
    = O(\gamma L^{-d}) .
\enalign
This proves the first bound of \refeq{E2res.1}.
Also, when $0 < |x| \leq L^{1+\alpha/(d+\kappa_2)}$, \refeq{5-Epbd.2} implies
\eq
    | (E_{z}*S_{\mu_z})(x) | = O(\gamma L^{-d})
    = \frac{|x|^{d+\kappa_2-\alpha}}{L^{d}} 
    O \Bigl ( \frac{\gamma}{|x|^{d+\kappa_2-\alpha}} \Bigr )
    \leq
    O \Bigl ( \frac{\gamma L^{\kappa_2}}
    {|x|^{d+\kappa_2-\alpha}} \Bigr ) .
\en

\bigskip\noindent \emph{Case~3:  $|x| > L^{1+\alpha/(d+\kappa_2)}$.}
We fix
$T = (\frac{|x|}{2\sigma})^{2-2\alpha/(d+\kappa_2)}$,
which is equal to $T_{x/2}$ of \refeq{ab-T-choice} with a smaller
$\alpha$.  We then define $X_1$ and $X_2$ by
\eq
    (E_{z}*S_{\mu_{z}})(x)
    = \sum_{y} E_{z}(x-y) S^{<}_{\mu_z}(y;T)  +
    \sum_{y} E_{z}(x-y) S^{>}_{\mu_z}(y;T)
    = X_{1} + X_{2}.
\en
The contribution $X_{1}$ is further divided as
\eq
    X_{1}
    = \sum_{y: |y| \leq |x|/2} E_{z}(x-y) S^{<}_{\mu_z}(y;T)
    + \sum_{y: |y| > |x|/2} E_{z}(x-y) S^{<}_{\mu_z}(y;T)
    = X_{11} + X_{12}.
\en
It remains to estimate $X_{11}$, $X_{12}$ and $X_2$.

For $X_{12}$, by our
choice of $T$ we can use \refeq{Cmusmalltbd}.  Since 
$\sum_{y} | E_{z}(y) | = O(\gamma)$, we obtain
\eq
\lbeq{X12z}
    | X_{12} | \leq \sum_{y: |y| > |x|/2} | E_{z} (x-y) |
    \frac{1} {(|y|+1)^{d+2}}
    \leq \sum_{y}| E_{z} (x-y) | 
    O \Bigl ( \frac{1}{|x|^{d+2} }\Bigr )
    =   O \Bigl ( \frac{\gamma}{|x|^{d+2} }\Bigr ).
\en
For $X_{11}$, we
use \refeq{CL-uni} to obtain
\eq
    S^{<}_{\mu_z}(y; T) \leq S_{\mu_z}(y) \leq \delta_{0,y} +
    O \Bigl ( \frac{1}{L^{2-\alpha} (|y|+1)^{d-2}} \Bigr ) .
\en
Since $|x-y| \geq |x|/2 \geq 2L$ (for large $L$), the third bound of 
Lemma~\ref{lem-Epbd} gives  
\eq
\lbeq{X11z}
    | X_{11} | \leq \biggl [
    1 + \sum_{y: |y| \leq |x|/2}
    O \Bigl ( \frac{1}{L^{2-\alpha} (|y|+1)^{d-2}} \Bigr )
    \biggr ]
    O \Bigl ( \frac{\gamma}{|x|^{d+2+\kappa}} \Bigr )
    \leq
    O \Bigl ( \frac{\gamma}{|x|^{d+\kappa}} \Bigr ).
\en

To control $X_{2}$, we use the integral representation \refeq{G><-def} for
$S^{>}_{\mu_z}$ to write
\eq
    X_{2} = (E_{z}*S^{>}_{\mu_z})(x) =  \int_{T}^{\infty} dt
    (I_{t, \mu_z} * E_{z})(x)
    = \int_{T}^{\infty} dt
    \int_{[-\pi,\pi]^d} \ddk e^{-i k \cdot x} \hat{I}_{t, \mu_z}(k) 
    \hat{E}_{z}(k).
\en
By \refeq{Idef}, $\hat{I}_{t,\mu_z}(k) = e^{-t[1-\mu_z\Dhat(k)]}$.
Since $1-\mu_z\Dhat \geq [1-\Dhat]/2$, it follows from 
Lemma~\ref{lem-Ezhatbd} that
\eq
\lbeq{63.11}
    | X_2 |
    \leq
    \int_{T}^{\infty}  \! \! dt
    \int_{[-\pi,\pi]^d} \ddk e^{-t[1-\Dhat(k)]/2}\,
    c\gamma  L^{2+\kappa_2} |k|^{2+\kappa_2}.
\en
We divide the $k$-integral according to whether $|k|$ is greater or
less than $L^{-1}$, as in the analysis of
$I^{(4)}_{t}(x)$ in \refeq{ab-I7.1}--\refeq{ab-I7.2}.  This gives
\eqalign
\lbeq{63.12}
    \int_{[-\pi,\pi]^d} \ddk e^{-t[1-\Dhat(k)]/2}\, |k|^{2+\kappa_2}
    &
    \leq O(L^{-(d+2+\kappa_2)}) t^{-(d+2+\kappa_2)/2} + O(e^{-\delta_{3} t}).
\enalign
The second error term can be absorbed into the first
for $t \geq T$ and sufficiently large $L$,
by arguing exactly as was done for \refeq{ab-RB1bd-1}.
Performing the $t$-integral then gives
\eqalign
\lbeq{X2z}
    | X_2 |
    & \leq
    O(\gamma  L^{-d}) \int_{T}^{\infty}
    t^{-(d+2+\kappa_2)/2}  \, dt
    \leq
     \frac{O(\gamma L^{\kappa_2 - \alpha})}{|x|^{d+\kappa_2-\alpha}}
    .
\enalign
Combining \refeq{X12z}, \refeq{X11z} and \refeq{X2z} gives the desired
estimate
\eq
\lbeq{63.14}
	\bigl | (E_z * S_{\mu_z})(x) \bigr |
	\leq
	O\big( 
	\frac{\gamma L^{\kappa_2 - \alpha}}{|x|^{d+\kappa_2-\alpha}}
	\big).
\en
for $|x| > L^{1+\alpha/(d+\kappa_2)}$.

\smallskip

The case $\kappa =2$ adds extra factors $\log |k|^{-1}$, 
$|\log t|$, and $\log(|x|/L)$ in \refeq{63.11}, 
\refeq{63.12}, \refeq{X2z}, and \refeq{63.14}.  However, the extra logarithm
of \refeq{63.14} can be absorbed in $|x|^{d+\kappa_2-\alpha}$, 
by slightly increasing the exponent $\alpha$ of \refeq{E2res.1}. 
\qed


\section*{Acknowledgements}
This work began while all three authors were visiting
the Fields Institute and was
supported in part by NSERC of Canada.
The work of T.H.\ was also supported in part by a \kakenhi\ of \monbushou\ 
of Japan.


\begin{thebibliography}{10}

\bibitem{AMAH90}
J.~Adler, Y.~Meir, A.~Aharony, and A.B. Harris.
\newblock Series study of percolation moments in general dimension.
\newblock {\em Phys.\ Rev.\ B}, {\bf 41}:9183--9206, (1990).

\bibitem{Aize97}
M.~Aizenman.
\newblock On the number of incipient spanning clusters.
\newblock {\em Nucl. Phys. B [FS]}, {\bf 485}:551--582, (1997).

\bibitem{AB87}
M.~Aizenman and D.J. Barsky.
\newblock Sharpness of the phase transition in percolation models.
\newblock {\em Commun. Math. Phys.}, {\bf 108}:489--526, (1987).

\bibitem{AKN87}
M.~Aizenman, H.~Kesten, and C.M. Newman.
\newblock Uniqueness of the infinite cluster and continuity of connectivity
  functions for short and long range percolation.
\newblock {\em Commun. Math. Phys.}, {\bf 111}:505--531, (1987).

\bibitem{AN84}
M.~Aizenman and C.M. Newman.
\newblock Tree graph inequalities and critical behavior in percolation models.
\newblock {\em J. Stat. Phys.}, {\bf 36}:107--143, (1984).

\bibitem{BA91}
D.J. Barsky and M.~Aizenman.
\newblock Percolation critical exponents under the triangle condition.
\newblock {\em Ann. Probab.}, {\bf 19}:1520--1536, (1991).

\bibitem{BFG86}
A.~Bovier, J.~Fr\"{o}hlich, and U.~Glaus.
\newblock Branched polymers and dimensional reduction.
\newblock In K.~Osterwalder and R.~Stora, editors, {\em Critical Phenomena,
  Random Systems, Gauge Theories}, Amsterdam, (1986). North-Holland.

\bibitem{BEI92}
D.~Brydges, S.N. Evans, and J.Z. Imbrie.
\newblock Self-avoiding walk on a hierarchical lattice in four dimensions.
\newblock {\em Ann. Probab.}, {\bf 20}:82--124, (1992).

\bibitem{BS85}
D.C. Brydges and T.~Spencer.
\newblock Self-avoiding walk in 5 or more dimensions.
\newblock {\em Commun. Math. Phys.}, {\bf 97}:125--148, (1985).

\bibitem{DS98}
E.~Derbez and G.~Slade.
\newblock The scaling limit of lattice trees in high dimensions.
\newblock {\em Commun.\ Math.\ Phys.}, {\bf 193}:69--104, (1998).

\bibitem{Grim99}
G.~Grimmett.
\newblock {\em Percolation}.
\newblock Springer, Berlin, 2nd edition, (1999).

\bibitem{HM54}
J.M. Hammersley and K.W. Morton.
\newblock Poor man's {Monte} {Carlo}.
\newblock {\em J. Roy. Stat. Soc. B}, {\bf 16}:23--38, (1954).

\bibitem{Hara00}
T.~Hara.
\newblock Critical two-point functions for nearest-neighbour high-dimensional
  self-avoiding walk and percolation.
\newblock In preparation.

\bibitem{HS90a}
T.~Hara and G.~Slade.
\newblock Mean-field critical behaviour for percolation in high dimensions.
\newblock {\em Commun. Math. Phys.}, {\bf 128}:333--391, (1990).

\bibitem{HS90b}
T.~Hara and G.~Slade.
\newblock On the upper critical dimension of lattice trees and lattice animals.
\newblock {\em J. Stat. Phys.}, {\bf 59}:1469--1510, (1990).

\bibitem{HS94}
T.~Hara and G.~Slade.
\newblock Mean-field behaviour and the lace expansion.
\newblock In G.\ Grimmett, editor, {\em Probability and Phase Transition},
  Dordrecht, (1994). Kluwer.

\bibitem{HS95}
T.~Hara and G.~Slade.
\newblock The self-avoiding-walk and percolation critical points in high
  dimensions.
\newblock {\em Combinatorics, Probability and Computing}, {\bf 4}:197--215,
  (1995).

\bibitem{HS00b}
T.~Hara and G.~Slade.
\newblock The scaling limit of the incipient infinite cluster in
  high-dimensional percolation. {II}. {Integrated} super-{Brownian} excursion.
\newblock {\em J.\ Math.\ Phys.}, {\bf 41}:1244--1293, (2000).

\bibitem{HS01a}
R.~van~der Hofstad and G.~Slade.
\newblock A generalised inductive approach to the lace expansion.
\newblock Preprint, (2000).

\bibitem{HS01b}
R.~van~der Hofstad and G.~Slade.
\newblock The lace expansion on a tree with application to networks of
  self-avoiding walks.
\newblock In preparation.

\bibitem{Hugh95}
B.D. Hughes.
\newblock {\em Random Walks and Random Environments}, volume 1: Random Walks.
\newblock Oxford University Press, Oxford, (1995).

\bibitem{Hugh96}
B.D. Hughes.
\newblock {\em Random Walks and Random Environments}, volume 2: Random
  Environments.
\newblock Oxford University Press, Oxford, (1996).

\bibitem{IM94}
D.~Iagolnitzer and J.~Magnen.
\newblock Polymers in a weak random potential in dimension four: rigorous
  renormalization group analysis.
\newblock {\em Commun. Math. Phys.}, {\bf 162}:85--121, (1994).

\bibitem{Kest82}
H.~Kesten.
\newblock {\em Percolation Theory for Mathematicians}.
\newblock Birkh\"{a}user, Boston, (1982).

\bibitem{Klar67}
D.A. Klarner.
\newblock Cell growth problems.
\newblock {\em Canad. J. Math.}, {\bf 19}:851--863, (1967).

\bibitem{Klei81}
D.J. Klein.
\newblock Rigorous results for branched polymer models with excluded volume.
\newblock {\em J. Chem. Phys.}, {\bf 75}:5186--5189, (1981).

\bibitem{LI79}
T.C. Lubensky and J.~Isaacson.
\newblock Statistics of lattice animals and dilute branched polymers.
\newblock {\em Phys. Rev.}, {\bf A20}:2130--2146, (1979).

\bibitem{MS93}
N.~Madras and G.~Slade.
\newblock {\em The Self-Avoiding Walk}.
\newblock Birkh{\"a}user, Boston, (1993).

\bibitem{Mens86}
M.V. Menshikov.
\newblock Coincidence of critical points in percolation problems.
\newblock {\em Soviet Mathematics, Doklady}, {\bf 33}:856--859, (1986).

\bibitem{NY93}
B.G. Nguyen and W-S. Yang.
\newblock Triangle condition for oriented percolation in high dimensions.
\newblock {\em Ann.\ Probab.}, {\bf 21}:1809--1844, (1993).

\bibitem{PS81}
G.~Parisi and N.~Sourlas.
\newblock Critical behavior of branched polymers and the {L}ee--{Y}ang edge
  singularity.
\newblock {\em Phys. Rev. Lett.}, {\bf 46}:871--874, (1981).

\bibitem{Penr94}
M.D. Penrose.
\newblock Self-avoiding walks and trees in spread-out lattices.
\newblock {\em J. Stat. Phys.}, {\bf 77}:3--15, (1994).

\bibitem{Spit76}
F.~Spitzer.
\newblock {\em Principles of Random Walk}.
\newblock Springer, New York, 2nd edition, (1976).

\bibitem{TH87}
H.~Tasaki and T.~Hara.
\newblock Critical behaviour in a system of branched polymers.
\newblock {\em Prog. Theor. Phys. Suppl.}, {\bf 92}:14--25, (1987).

\bibitem{Uchi98}
K.~Uchiyama.
\newblock Green's functions for random walks on ${Z}^{N}$.
\newblock {\em Proc.\ London Math.\ Soc.}, {\bf 77}:215--240, (1998).

\end{thebibliography}
%

\end{document}